\documentclass[10pt,a4paper]{article}
\usepackage[english]{babel}
\usepackage[latin1]{inputenc}
\usepackage{amsfonts,amsbsy,bm,euscript,mathrsfs}
\usepackage{amssymb,stmaryrd,faktor}
\usepackage[tbtags]{amsmath}
\usepackage{bbm}
\usepackage{graphicx}
\usepackage[title,titletoc]{appendix}
\usepackage[bookmarks=true,colorlinks=true,linkcolor=blue,citecolor=blue,urlcolor=blue,bookmarksnumbered]{hyperref}

\usepackage{dsfont}
\usepackage{collref}
\usepackage{lmodern}
\usepackage{mathrsfs}
\usepackage{mathtools}
\usepackage{bbm}
\usepackage{braket}
\usepackage{slashed}

\usepackage{graphicx}
\usepackage{booktabs}
\usepackage{subfig}
\usepackage{tikz}
\usetikzlibrary{plotmarks,calc,decorations,decorations.pathmorphing}

\numberwithin{equation}{section}

\makeatletter
\renewcommand\section{\@startsection {section}{1}{\z@}
{-3.5ex \@plus -1ex \@minus -.2ex}
{2.3ex \@plus.2ex}
{\normalfont\Large\bfseries}}
\renewcommand\subsection{\@startsection{subsection}{2}{\z@}
{-3.25ex\@plus -1ex \@minus -.2ex}
{1.5ex \@plus.2ex}
{\normalfont\large\bfseries}}
\makeatother

\newcommand{\arXivlink}[1]{\href{http://arXiv.org/abs/#1}{arXiv:#1}}

\newcommand{\alg}[1]{\mathfrak{#1}}

\usepackage{blkarray}

\begin{document}
%%%%%%%%%%%%%%%%%%%%%%%%%%%%%%%%%%%%%%

\thispagestyle{empty}
\begin{flushright}\footnotesize\ttfamily
DMUS-MP-21/08
\end{flushright}
\vspace{2em}

\begin{center}

{\Large\bf \vspace{0.2cm}
{\color{black} \large A study of integrable form factors in massless relativistic $AdS_3$}} %v2ref
\vspace{1.5cm}

\textrm{Alessandro Torrielli\footnote{\texttt{a.torrielli@surrey.ac.uk}}}

\vspace{2em}

\vspace{1em}
\begingroup\itshape
Department of Mathematics, University of Surrey, Guildford, GU2 7XH, UK
\par\endgroup

\end{center}

\vspace{2em}

\begin{abstract}\noindent %v2ref
We show that the massless integrable sector of the $AdS_3 \times S^3 \times T^4$ superstring theory, which admits a non-trivial relativistic limit, provides a setting where it is possible to determine exact minimal solutions to the form factor axioms, in integral form, based on analyticity considerations, along the same lines of ordinary relativistic integrable models. We construct in full detail the formulas for the two- and three-particle case, and show the similarities as well as the differences with respect to the off-shell Bethe ansatz procedure of Babujian et al. We show that our expressions pass a series of non-trivial consistency checks which are substantially more involved than in the traditional case. We speculate on the problems concerned in a possible generalisation to an arbitrary number of particles, and on a possible connection with the hexagon programme.

\end{abstract}

\newpage

\overfullrule=0pt
\parskip=2pt
\parindent=12pt
\headheight=0.0in \headsep=0.0in \topmargin=0.0in \oddsidemargin=0in

\vspace{-3cm}
\thispagestyle{empty}
\vspace{-1cm}

\tableofcontents

\setcounter{footnote}{0}

\section{\label{sec:Intro}Introduction}

\subsection{Integrability in $AdS_3$ and $AdS_2$ backgrounds}

Integrability in $AdS_3 \times S^3 \times S^3 \times S^1$ and $AdS_3 \times S^3 \times T^4$ string backgrounds  \cite{Bogdan,Sundin:2012gc} (see also \cite{rev3,Borsato:2016hud}), has advanced a great deal, taking the moves from the analysis of the infinite-length solution of the $AdS_5/CFT_4$ spin-chain. A significant amount of work has been produced \cite{OhlssonSax:2011ms,seealso3,Borsato:2012ss,Borsato:2013qpa,Borsato:2013hoa,Rughoonauth:2012qd,PerLinus,CompleteT4,Borsato:2015mma,Beccaria:2012kb,Sundin:2013ypa,Bianchi:2013nra,Bianchi:2013nra1,Bianchi:2013nra2}. The new features brought into the game by the presence of the massless modes have led to the development of a new paradigm \cite{Sax:2012jv,Borsato:2016xns,Sax:2014mea,Baggio:2017kza} reminiscent of the connection between the study of massless $S$-matrices and conformal field theories \cite{Zamol2,Fendley:1993jh,FI1,FI2,Fioravanti:1996rz,DiegoBogdanAle}. This proceeds into a region beyond the realm of applicability of perturbation theory \cite{Lloyd:2013wza,Abbott:2012dd,Abbott:2014rca,MI,Abbott:2020jaa} (see also \cite{Eberhardt:2017fsi,Gaber1,Gaber2,Gaber3,MCAAP,MCAAP2,Gaber4,Gaber5,
GaberdielUltimo,Dei:2021xgh,Prin,Prin1,Abbott:2015mla,Per9,Hoare:2018jim,AntonioMartin,Regelskis:2015xxa,JuanMiguelAle}).

The massless sector shows very clearly the traits of the integrable string sigma model as a quantum-group type deformation of $1+1$-dimensional Poincar\'e supersymmetry. This idea was carried over in \cite{CesarRafa,Charles} to the case of $AdS_5$, see also \cite{Riccardo} in the same context. In the situation where massless modes are present, such as the one offered by $AdS_3$, one can push this approach to a greater extent \cite{JoakimAle,BorStromTorri,Andrea,Garcia:2020vbz} and one can prove \cite{gamma1,gamma2} that there is a change of variables which recasts the massless non-relativistic $S$-matrix, inclusive of its dressing factor, in a manifest difference form. For right-right and left-left movers - which is still a well-defined concept even away from the relativistic limit \cite{Borsato:2016xns} - this form is truly the exact same functional form which one observes in the (non-trivial) BMN limit \cite{DiegoBogdanAle}. The construction of the massless thermodynamic Bethe ansatz (TBA) performed in \cite{DiegoBogdanAle} is then directly adapted to the complete (massless) non-relativistic $S$-matrix, as done in \cite{gamma2}. 

The $AdS_3$ background is one which permits an extension to mixed fluxes \cite{Cagnazzo:2012se,s1,s2,Babichenko:2014yaa,seealso12}, and this changes the usual magnon dispersion relation \cite{ArkadyBenStepanchuk,Lloyd:2014bsa} - see \cite{Gaber3} for a complete analysis of the moduli space of the theory. A special relativistic limit has been introduced in \cite{gamma2}, and in that instance another example of nontrivial scattering for right-right and left-left movers on the world-sheet was obtained, describing a class of conformal field theories via their exact TBA description. A significant body of work on this issue and on related deformations has now become available \cite{Baggio,Majumder:2021zkr}.

\subsection{Form factors}

The calculation of form factors is an essential component of the so-called {\it bootstrap programme} to completely solve integrable quantum field theories. This programme of course begins with the establishment of the exact $S$-matrix and the derivation of the spectrum of states, which is the content of the literature which we have quoted in the previous subsection, and then continues with the form factor analysis \cite{Karo1}. From the knowledge of all the form factors of the theory, one gains access to the $n$-point correlators (Wightman functions). This is sufficient to define the theory, even if one did not have a Lagrangian to begin with. The Lagrangian, when present, does of course allow for a perturbative check of the results via Feynman-graph expansion. We refer to \cite{Babu} for an excellent review and an extensive list of references. 

In a relativistic field theory, such as the one we are dealing with in what follows, the $n$-particle form factor associated with a local operator $\cal{O}$ is the matrix element
\begin{equation}
\label{in}
F^{\cal{O}}_{\alpha_1 ... \alpha_n} (\theta_1,...,\theta_n) = \langle 0| {\cal{O}}(0)|\theta_1,...,\theta_n\rangle_{\alpha_1 ... \alpha_n}
\end{equation}
where $\theta_i$ is the rapidity of the $i$-th particle in the multi-particle {\it in} state appearing in (\ref{in}), and $\alpha_i$, respectively, label the internal degrees of freedom. For definiteness, we shall deal in the following with massless right-moving particles forming a boson-fermion doublet:
\begin{equation}
p_i = E_i = e^{\theta_i}, \qquad \alpha_i \in \{b,f\}, \qquad b = \mbox{boson}, \qquad f = \mbox{fermion}. 
\end{equation}
We have suppressed a mass scale $M$ in the dispersion relation since it will not play any role in the discussion. The theory will be at the critical point, and the only way the mass parameter features is when coupled with the natural finite cylinder length of the TBA \cite{DiegoBogdanAle}.   

The form factors are controlled by a series of conditions, which can be considered as the axioms of the form factor programme. We refer to \cite{Babu} and \cite{BabuF} for a complete list of such conditions, and restrict here to the ones which will be directly relevant to our case. In particular, no bound-states exist in the spectrum for massless particles, therefore we shall not need the condition on the residues of the form factors which are traditionally associated with bound states. Besides requiring relativistic invariance, we shall focus on the following conditions:
\begin{itemize}

\item {\it Permutation}
\begin{eqnarray}
&&F^{\cal{O}}_{\alpha_1 ... \alpha_{j-1} \, \beta_{j} \, \beta_{j+1} \, \alpha_{j+2} ... \alpha_n} (\theta_1,...,\theta_{j-1} , \theta_{j} , \theta_{j+1} , \theta_{j+2}, ... \theta_n) = \\
&&\qquad \qquad F^{\cal{O}}_{\alpha_1 ... \alpha_{j-1} \, \alpha_j \, \alpha_{j+1} \, \alpha_{j+2} ... \alpha_n} (\theta_1,...,\theta_{j-1} , \theta_{j+1} , \theta_j, \theta_{j+2}, ... \theta_n) \, S^{\alpha_j \alpha_{j+1}}_{\beta_j \beta_{j+1}} (\theta_j - \theta_{j+1}),\nonumber
\end{eqnarray}
also called Watson's equations, where the $S$-matrix entries are defined as
\begin{equation}
S : V_1 \otimes V_2 \longrightarrow V_2 \otimes V_1, \qquad S |v_\alpha (\theta_1) \rangle \otimes |v_\beta (\theta_2)\rangle = S^{\rho \sigma}_{\alpha \beta} (\theta_1 - \theta_2) |v_\rho (\theta_2) \rangle \otimes |v_\sigma (\theta_1)\rangle.
\end{equation}
In this paper we will often be working with the $R$-matrix, which is related to the $S$-matrix in the following fashion 
\begin{eqnarray}
S = \Pi \circ R,
\end{eqnarray}
where $\Pi$ is the graded permutation on states:
\begin{eqnarray}
\Pi |a\rangle \otimes |b\rangle = (-)^{\sigma} \, |b\rangle \otimes |a\rangle,
\end{eqnarray}
$\sigma$ being $1$ only when the two states are both fermions, and $0$ otherwise.
The $R$-matrix is a map between vector spaces
\begin{equation}
R : V_1 \otimes V_2 \longrightarrow V_1 \otimes V_2.
\end{equation}
\item {\it Periodicity}
\begin{eqnarray}
&&F^{\cal{O}}_{\alpha_1 \, \alpha_{2} ... \alpha_{n-1} \, \alpha_n} (\theta_1 + 2i \pi,\theta_{2}, ... \theta_{n-1}, \theta_n) = (-)^\sigma\, F^{\cal{O}}_{\alpha_2 \, \alpha_{3} ... \alpha_{n} \, \alpha_1} (\theta_2 , \theta_{3}, ... \theta_{n} , \theta_1),
\end{eqnarray}
where by $\sigma$ we shall always be denoting statistical factor appropriate to the situation - in this case it is obtained by permuting the first particle across the operator. All the form factors respect total fermion number conservation.
%\begin{eqnarray}
%(-)^\sigma = (-)^{\mbox{deg}(\alpha_1)\sum_{i=2}^n \mbox{deg}(\alpha_i)}=(-)^{\mbox{deg}(\alpha_1)\big[\mbox{deg}({\cal{O}})+1\big]}
%\end{eqnarray} 
%where deg is the fermionic degree: $0 \, \mbox{mod} \, 2$ for bosons and $1 \, \mbox{mod} \, 2$ for fermions.

\item {\it Lorentz boost}
\begin{eqnarray}
F^{\cal{O}}_{\alpha_1 \, \alpha_{2} ... \alpha_{n-1} \, \alpha_n} (\theta_1 + \Lambda,\theta_{2}+\Lambda, ... \theta_{n-1}+\Lambda, \theta_n+\Lambda) = e^{s \Lambda}F^{\cal{O}}_{\alpha_1 \, \alpha_{2} ... \alpha_{n-1} \, \alpha_n} (\theta_1,\theta_{2}, ... \theta_{n-1}, \theta_n), 
\end{eqnarray}
where $s$ is the spin of the operator $\cal{O}$.

\item {\it Residue at the kinematical singularities}

The form factors shall be meromorphic functions of their arguments, except for the presence of poles. The condition for the residue at the kinematical simple poles can be recast in the form
\begin{eqnarray}
\label{reso}
&&-\frac{i}{2} \, \mbox{Res}_{\theta_1 = \theta_2+i\pi} F^{\cal{O}}_{\bar{\alpha}_2 \, \alpha_{2} ... \alpha_{n-1} \, \alpha_n} (\theta_1,\theta_{2}, ... \theta_{n-1}, \theta_n) = \\
&&\qquad \qquad {\bf C}_{\bar{\alpha}_2\beta_2} \, \Big[\mathbbmss{1} - (-)^\sigma S_{\alpha_n \rho_{n-3}}^{\beta_n \beta_2}(\theta_2 - \theta_n)...S^{\beta_3\rho_1}_{\alpha_3\alpha_2}(\theta_2-\theta_3)\Bigg]F^{\cal{O}}_{\beta_3 \, \beta_{4} ... \beta_{n-1} \, \beta_n} (\theta_3,... \theta_n), \nonumber
\end{eqnarray}
where $\bar{\alpha}$ denotes the anti-particle of $\alpha$ and ${\bf C}_{\alpha \beta}$ is the charge-conjugation matrix, as we will clarify in greater detail later on. The symbol $\mathbbmss{1}$ denotes the same tensor as the second term inside the brackets where instead of the $S$-matrices one has placed all Kronecker deltas. The statistical factor $(-)^\sigma$ in (\ref{reso}) is the one picked up by the left-most state cycling through the operator, and it will play an important role in our analysis. 

\end{itemize}

In a massive theory there are also bound-state singularities, and an associated equation that the form factors have to satisfy, however in our massless case we shall not need to consider them \cite{DelfinoMussSimo}. The kinematical singularities are still relevant for scattering between purely left movers, and scattering between purely right movers, since these $S$-matrices typically have the same functional form
%\footnote{As discussed in \cite{DiegoBogdanAle}, our $S$-matrix resembles the sine-Gordon $S$-matrix at the supersymmetric point if it were not for some crucial factors of $i$ and minuses that come from a different statistics of the particles in the two situations. The supersymmetric point is also in the repulsive regime, which confirms the absence of bound states.} 
as the massive ones \cite{Fendley:1993jh}. We are not entirely sure whether this necessarily has to hold for our situation as well, given that the massive theory we can compare to in the context of $AdS_3/CFT_2$ is non-relativistic \cite{rev3}. Strictly speaking, therefore, we might not have any term of comparison for our kinematical singularities. Nevertheless, we will have them in our formulas and they will respect the requirement (\ref{reso}), in particular we will see that they are rather crucial in our analysis. For mixed left-right scattering (which we do not have) the kinematical singularities would most likely not be required, since the traditional reasoning behind them would not apply in the massless left-right and right-left kinematics.  

The power, and at the same time the drawback, of the programme so established is that one does not know exactly which operator one is describing, beyond a few basic structural requisites such as for instance the spin. The axioms are supposed to hold in general, and one proceeds to find solutions to these requirement as one would do to a purely mathematical problem. It is then rather a painful work to try and connect the solutions found to the actual operators in the theory, and one has to use a series of hints, collateral arguments and auxiliary intuition to precisely connect the abstract formulas to a specific operator which may exist in the model.   

The bootstrap programme is thoroughly described in \cite{Karo} (see also \cite{Weisz:1977ij}), with the determination of a set of recursive equations satified by the form factors. There is a gigantic literature which took the moves from there, a cross-section of which is contained in \cite{Babu}. The calculation of Wightman functions, using the insertion of the resolution of the identity and the summation over intermediate states, remains in almost all cases a formidable task, with significant progress having been made in a subset of models. We shall refer to \cite{Babu} and \cite{Mussardo} for the appropriate references - see also the recent \cite{Sergey} for a detailed exposition.

The form factor programme has been extensively studied in the context of $AdS/CFT$, focusing primarily on $AdS_5/CFT_4$, where the complications of the supersymmetric setup make it a steep climb. Remarkable progress has been made in this and closely related setup \cite{Thomas}. No systematic attempt along these strictly traditional lines has yet been made in $AdS_3/CFT_2$ and $AdS_2/CFT_1$, mostly because, at odds with the $S$-matrix programme, exact solutions are very difficult to come by. However recently, the very powerful approach of hexagon form factors \cite{Basso:2013vsa} - see also \cite{hexagon} - has been extended to $AdS_3$ \cite{Eden:2021xhe}. Such an approach is very much related to the one which we adopt here, although it follows a different path. We hope that in the future it will be possible to compare the results of \cite{Eden:2021xhe} with the route which we are taking in this paper. To be able to draw a direct analogy, a sort of {\it conformal} hexagon might have to be considered, tailored to the relativistic (BMN) limit of the massless purely left-left and right-right moving $AdS_3$ $S$-matrix.

\subsection{This paper}

This is what makes the relativistic massless limit particularly interesting and unique in this respect. What we have set out to do in this paper is to show that in this particular limit one recovers the nice features of relativistic models, {\it i.e.} one can solve for the form factors exactly and obtain close formulas with standard analycity properties following the traditional approach. The crucial feature will in fact be that we can explicitly provide the analytic continuation of the formulas involved and proceed with manipulations which are immune from issues of branch cuts. We recover in this way the inner simplicity of the analytic methods which are at work in the relativistic bootstrap.

We derive here the right-right (and the identical left-left) two- and three-particle exact minimal form factor solutions, and study their properties. Minimal solutions for the form factors are very difficult to obtain, as it is the case in other supersymmetric settings \cite{superMussardo}. As we have already remarked, the form factors we derive in this paper should pertain to a critical theory - see also \cite{masslessform}, and the consequent study of correlators should fall into the category of CFT correlators and reveal specific properties of this critical point. We will find that the ordinary approach is confronted with the convoluted spectrum of particles and anti-particles, and that organising the formulas and proving the form-factor requirements rapidly becomes a rather non-trivial task, which we will try to describe in as much detail as we will be able to. We will point out at each stage the novelties which we shall need to deal with, and how our formulas will have to be quite non-trivially adapted to the scattering theory at hand.

The organisation of the work is as follows: in section \ref{sec:Dressing} we will review the scattering theory relevant to our problem, and summarise the analytic properties of the dressing factor. In section \ref{sec:Min} we will derive the minimal two-particle form factor following Karowski and Weisz. In section \ref{sec:Minimo} we will display the integral representation of the dressing factor which is most suitable to our goals. In section \ref{sec:Watson} we will give both an integral representation for the minimal form factor, and a manifestly meromorphic representation in terms of products of gamma functions. We will obtain a complete portrait of the singularities of the minimal solution, which will be of constant use in the later sections. In section \ref{fourfour} we shall begin to study the three-particle form factors. We will construct the off-shell Bethe states necessary to employ the formulas of \cite{BabuF}, and provide the integrals for the three basic form factors, which will constitute the basis on which to construct all the other ones. In particular, in section \ref{four1} we will discuss the prescriptions for the integration contour and the singularities of the integrand. In section \ref{graphs} we shall provide some graphs for such singularities and contours. In section \ref{conv} we will discuss the issue of convergence.       
In section \ref{prop} we will provide the proofs of how the three basic functions respect the form factor axioms. We will begin with permutation in section \ref{permutalo}, then proceed to periodicity in section \ref{perrio}, and boost symmetry in section \ref{boos}. The kinematical singularities will be extensively dealt with in section \ref{singo}. In section \ref{antipo} we will introduce anti-particles, give the full set of $S$-matrix entries and their dressing factors, and construct mixed-representations minimal form factors. We will then discuss the relative statistics of the particles in connection with the periodicity requirement. In section \ref{resd} we will perform the resiude calculation necessary for the requirement (\ref{reso}), by tacking the poles which are pinching the contour as $\theta_1 \to \theta_2 + i \pi$, as instructed by \cite{BabuF}. In section \ref{specola} we will speculate on a possible generalisation to an arbitrary number of particles of our formulas, and then in section \ref{dufe} we will go back to the two-particle case in this light. We will provide some remarks and more words of caution in the conclusions, while in the appendix we will present a curious extremely simple solution to the form factor axioms without kinematical singularities - whose presence in the massless case, as we will amply discussed, is in any case a subtle matter.    

\medskip

{\it Note Added:} Some time after this paper appeared on the archive, three papers were posted \cite{AleSSergey} providing an extremely interesting revisitation of the entire scattering theory pertaining to the $AdS_3 \times S^3 \times T^4$ superstring theory background. For what concerns the massless sector in its same-chirality BMN limit, which is what this paper is focusing on, the new proposal differs in a number of aspects, which we think we can distill as follows. One general remark is that the Hopf algebra which \cite{AleSSergey} utilises, including the antipode realisation via the charge conjugation matrix, is constructed directly on two copies of $\mathfrak{psu}(1|1)^2$, while we rely on a Hopf algebra which is mathematically consistent on its own right already within each individual copy of $\mathfrak{psu}(1|1)^2$. The condition which we have to adopt in order to be able to focus on a single copy is that our charge-conjugation matrix might not satisfy the additional condition which is required of it by \cite{AleSSergey}. We also allow the antiparticle representation to be isomorphic but not exactly identical to the particle one, which allows us to accommodate two (in fact four) in principle independent dressing factors - which however turn out to be consistently chosen to be proportional to each other. The mixed crossing relations fix the proportionality factor to be a factor of $i$. This precisely accounts for the factor of $i$  difference on the r.h.s. of the crossing relation, which was noticed in \cite{AleSSergey}. This is a reason for \cite{AleSSergey} to supplement the sine-Gordon factor with an Ising-type dressing factor, which will survive the BMN limit in the same-chirality case. Such factor only makes sense analytically ({\it i.e.} appears with an integer power) when dealing with the two copies at once. An additional distinction is that the sine-Gordon factor appears in \cite{AleSSergey} with an inverse power w.r.t. what we have here. This is once again due to a different sign on the r.h.s. of the crossing relation, connected with the different assumptions in building the Hopf-algebra/charge-conjugation matrix. Yet another difference is that the massless dressing factor is equipped with an extra CDD factor, given by a suitably regularised/prescribed BES integral. This is initially not (immediately recast, whether at all) in difference-form, however its BMN limit in the same-chirality case, if non-trivial, must be of difference form. We have not worked out what this limit is, and whether it will reduce in some way to some integer powers of the sine-Gordon factor. 

We believe that, in the absence of any tests - which is the problem we will be highlighting further in the Conclusions - it is still to a certain extent open to debate what the final structure will be. We have decided to still present here the form-factor analysis performed with the old proposal, whether or not the very interesting new proposal of \cite{AleSSergey} be nearer the mark in the context of the real superstring integrable model. Our choice may serve a number of purposes. First, the theory we present here is mathematically consistent, even if one were for a moment to disregard any string theory origin. Whether or not it describes a worldsheet CFT, it does describe a CFT in terms of a massless integrable scattering, and it has therefore in our opinion intrinsic value. It is simpler (though complicated as it is) then dealing with two copies at once, although that could probably be arranged. In this respect, the formalism we develop in this paper provides the basic building blocks for any future attempt to extract the form factors for a theory based directly on the two copies. Adding the Ising factor might not be too difficult, as it has an enormously simpler functional form than its sine-Gordon counterpart, and the minimal solution is well-known. The relative particle statistics will most likely suitably rearrange itself when including the two copies and all the appropriate dressing factors. Finally, different integer powers in the sine-Gordon part might be manageable starting from our formulas with minor tweaks, once we have derived the minimal solution to Watson's equations - which we will do shortly. 

\section{\label{sec:Dressing}Massless relativistic $AdS_3$ $S$-matrix}

We begin by reviewing the basic scattering theory of the model. We will start with the $R$-matrix, which takes the difference form \cite{DiegoBogdanAle}
\begin{equation}\label{eq:RLLlimtheta}
  \begin{aligned}
    &R |b\rangle \otimes |b\rangle\ = {} \, |b\rangle \otimes |b\rangle, \\
    &R |b\rangle \otimes |f\rangle\ = -\tanh\frac{\theta}{2} |b\rangle \otimes |f\rangle + {\rm sech}\frac{\theta}{2}|f\rangle \otimes |b\rangle, \\
    &R |f\rangle \otimes |b\rangle\ = {} \, {\rm sech}\frac{\theta}{2} |b\rangle \otimes |f\rangle + \tanh\frac{\theta}{2} |f\rangle \otimes |b\rangle, \\
    &R |f\rangle \otimes |f\rangle\ = - {} \, |f\rangle \otimes |f\rangle\,,
\end{aligned}
\end{equation}
where 
\begin{equation}
\label{eq:diff-form}
\theta \equiv \theta_1 - \theta_2\,
\end{equation}
and $(b,f)$ are our (boson,fermion) doublet, whose bare statistics is $(-)^{(0,1)}$ as elements of the graded vector space $\mathbbmss{C}^{1|1}$.
One should then multiply $R$ by the dressing factor
\begin{eqnarray}
\label{zamo}
\Phi(\theta) = \prod_{\ell=1}^\infty \frac{\Gamma^2(\ell - \tau) \, \Gamma(\frac{1}{2} + \ell + \tau) \,\Gamma(- \frac{1}{2} + \ell + \tau)}{\Gamma^2(\ell + \tau) \, \Gamma(\frac{1}{2} + \ell - \tau) \,\Gamma(- \frac{1}{2} + \ell - \tau)},
\end{eqnarray}
where  
\begin{eqnarray}
\tau \equiv \frac{\theta}{2 \pi i}.
\end{eqnarray}
Expression (\ref{zamo}) solves the crossing equation
\begin{eqnarray}
\label{crosszamo}
\Phi(\theta) \, \Phi(\theta + i \pi) = i \tanh \frac{\theta}{2}\,,
\end{eqnarray}
As discussed in \cite{DiegoBogdanAle}, this is Zamolodchikov's famous dressing factor for soliton-antisoliton scattering at the supersymmetric point, which is in the repulsive regime - accordingly, there are no bound-states poles in the physical strip $\mbox{Im}(\theta)\in [0,\pi]$, as it should be for a massless two-body S-matrix. 

Despite the similarities, the $S$-matrix $S = \Pi \circ R$ differs from the ${\cal{N}}=2$ $S$-matrix, studied in particular by \cite{FI1,FI2}, in a number of ways, most importantly it differs from it in its entries by some crucial factors of $i$ and $-1$, which are related to a different statistics of the scattering particles. This is encoded in a different coproduct for the supercharges, as explained in \cite{DiegoBogdanAle}. This also means that the physics is expected to be quite different with respect to the one analysed in \cite{FI1,FI2}. We will see in later sections how rather more convoluted in fact the model under consideration here will become.  

The function $\Phi(\theta)$ is meromorphic in the entire complex plane, with poles and zeros on the imaginary axis. For future purposes, let us list the poles and zeros:

\begin{enumerate}

\item {\it Poles}

The poles come from the poles of the gamma functions at the numerator, namely:
\begin{eqnarray}
\label{polZamo}
&&\ell - \tau = - n, \qquad n=0,1,2,..., \qquad \theta = 2 \pi i \, \, \mbox{[double]}, \, \, 4 \pi i \, \, [4^{th} \mbox{order}], \, \, ...,  \nonumber\\
&&\frac{1}{2} +\ell + \tau = - n, \qquad n=0,1,2,..., \qquad \theta = -3 \pi i \, \, \mbox{[simple]}, \, \, -5 \pi i \, \, \mbox{[double]}, \, \, ...,  \nonumber\\
&&-\frac{1}{2} +\ell + \tau = - n, \qquad n=0,1,2,..., \qquad \theta = - \pi i \, \, \mbox{[simple]}, \, \, -3 \pi i \, \, \mbox{[double]}, \, \, ... \, .
\end{eqnarray}

\item {\it Zeros}

The zeros come from the poles of the gamma functions at the denominator, namely:
\begin{eqnarray}
\label{zerZamo}
&&\ell + \tau = - n, \qquad n=0,1,2,..., \qquad \theta = -2 \pi i \, \, \mbox{[double]}, \, \, -4 \pi i \, \, [4^{th} \mbox{order}], \, \, ...,  \nonumber\\
&&\frac{1}{2} +\ell - \tau = - n, \qquad n=0,1,2,..., \qquad \theta = 3 \pi i \, \, \mbox{[single]}, \, \, 5 \pi i \, \, \mbox{[double]}, \, \, ...,  \nonumber\\
&&-\frac{1}{2} +\ell - \tau = - n, \qquad n=0,1,2,..., \qquad \theta = \pi i \, \, \mbox{[single]}, \, \, 3 \pi i \, \, \mbox{[double]}, \, \, ... \, .
\end{eqnarray}

\end{enumerate}

Notice that the S-matrix, as we will have the opportunity to discuss more fully in later sections, satisfies
\begin{equation}
S_{bb}^{bb} = 1 = S_{ff}^{ff},
\end{equation}
since $R$ and $S$ are related by the graded permutation, and $R$ has a $-1$ in the fermion-fermion entry. 

We will only focus on right movers throughout the paper, since the left movers will behave in the exact same way. The conformal point we are sitting at enjoys a factorisation of the form factors in a purely left- times a purely right-moving part \cite{DelfinoMussSimo}, therefore it is sufficient to deal with only one of the two types.

\section{\label{sec:Min}Minimal two-particle form factors}

We shall now turn to the study of minimal form factors based on the $S$-matrix which we have described in the previous section.

\subsection{\label{sec:Minimo}Integral representation of the dressing factor}

An important step which is required to be able to apply the familiar procedure of relativistic models is to identify the most useful integral representation for the dressing factor. Naturally this representation will be defined only in an appropriate region in the complex $\theta$-plane. In this case, following \cite{FI1,FI2,DiegoBogdanAle}, one finds
\begin{equation}
\label{integrorep}
\Phi(\theta) = \exp \int_0^\infty dx \, f(x) \, \sinh \frac{x \theta}{i \pi}, \qquad f(x) = -\frac{1}{2 x \cosh^2 \frac{x}{2}},
\end{equation}
in the strip $\mbox{Im}(\theta) \in (-\pi,\pi)$. In this form we can simply use the Karowski-Weisz theorem \cite{Karo1} to find the minimal two-particle form factor, as we shall now describe.

\subsection{\label{sec:Watson}Watson's equations}

Following the theorem proven in \cite{Karo1}, let us provide a minimal solution to Watson's equations specialised to the two-particle form factor, in the case of a bosonic local operator $\cal{O}$, for two bosonic particles: 
\begin{equation}
F_{bb}^{\cal{O}} (\theta_1, \theta_2) \equiv F(\theta), \qquad \theta \equiv \theta_1 - \theta_2.
\end{equation}
The case of fermionic operators will be dealt with more conveniently in a uniform fashion when we will derive our general formula later on. The function $F$ is required to be meromorphic in the complex $\theta$-plane. Thanks to this requirement, one can easily obtain from the axioms that
\begin{equation}
\label{Watson}
F(\theta) = F(-\theta) \, \Phi(\theta), \qquad F (i \pi - \theta) = F (i \pi + \theta). 
\end{equation}
From \cite{Karo1} we simply get
\begin{equation}
\label{eff}
F(\theta) = \exp \int_0^\infty dx \, f(x) \, \frac{\sin^2 \frac{x(i \pi - \theta)}{2 \pi}}{\sinh x}, \qquad f(x) = -\frac{1}{2 x \cosh^2 \frac{x}{2}},
\end{equation}
in a sufficiently small open strip in the $\theta$-plane containing the real axis\footnote{Naively, one can see that
\begin{eqnarray}
\frac{F(-\theta)}{F(\theta)} = \exp \int_0^\infty dx \frac{f(x)}{\sinh x}\Big[\sin^2 \big(\frac{ix}{2}+\frac{\theta x}{2\pi}\big)-\sin^2 \big(\frac{ix}{2}-\frac{\theta x}{2\pi}\big)\Big] = \exp \int_0^\infty dx \frac{f(x)}{\sinh x} \sin ix \, \sin \frac{\theta x}{\pi}=\Phi(\theta)^{-1},
\end{eqnarray}
where we have used a trigonometric identity and the integral form (\ref{integrorep}). This manipulation is not however very mindful of branch-cuts, therefore one should either rely on the analysis of \cite{Karo1}, or on the manifestly meromorphic expression which we will provide in what follows. The meromorphic form will also be prone to numeric experimentation with {\it Mathematica}, which is a way in which we shall cross-check a number of formulas. For completeness, let us mention that \cite{Karo1} provide an alternative integral representation, obtained under suitable assumptions as follows. One can start by writing  
\begin{eqnarray}
\frac{d}{d\theta}\log F(\theta) = \frac{1}{8 \pi i} \int_{C_s} \frac{dz}{\sinh^2 \frac{1}{2} (z-\theta)} \log F(z) =  \frac{1}{8 \pi i} \int_{-\infty}^\infty \frac{dz}{\sinh^2 \frac{1}{2} (z-\theta)} \log \frac{F(z)}{F(z+2 \pi i)}.
\end{eqnarray}
Here $C_s$ is a rectangular contour perfectly surrounding the strip Im$(\theta)\in[0,2\pi]$, and the function $\log F$ does not grow too fast - in this way one can reduce the integral to the two infinite horizontal lines. In our case one can for instance explicitly verify, as we will do in a later section on convergence of the three-particle form factor integrals, that $F$ is such that $\log F$ grows linearly in $\theta$. By using (\ref{Watson}) one then concludes that $\frac{F(z)}{F(z+2 \pi i)} = \frac{F(z)}{F(-z)} = \Phi(z)$, hence
\begin{eqnarray}
\frac{d}{d\theta}\log F(\theta)=\frac{1}{8 \pi i} \int_{-\infty}^\infty \frac{dz}{\sinh^2 \frac{1}{2} (z-\theta)} \log \Phi(z).
\end{eqnarray}
Although it shows that the dressing factor uniquely determines the minimal form factor, this form of the function $F$ is however less convenient in practice than the one we shall utilise in the main text, as remarked in \cite{Karo1} as well.}.  
The same way as for the dressing factor, we will use the analytic continuation of (\ref{eff}) to the entire complex plane, in the form of an infinite double product of gamma functions. This formula makes it clear that the only singularities at finite distance from the origin are zeros and poles (of increasing order). The expression we have found is given by
\begin{eqnarray}
\label{pro}
F(\theta) = \prod_{i,j=1}^\infty \frac{\Gamma(i+j-2\lambda)\Gamma(i+j+2\lambda)\Gamma(i+j-2\lambda-1)\Gamma(i+j+2\lambda-1)\Gamma^4(i+j-\frac{1}{2})}{\Gamma^2(i+j)\Gamma^2(i+j-1)\Gamma^2(i+j-2\lambda-\frac{1}{2})\Gamma^2(i+j+2\lambda-\frac{1}{2})},
\end{eqnarray}  
where 
\begin{eqnarray}
\lambda = \frac{\pi + i \theta}{4 \pi}.
\end{eqnarray}
We have derived this formula analytically by a reverse-engineering method: using the Malmst\'en representation of the gamma function
\begin{eqnarray}
\label{Mal}
\Gamma(z) = \exp \int_0^\infty \frac{dx}{x} \, e^{-x} \, \Big[z - 1 -\frac{1-e^{-x(z-1)}}{1-e^{-x}}\Big],
\end{eqnarray}
valid for Re$(z)>0$ \cite{Weisz:1977ij}, it is possible to organise a ratio of gamma functions in such a way that, plugging in (\ref{Mal}) into (\ref{pro}) one ends up precisely with (\ref{eff}). Notice that the ratio in (\ref{pro}) is concocted in such a way to be balanced, namely the sum of all arguments at the numerator equal the sum of all arguments at the denominator, therefore the linear $z-1$ part of (\ref{Mal}) cancels out entirely. The $1$ at the numerator of the remaining piece inside the square bracket of (\ref{Mal}) also cancels out, since there are as many gammas at the numerator than at the denominator, and the final part is arranged with some effort so as to actually produce the desired integrand. The infinite products become sums at the exponent, and factorise in two equal geometric series of easy resummation. We finally have verified our formula numerically with {\it Mathematica} (notwithstanding the slow convergence of the double product). 

From the analytic continuation (\ref{pro}) it is again possible to provide the list the poles and zeros of $F(\theta)$. We shall display the first few nearer to the origin, which will be the ones relevant to our analysis, and just indicate with dots the easily reconstructible pattern: 

\begin{enumerate}

\item {\it Poles}

The poles come from the poles of the gamma functions at the numerator, namely:
\begin{eqnarray}
\label{polF}
&&i+j - 2\lambda = - n, \qquad n=0,1,2,..., \qquad \theta = -3 \pi i \, \, \mbox{[simple]}, \, \, -5 \pi i \, \, [\mbox{triple}], \, \, ...,  \nonumber\\
&&i+j + 2\lambda = - n, \qquad n=0,1,2,..., \qquad \theta = 5 \pi i \, \, \mbox{[simple]}, \, \, 7 \pi i \, \, [\mbox{triple}], \, \, ...,  \nonumber\\
&&i+j - 2\lambda -1 = - n, \qquad n=0,1,2,..., \qquad \theta = - \pi i \, \, \mbox{[simple]}, \, \, -3 \pi i \, \, [\mbox{triple}], \, \, ...,  \nonumber\\
&&i+j + 2\lambda -1 = - n, \qquad n=0,1,2,..., \qquad \theta = 3 \pi i \, \, \mbox{[simple]}, \, \, 5 \pi i \, \, [\mbox{triple}], \, \, ... \, .
\end{eqnarray}

\item {\it Zeros}

The zeros come from the poles of the gamma functions at the denominator, namely:
\begin{eqnarray}
\label{zerF}
&&i+j - 2\lambda -\frac{1}{2}= - n, \qquad n=0,1,2,..., \qquad \theta = -2 \pi i \, \, \mbox{[double]}, \, \, -4 \pi i \, \, [6^{th} \mbox{order}], \, \, ...,  \nonumber\\
&&i+j + 2\lambda -\frac{1}{2}= - n, \qquad n=0,1,2,..., \qquad \theta = 4 \pi i \, \, \mbox{[double]}, \, \, 6 \pi i \, \, [6^{th} \mbox{order}], \, \, ... \, .
\end{eqnarray}

\end{enumerate}

As we can see the function $F(\theta)$ is analytic in the physical strip Im$(\theta)\in [0,\pi]$ in accordance with the Karowski-Weisz theorem \cite{Karo1}. We can introduce a simplification by exploiting the fact that the arguments of the gamma functions only depend on $i+j$:
\begin{eqnarray}
\label{prodeg}
F(\theta) = \prod_{M=2}^\infty \Bigg[\frac{\Gamma(M-2\lambda)\Gamma(M+2\lambda)\Gamma(M-2\lambda-1)\Gamma(M+2\lambda-1)\Gamma^4(M-\frac{1}{2})}{\Gamma^2(M)\Gamma^2(M-1)\Gamma^2(M-2\lambda-\frac{1}{2})\Gamma^2(M+2\lambda-\frac{1}{2})}\Bigg]^{M-1}.
\end{eqnarray}  

The boost axiom will also be dealt with later on in a uniform fashion via the general formula. This requirement typically forces the decoration of the function $F(\theta)$ by $e^{s(\theta_1+\theta_2)}$, although we will see that in our cases we will have to embark on a significantly more complicated journey.

We postpone the complete formulation of the two-particle form factors, since it is not possible to unfold all their parts without taking a look at the three-particle case. This case first displays a series of subtleties which are not present in the two-particle case, and this is why we need to delve into it at this point. However, the basic function $F$ which we have discussed here, as well as the dressing factor $\Phi$, will keep recurring over and over again, and will constitute some of the principal ingredients of the complete expressions. 

\section{\label{fourfour}Three-particle form factors}

In this section we begin the construction of the minimal three-particle form factors, starting with a fermionic operator. We know that fermionic operators must exist in our theory since supersymmetry is preserved at this critical point \cite{DiegoBogdanAle}. 
We will be adopting the tecnique of the {\it off-shell} Bethe ansatz, particularly as displayed in \cite{BabuF}. Given the differences in the state-space - due to the presence of (boson,fermion) doublets - and in the analytic structure of the dressing factor, the procedure that follows is not a simple application of the formulas of \cite{BabuF}, which means that we cannot rely on their proofs but instead we need to explicitly demonstrate each properties for the  expressions which we will write. In fact in several crucial points the procedure will need substantial adaptation, especially, as we shall see, in the contributions from the contour integration, and also for what concerns the implementation of crossing symmetry.  

In order to begin the procedure we need a few ingredients from the algebraic Bethe ansatz \cite{DiegoBogdanAle}. In particular, we shall require the definition of the three-particle pseudo-vacuum 
\begin{eqnarray}
|\Omega\rangle = |b\rangle \otimes |b\rangle \otimes |b\rangle,
\end{eqnarray}
and the $C$ operator from the monodromy matrix:
\begin{eqnarray}
{\cal{M}} \equiv \prod_{i=1}^3 R_{i0}(\theta_i - \theta_0) = A \otimes E_{11} + B \otimes E_{12} + C \otimes E_{21} + D \otimes E_{22}.
\end{eqnarray} 
In this formula, $R$ is the $R$-matrix (\ref{eq:RLLlimtheta}), inclusive of the dressing factor, and $\theta_0$ is the spectral parameter associated with the auxiliary space $0$ in the graded tensor product. For three sites the formula for $A,B,C$ and $D$ as $8 \times 8$ matrices can be obtained explicitly without difficulties using {\it Mathematica}.

One of the components one needs to assemble following \cite{BabuF} is the particular bra-vector
\begin{eqnarray}
\langle \Omega | C.
\end{eqnarray}
If we arrange the quantum space of the three physical particles in an $8$-dimensional vector, we find that
\begin{eqnarray}
&&\langle \Omega | C = (1,0,0,0,0,0,0,0) C = \Big(0,\frac{\tanh \frac{\theta_1-\theta_0}{2} \tanh \frac{\theta_2-\theta_0}{2}}{\cosh \frac{\theta_3-\theta_0}{2}},-\frac{\tanh \frac{\theta_1-\theta_0}{2}}{\cosh \frac{\theta_2-\theta_0}{2}},0,\frac{1}{\cosh \frac{\theta_1-\theta_0}{2}},0,0,0\Big)=\nonumber\\
&&\frac{\tanh \frac{\theta_1-\theta_0}{2} \tanh \frac{\theta_2-\theta_0}{2}}{\cosh \frac{\theta_3-\theta_0}{2}} \langle b| \otimes \langle b| \otimes \langle f| -\frac{\tanh \frac{\theta_1-\theta_0}{2}}{\cosh \frac{\theta_2-\theta_0}{2}} \langle b| \otimes \langle f| \otimes \langle b| + \frac{1}{\cosh \frac{\theta_1-\theta_0}{2}}\langle f| \otimes \langle b| \otimes \langle b|.
\end{eqnarray}
In the ordering used, the entry nr. 1 of the vector is $\langle b| \otimes \langle b| \otimes \langle b|$, the entry nr. 2 is $\langle b| \otimes \langle b| \otimes \langle f|$, the entry nr. 3 is $\langle b| \otimes \langle f| \otimes \langle b|$ and the entry nr. 5 is $\langle f| \otimes \langle b| \otimes \langle b|$, etc. The entries 2, 3 and 5 are in one-to-one correspondence with the non-zero fermionic form factors. The way they eventually will enter the formulas for the form factors is via three functions which we define here below, where the bra-vector entries are easily recognisable as suggested by the labels: 
\begin{eqnarray}
\label{fermiFF}
&&G_{bbf}(\theta_1, \theta_2,\theta_3) = N_{bbf} \, \sigma(\theta_1,\theta_2,\theta_3) \, \prod_{1\leq i<j\leq3}F(\theta_i-\theta_j) \times\nonumber\\
&&\qquad \qquad \qquad \qquad \times \int_{C1;123} du \, e^{\big[\ell \sum_{i=1}^3 \theta_i + (s - 3 \ell) u\big]} \, \frac{\tanh \frac{\theta_1-u}{2} \tanh \frac{\theta_2-u}{2}}{\cosh \frac{\theta_3-u}{2}}\prod_{i=1}^3 Z(\theta_i-u) \Phi(\theta_i - u),\nonumber\\
&&G_{bfb}(\theta_1, \theta_2,\theta_3) = - N_{bfb} \, \sigma(\theta_1,\theta_2,\theta_3) \, \prod_{1\leq i<j\leq3}F(\theta_i-\theta_j)\times\nonumber\\
&&\qquad \qquad \qquad \qquad \times \int_{C2;123} du \, e^{\big[\ell \sum_{i=1}^3 \theta_i + (s - 3 \ell) u\big]} \, \frac{\tanh \frac{\theta_1-u}{2}}{\cosh \frac{\theta_2-u}{2}}\prod_{i=1}^3 Z(\theta_i-u) \Phi(\theta_i - u),\nonumber\\
&&G_{fbb}(\theta_1, \theta_2,\theta_3) = N_{fbb} \, \sigma(\theta_1,\theta_2,\theta_3) \, \prod_{1\leq i<j\leq3}F(\theta_i-\theta_j)\times\nonumber\\
&&\qquad \qquad \qquad \qquad \times \int_{C3;123} du \, e^{\big[\ell \sum_{i=1}^3 \theta_i + (s - 3 \ell) u\big]} \, \frac{1}{\cosh \frac{\theta_1-u}{2}}\prod_{i=1}^3 Z(\theta_i-u) \Phi(\theta_i - u).
\end{eqnarray}
We will adopt the symbol $G$ to stand for $F^{\cal{O}}$ for a lighter notation. In these formulas the spectral parameter is now called $u$ and is integrated along a contour to be specified momentarily. The function $\sigma$ is given by
\begin{eqnarray}
\label{sigma}
\sigma(\theta_1,\theta_2,\theta_3) = -\frac{1}{4}\Bigg[1+\frac{\sigma^{(3)}_2(x_1,x_2,x_3)}{\Big(\sigma^{(3)}_1(x_1,x_2,x_3)\Big)^2}\Bigg], \qquad x_i = e^{\theta_i}, \qquad i=1,2,3,
\end{eqnarray}
where $\sigma^{(n)}_k$ are the elementary homogeneous completey symmetric polynomials of degree $k$ in $n$ variables, in particular
\begin{eqnarray}
\sigma^{(3)}_3 = x_1 x_2 x_3, \qquad \sigma^{(3)}_2 = x_1 x_2 + x_1 x_3 + x_2 x_3, \qquad \sigma^{(3)}_1 = x_1 + x_2 + x_3, \qquad \sigma^{(3)}_0 = 1.   
\end{eqnarray} 
The factor of $- \frac{1}{4}$ is for later convenience. This function will be instrumental at the end when dealing with kinematical singularities. 
  
The quantities $N_{\alpha \beta \gamma}$ are normalisation constants depending on which operator in under consideration. The function $F$ is the minimal form factor (\ref{pro}), the function $\Phi$ is the Zamolodchikov dressing factor (\ref{zamo}), and $s$ is the spin of the operator. We have suppressed any spinor label of the operator, which is understood as a label of $N_{\alpha_1 \alpha_2 \alpha_3}$. The overall signs have been kept (and not reabsorbed in a redefinition of the constants $N$) to make it easier to identify exactly the integrand with the entries of the Bethe bra-vector. The real number $\ell$ will be fixed later on. Finally, the function $Z$ is given by
\begin{eqnarray}
Z(\theta) \equiv \frac{1}{F(\theta)F(\theta+i\pi) \sinh \theta}.
\end{eqnarray}

\subsection{\label{four1}Specification of the contour and analytic structure}

In order to specify the contour we need to elucidate the analytic structure of the integrand.

The function $Z$ has the following poles and zeros, easily obtained from (\ref{polF}) and (\ref{zerF}) and from the fact that $\sinh \theta$ has zeros at $\theta = n \pi i$, $n$ integer:

\begin{enumerate}

\item {\it Poles}
\begin{eqnarray}
\label{polZ}
&&\mbox{from} \, \, F(\theta) \qquad \qquad \qquad \qquad\qquad \qquad\qquad \mbox{from} \, \, F(\theta+i\pi)\nonumber\\
&&\theta = -2 \pi i \, \, \mbox{[double]}, \, \, -4 \pi i \, \, [6^{th} \mbox{order}] \, \, ... , \qquad  \theta = -3 \pi i \, \, \mbox{[double]}, \, \, -5 \pi i \, \, [6^{th} \mbox{order}] \, \, ... , \nonumber\\
&&\theta = 4 \pi i \, \, \mbox{[double]}, \, \, 6 \pi i \, \, [6^{th} \mbox{order}], \, \, ... , \qquad\qquad \theta = 3 \pi i \, \, \mbox{[double]}, \, \, 5 \pi i \, \, [6^{th} \mbox{order}], \, \, ..., \nonumber\\
&&\mbox{and from} \, \, \sinh \theta\nonumber\\
&&\theta = ... - 2 \pi i \, \, \mbox{[simple]}, \, \, -\pi i \, \, \mbox{[simple]}, \, \, 0 \, \, \mbox{[simple]}, \, \, \pi i \, \, \mbox{[simple]}, \, \, 2 \pi i \, \, \mbox{[simple]}, \, \, ... \, ,
\end{eqnarray}
namely in total
\begin{eqnarray}
\theta = ... - 2 \pi i \, \, \mbox{[triple]}, \, \, - \pi i \, \, \mbox{[simple]}, \, \, 0 \, \, \mbox{[simple]}, \, \, \pi i \, \, \mbox{[simple]}, \, \, 2 \pi i \, \, \mbox{[simple]}, \, \, ... \, . 
\end{eqnarray}

\item {\it Zeros}
\begin{eqnarray}
\label{zerZ}
&&\mbox{from} \, \, F(\theta) \qquad \qquad \qquad \qquad\qquad \qquad\qquad \mbox{from} \, \, F(\theta+i\pi)\nonumber\\
&&\theta = -3 \pi i \, \, \mbox{[single]}, \, \, -5 \pi i \, \, [\mbox{triple}], \, \, ..., \qquad \theta = -4 \pi i \, \, \mbox{[single]}, \, \, -6 \pi i \, \, [\mbox{triple}], \, \, ..., \nonumber\\
&&\theta = 5 \pi i \, \, \mbox{[single]}, \, \, 7 \pi i \, \, [\mbox{triple}], \, \, ..., \qquad \qquad \theta = 4 \pi i \, \, \mbox{[single]}, \, \, 6 \pi i \, \, [\mbox{triple}], \, \, ..., \nonumber\\
&&\theta = - \pi i \, \, \mbox{[single]}, \, \, -3 \pi i \, \, [\mbox{triple}], \, \, ..., \qquad \theta = - 2\pi i \, \, \mbox{[single]}, \, \, -4 \pi i \, \, [\mbox{triple}], \, \, ...,\nonumber\\
&&\theta = 3 \pi i \, \, \mbox{[single]}, \, \, 5 \pi i \, \, [\mbox{triple}], \, \, ..., \qquad \qquad \theta = 2 \pi i \, \, \mbox{[single]}, \, \, 4 \pi i \, \, [\mbox{triple}], \, \, ...\, ,
\end{eqnarray}
namely in total
\begin{eqnarray}
\theta = ... - 2 \pi i \, \, \mbox{[single]}, \, \, - \pi i \, \, \mbox{[single]}, \, \, 2 \pi i \, \, \mbox{[single]}, \, \, ... \, . 
\end{eqnarray}

\end{enumerate}

Let us describe the analytic structure of the various components of the integrands of (\ref{fermiFF}) in detail, paying attention to the fact that the variable $u$ is the one integrated over.

\begin{enumerate}

\item The function 
\begin{eqnarray}
Z(\theta_i-u) \equiv \frac{1}{F(\theta_i-u)F(\theta_i-u+i\pi) \sinh (\theta_i-u)}
\end{eqnarray}
has poles and zeros in the variable
\begin{eqnarray}
v \equiv u - \theta_i
\end{eqnarray}
according to:

\begin{enumerate}

\item {\it Poles}
\begin{eqnarray}
v = ... - 2 \pi i \, \, \mbox{[simple]}, \, \, - \pi i \, \, \mbox{[simple]}, \, \, 0 \, \, \mbox{[simple]}, \, \, \pi i \, \, \mbox{[simple]}, \, \, 2 \pi i \, \, \mbox{[triple]}, \, \, ... \, . 
\end{eqnarray}

\item {\it Zeros}
\begin{eqnarray}
v = ... - 2 \pi i \, \, \mbox{[single]}, \, \,  \pi i \, \, \mbox{[single]}, \, \, 2 \pi i \, \, \mbox{[single]}, \, \, ... \, . 
\end{eqnarray}

\end{enumerate}

\item The function $\Phi(\theta_i - u)$: the relevant poles and zeros are listed in (\ref{polZamo}) and (\ref{zerZamo}). Here we get as a total:

\begin{enumerate}

\item {\it Poles}
\begin{eqnarray}
\label{polZamob}
&&v = ...-2\pi i \, \, \mbox{[double]}, \, \, \pi i \, \, [\mbox{simple}], \, \, ...\, .
\end{eqnarray}

\item {\it Zeros}
\begin{eqnarray}
\label{zerZamob}
&&v = ...-\pi i \, \, \mbox{[single]}, \, \, 2\pi i \, \, [\mbox{double}], \, \, ... \, .
\end{eqnarray}

\end{enumerate}

\item The function $\sinh \frac{\theta}{2}$, with zeros at $\theta = 2 n \pi i$, $n$ integer, and the function $\frac{1}{\cosh \frac{\theta}{2}}$, with poles at $\theta = -\pi i + 2 n \pi i$, $n$ integer. Therefore we get

\begin{enumerate}

\item {\it Poles} of $\frac{1}{\cosh \frac{\theta_i-u}{2}}$:
\begin{eqnarray}
v = ... - \pi i \, \, \mbox{[simple]}, \, \, \pi i \, \, \mbox{[simple]}, \, \, ... \, . 
\end{eqnarray}

\item {\it Zeros} of $\sinh \frac{\theta_i-u}{2}$:
\begin{eqnarray}
v = ... - 2 \pi i \, \, \mbox{[single]}, \, \, 0 \, \, \mbox{[single]}, \, \, 2 \pi i \, \, \mbox{[single]}, \, \, ... \, . 
\end{eqnarray}

\end{enumerate}

\end{enumerate}

Let us put everything together. 

\begin{enumerate}

\item If we focus on $G_{bbf}$, we have

\begin{enumerate}

\item {\it Poles} 
\begin{eqnarray}
&&u - \theta_1 = ... - \pi i \, \, \mbox{[simple]}, \, \, \pi i \, \, \mbox{[simple]}, \, \, ... \, , \nonumber\\
&&u - \theta_1 = ... - 2 \pi i \, \, \mbox{[simple]}, \, \, - \pi i \, \, \mbox{[simple]}, \, \, 0 \, \, \mbox{[simple]}, \, \, \pi i \, \, \mbox{[simple]}, \, \, 2 \pi i \, \, \mbox{[triple]}, \, \, ... \, ,\nonumber\\
&&u - \theta_1 = ...-2\pi i \, \, \mbox{[double]}, \, \, \pi i \, \, [\mbox{simple}], \, \, ...\, ,\nonumber\\
&&u - \theta_2 = ... - \pi i \, \, \mbox{[simple]}, \, \, \pi i \, \, \mbox{[simple]}, \, \, ... \, , \nonumber\\
&&u - \theta_2 = ... - 2 \pi i \, \, \mbox{[simple]}, \, \, - \pi i \, \, \mbox{[simple]}, \, \, 0 \, \, \mbox{[simple]}, \, \, \pi i \, \, \mbox{[simple]}, \, \, 2 \pi i \, \, \mbox{[triple]}, \, \, ... \, ,\nonumber\\
&&u - \theta_2 = ...-2\pi i \, \, \mbox{[double]}, \, \, \pi i \, \, [\mbox{simple}], \, \, ...\, ,\nonumber\\
&&u - \theta_3 = ... - \pi i \, \, \mbox{[simple]}, \, \, \pi i \, \, \mbox{[simple]}, \, \, ... \, , \nonumber\\
&&u - \theta_3 = ... - 2 \pi i \, \, \mbox{[simple]}, \, \, - \pi i \, \, \mbox{[simple]}, \, \, 0 \, \, \mbox{[simple]}, \, \, \pi i \, \, \mbox{[simple]}, \, \, 2 \pi i \, \, \mbox{[triple]}, \, \, ... \, , \nonumber\\
&&u - \theta_3 = ...-2\pi i \, \, \mbox{[double]}, \, \, \pi i \, \, [\mbox{simple}], \, \, ...\, .
\end{eqnarray}
In total, one gets
\begin{eqnarray}
&&u - \theta_1 = ... - 2 \pi i \, \, \mbox{[triple]}, \, \, - \pi i \, \, \mbox{[double]}, \, \, 0 \, \, \mbox{[simple]}, \, \, \pi i \, \, \mbox{[triple]}, \, \, 2 \pi i \, \, \mbox{[triple]}, \, \, ... \, ,\nonumber\\ 
&&u-\theta_2 = ... - 2 \pi i \, \, \mbox{[triple]}, \, \, - \pi i \, \, \mbox{[double]}, \, \, 0 \, \, \mbox{[simple]}, \, \, \pi i \, \, \mbox{[triple]}, \, \, 2 \pi i \, \, \mbox{[triple]}, \, \, ... \, ,\nonumber\\
&&u-\theta_3 = ... - 2 \pi i \, \, \mbox{[triple]}, \, \, - \pi i \, \, \mbox{[double]}, \, \, 0 \, \, \mbox{[simple]}, \, \, \pi i \, \, \mbox{[triple]}, \, \, 2 \pi i \, \, \mbox{[triple]}, \, \, ... \, .\nonumber\\ 
\end{eqnarray}

\item {\it Zeros}
\begin{eqnarray}
&&u - \theta_1  = ... - 2 \pi i \, \, \mbox{[single]}, \, \, 0 \, \, \mbox{[single]}, \, \, 2 \pi i \, \, \mbox{[single]}, \, \, ... \, , \nonumber\\
&&u - \theta_1  = ... - 2 \pi i \, \, \mbox{[single]}, \, \,  \pi i \, \, \mbox{[single]}, \, \, 2 \pi i \, \, \mbox{[single]}, \, \, ... \, ,\nonumber\\
&&u - \theta_1  = ...-\pi i \, \, \mbox{[single]}, \, \, 2\pi i \, \, [\mbox{double}], \, \, ... \, ,\nonumber\\
&&u - \theta_2  = ... - 2 \pi i \, \, \mbox{[single]}, \, \, 0 \, \, \mbox{[single]}, \, \, 2 \pi i \, \, \mbox{[single]}, \, \, ... \, ,\nonumber\\
&&u - \theta_2  = ... - 2 \pi i \, \, \mbox{[single]}, \, \,  \pi i \, \, \mbox{[single]}, \, \, 2 \pi i \, \, \mbox{[single]}, \, \, ... \, ,\nonumber\\
&&u - \theta_2 = ...-\pi i \, \, \mbox{[single]}, \, \, 2\pi i \, \, [\mbox{double}], \, \, ... \, ,\nonumber\\
%&&u - \theta_3 = \nonumber\\
&&u - \theta_3  = ... - 2 \pi i \, \, \mbox{[single]}, \, \,  \pi i \, \, \mbox{[single]}, \, \, 2 \pi i \, \, \mbox{[single]}, \, \, ... \, ,\nonumber\\
&&u - \theta_3 = ...-\pi i \, \, \mbox{[single]}, \, \, 2\pi i \, \, [\mbox{double}], \, \, ... \, .
\end{eqnarray}
In total, one gets
\begin{eqnarray}
&&u - \theta_1 = ... - 2 \pi i \, \, \mbox{[double]}, \, \, - \pi i \, \, \mbox{[single]}, \, \, 0 \, \, \mbox{[single]}, \, \, \pi i \, \, \mbox{[single]}, \, \, 2 \pi i \, \, [4^{th} \mbox{order}], \, \, ... \, ,\nonumber\\ 
&&u - \theta_2 = ... - 2 \pi i \, \, \mbox{[double]}, \, \, - \pi i \, \, \mbox{[single]}, \, \, 0 \, \, \mbox{[single]}, \, \, \pi i \, \, \mbox{[single]}, \, \, 2 \pi i \, \, [4^{th} \mbox{order}], \, \, ... \, ,\nonumber\\
&&u - \theta_3 = ... - 2 \pi i \, \, \mbox{[single]}, \, \, - \pi i \, \, \mbox{[single]}, \, \, \pi i \, \, \mbox{[single]}, \, \, 2 \pi i \, \, \mbox{[triple]}, \, \, ... \, . 
\end{eqnarray}

\item This means that in summary $G_{bbf}$ has
\begin{eqnarray}
&&u - \theta_1 = ... - 2 \pi i \, \, \mbox{[simple pole]}, \, \, - \pi i \, \, \mbox{[simple pole]}, \, \, \pi i \, \, \mbox{[double pole]}, \, \, 2 \pi i \, \, \mbox{[single zero]}, \, \, ... \, ,\nonumber\\ 
&&u-\theta_2 = ... - 2 \pi i \, \, \mbox{[simple pole]}, \, \, - \pi i \, \, \mbox{[simple pole]}, \, \, \pi i \, \, \mbox{[double pole]}, \, \, 2 \pi i \, \, \mbox{[single zero]}, \, \, ... \, ,\nonumber\\
&&u-\theta_3 = ... - 2 \pi i \, \, \mbox{[double pole]}, \, \, - \pi i \, \, \mbox{[simple pole]}, \, \, 0 \, \, \mbox{[simple pole]}, \, \, \pi i \, \, \mbox{[double pole]}, \, \, ... \, . \nonumber\\
\end{eqnarray}

\end{enumerate}

\item If we focus on $G_{bfb}$, we have

\begin{enumerate}

\item {\it Poles} 
\begin{eqnarray}
&&u - \theta_1 = ... - \pi i \, \, \mbox{[simple]}, \, \, \pi i \, \, \mbox{[simple]}, \, \, ... \, , \nonumber\\
&&u - \theta_1 = ... - 2 \pi i \, \, \mbox{[simple]}, \, \, - \pi i \, \, \mbox{[simple]}, \, \, 0 \, \, \mbox{[simple]}, \, \, \pi i \, \, \mbox{[simple]}, \, \, 2 \pi i \, \, \mbox{[triple]}, \, \, ... \, ,\nonumber\\
&&u - \theta_1 = ...-2\pi i \, \, \mbox{[double]}, \, \, \pi i \, \, [\mbox{simple}], \, \, ...\, ,\nonumber\\
&&u - \theta_2 = ... - \pi i \, \, \mbox{[simple]}, \, \, \pi i \, \, \mbox{[simple]}, \, \, ... \, , \nonumber\\
&&u - \theta_2 = ... - 2 \pi i \, \, \mbox{[simple]}, \, \, - \pi i \, \, \mbox{[simple]}, \, \, 0 \, \, \mbox{[simple]}, \, \, \pi i \, \, \mbox{[simple]}, \, \, 2 \pi i \, \, \mbox{[triple]}, \, \, ... \, ,\nonumber\\
&&u - \theta_2 = ...-2\pi i \, \, \mbox{[double]}, \, \, \pi i \, \, [\mbox{simple}], \, \, ...\, ,\nonumber\\
%&&u - \theta_3 =  \nonumber\\
&&u - \theta_3 = ... - 2 \pi i \, \, \mbox{[simple]}, \, \, - \pi i \, \, \mbox{[simple]}, \, \, 0 \, \, \mbox{[simple]}, \, \, \pi i \, \, \mbox{[simple]}, \, \, 2 \pi i \, \, \mbox{[triple]}, \, \, ... \, , \nonumber\\
&&u - \theta_3 = ...-2\pi i \, \, \mbox{[double]}, \, \, \pi i \, \, [\mbox{simple}], \, \, ...\, .
\end{eqnarray}
In total, one gets
\begin{eqnarray}
&&u - \theta_1 = ... - 2 \pi i \, \, \mbox{[triple]}, \, \, - \pi i \, \, \mbox{[double]}, \, \, 0 \, \, \mbox{[simple]}, \, \, \pi i \, \, \mbox{[triple]}, \, \, 2 \pi i \, \, \mbox{[triple]}, \, \, ... \, ,\nonumber\\ 
&&u-\theta_2 = ... - 2 \pi i \, \, \mbox{[triple]}, \, \, - \pi i \, \, \mbox{[double]}, \, \, 0 \, \, \mbox{[simple]}, \, \, \pi i \, \, \mbox{[triple]}, \, \, 2 \pi i \, \, \mbox{[triple]}, \, \, ... \, ,\nonumber\\
&&u-\theta_3 = ... - 2 \pi i \, \, \mbox{[triple]}, \, \, - \pi i \, \, \mbox{[simple]}, \, \, 0 \, \, \mbox{[simple]}, \, \, \pi i \, \, \mbox{[double]}, \, \, 2 \pi i \, \, \mbox{[triple]}, \, \, ... . \nonumber\\ 
\end{eqnarray}

\item {\it Zeros}
\begin{eqnarray}
&&u - \theta_1  = ... - 2 \pi i \, \, \mbox{[single]}, \, \, 0 \, \, \mbox{[single]}, \, \, 2 \pi i \, \, \mbox{[single]}, \, \, ... \, , \nonumber\\
&&u - \theta_1  = ... - 2 \pi i \, \, \mbox{[single]}, \, \,  \pi i \, \, \mbox{[single]}, \, \, 2 \pi i \, \, \mbox{[single]}, \, \, ... \, ,\nonumber\\
&&u - \theta_1  = ...-\pi i \, \, \mbox{[single]}, \, \, 2\pi i \, \, [\mbox{double}], \, \, ... \, ,\nonumber\\
%&&u - \theta_2  = \nonumber\\
&&u - \theta_2  = ... - 2 \pi i \, \, \mbox{[single]}, \, \,  \pi i \, \, \mbox{[single]}, \, \, 2 \pi i \, \, \mbox{[single]}, \, \, ... \, ,\nonumber\\
&&u - \theta_2 = ...-\pi i \, \, \mbox{[single]}, \, \, 2\pi i \, \, [\mbox{double}], \, \, ... \, ,\nonumber\\
%&&u - \theta_3 = \nonumber\\
&&u - \theta_3  = ... - 2 \pi i \, \, \mbox{[single]}, \, \,  \pi i \, \, \mbox{[single]}, \, \, 2 \pi i \, \, \mbox{[single]}, \, \, ... \, ,\nonumber\\
&&u - \theta_3 = ...-\pi i \, \, \mbox{[single]}, \, \, 2\pi i \, \, [\mbox{double}], \, \, ... \, .
\end{eqnarray}
In total, one gets
\begin{eqnarray}
&&u - \theta_1 = ... - 2 \pi i \, \, \mbox{[double]}, \, \, - \pi i \, \, \mbox{[single]}, \, \, 0 \, \, \mbox{[single]}, \, \, \pi i \, \, \mbox{[single]}, \, \, 2 \pi i \, \, [4^{th} \mbox{order}], \, \, ... \, ,\nonumber\\ 
&&u - \theta_2 = ... - 2 \pi i \, \, \mbox{[single]}, \, \, - \pi i \, \, \mbox{[single]}, \, \, \pi i \, \, \mbox{[single]}, \, \, 2 \pi i \, \, [\mbox{triple}], \, \, ... \, ,\nonumber\\
&&u - \theta_3 = ... - 2 \pi i \, \, \mbox{[single]}, \, \, - \pi i \, \, \mbox{[single]}, \, \, \pi i \, \, \mbox{[single]}, \, \, 2 \pi i \, \, \mbox{[triple]}, \, \, ... \, . 
\end{eqnarray}

\item This means that in summary $G_{bfb}$ has
\begin{eqnarray}
&&u - \theta_1 = ... - 2 \pi i \, \, \mbox{[simple pole]}, \, \, - \pi i \, \, \mbox{[simple pole]}, \, \, \pi i \, \, \mbox{[double pole]}, \, \, 2 \pi i \, \, \mbox{[single zero]}, \, \, ... \, ,\nonumber\\ 
&&u-\theta_2 = ... - 2 \pi i \, \, \mbox{[double pole]}, \, \, - \pi i \, \, \mbox{[simple pole]}, \, \, 0 \, \, \mbox{[simple pole]}, \, \, \pi i \, \, \mbox{[double pole]}, \, \, ...\, ,\nonumber\\
&&u-\theta_3 = ... - 2 \pi i \, \, \mbox{[double pole]}, \, \, 0 \, \, \mbox{[simple pole]}, \, \, \pi i \, \, \mbox{[simple pole]}, \, \, ... \, .  
\end{eqnarray}

\end{enumerate}

\item If we focus on $G_{fbb}$, we have

\begin{enumerate}

\item {\it Poles} 
\begin{eqnarray}
&&u - \theta_1 = ... - \pi i \, \, \mbox{[simple]}, \, \, \pi i \, \, \mbox{[simple]}, \, \, ... \, , \nonumber\\
&&u - \theta_1 = ... - 2 \pi i \, \, \mbox{[simple]}, \, \, - \pi i \, \, \mbox{[simple]}, \, \, 0 \, \, \mbox{[simple]}, \, \, \pi i \, \, \mbox{[simple]}, \, \, 2 \pi i \, \, \mbox{[triple]}, \, \, ... \, ,\nonumber\\
&&u - \theta_1 = ...-2\pi i \, \, \mbox{[double]}, \, \, \pi i \, \, [\mbox{simple}], \, \, ...\, ,\nonumber\\
%&&u - \theta_2 = \nonumber\\
&&u - \theta_2 = ... - 2 \pi i \, \, \mbox{[simple]}, \, \, - \pi i \, \, \mbox{[simple]}, \, \, 0 \, \, \mbox{[simple]}, \, \, \pi i \, \, \mbox{[simple]}, \, \, 2 \pi i \, \, \mbox{[triple]}, \, \, ... \, ,\nonumber\\
&&u - \theta_2 = ...-2\pi i \, \, \mbox{[double]}, \, \, \pi i \, \, [\mbox{simple}], \, \, ...\, ,\nonumber\\
%&&u - \theta_3 = \nonumber\\
&&u - \theta_3 = ... - 2 \pi i \, \, \mbox{[simple]}, \, \, - \pi i \, \, \mbox{[simple]}, \, \, 0 \, \, \mbox{[simple]}, \, \, \pi i \, \, \mbox{[simple]}, \, \, 2 \pi i \, \, \mbox{[triple]}, \, \, ... \, , \nonumber\\
&&u - \theta_3 = ...-2\pi i \, \, \mbox{[double]}, \, \, \pi i \, \, [\mbox{simple}], \, \, ...\, .
\end{eqnarray}
In total, one gets
\begin{eqnarray}
&&u - \theta_1 = ... - 2 \pi i \, \, \mbox{[triple]}, \, \, - \pi i \, \, \mbox{[double]}, \, \, 0 \, \, \mbox{[simple]}, \, \, \pi i \, \, \mbox{[triple]}, \, \, 2 \pi i \, \, \mbox{[triple]}, \, \, ... \, ,\nonumber\\ 
&&u-\theta_2 = ... - 2 \pi i \, \, \mbox{[triple]}, \, \, - \pi i \, \, \mbox{[simple]}, \, \, 0 \, \, \mbox{[simple]}, \, \, \pi i \, \, \mbox{[double]}, \, \, 2 \pi i \, \, \mbox{[triple]}, \, \, ... \, ,\nonumber\\
&&u-\theta_3 = ... - 2 \pi i \, \, \mbox{[triple]}, \, \, - \pi i \, \, \mbox{[simple]}, \, \, 0 \, \, \mbox{[simple]}, \, \, \pi i \, \, \mbox{[double]}, \, \, 2 \pi i \, \, \mbox{[triple]}, \, \, ... .  \nonumber\\
\end{eqnarray}

\item {\it Zeros}
\begin{eqnarray}
%&&u - \theta_1  =  \nonumber\\
&&u - \theta_1  = ... - 2 \pi i \, \, \mbox{[single]}, \, \,  \pi i \, \, \mbox{[single]}, \, \, 2 \pi i \, \, \mbox{[single]}, \, \, ... \, ,\nonumber\\
&&u - \theta_1  = ...-\pi i \, \, \mbox{[single]}, \, \, 2\pi i \, \, [\mbox{double}], \, \, ... \, ,\nonumber\\
%&&u - \theta_2  = \nonumber\\
&&u - \theta_2  = ... - 2 \pi i \, \, \mbox{[single]}, \, \,  \pi i \, \, \mbox{[single]}, \, \, 2 \pi i \, \, \mbox{[single]}, \, \, ... \, ,\nonumber\\
&&u - \theta_2 = ...-\pi i \, \, \mbox{[single]}, \, \, 2\pi i \, \, [\mbox{double}], \, \, ... \, ,\nonumber\\
%&&u - \theta_3 = \nonumber\\
&&u - \theta_3  = ... - 2 \pi i \, \, \mbox{[single]}, \, \,  \pi i \, \, \mbox{[single]}, \, \, 2 \pi i \, \, \mbox{[single]}, \, \, ... \, ,\nonumber\\
&&u - \theta_3 = ...-\pi i \, \, \mbox{[single]}, \, \, 2\pi i \, \, [\mbox{double}], \, \, ... \, .
\end{eqnarray}
In total one gets
\begin{eqnarray}
&&u - \theta_1 = ... - 2 \pi i \, \, \mbox{[single]}, \, \, - \pi i \, \, \mbox{[single]}, \, \, \pi i \, \, \mbox{[single]}, \, \, 2 \pi i \, \, \mbox{[triple]}, \, \, ... \, ,\nonumber\\ 
&&u - \theta_2 = ... - 2 \pi i \, \, \mbox{[single]}, \, \, - \pi i \, \, \mbox{[single]}, \, \, \pi i \, \, \mbox{[single]}, \, \, 2 \pi i \, \, [\mbox{triple}], \, \, ... \, ,\nonumber\\
&&u - \theta_3 = ... - 2 \pi i \, \, \mbox{[single]}, \, \, - \pi i \, \, \mbox{[single]}, \, \, \pi i \, \, \mbox{[single]}, \, \, 2 \pi i \, \, \mbox{[triple]}, \, \, ... \, . 
\end{eqnarray}

\item This means that in summary $G_{fbb}$ has
\begin{eqnarray}
&&u - \theta_1 = ... - 2 \pi i \, \, \mbox{[double pole]}, \, \, - \pi i \, \, \mbox{[simple pole]}, \, \, 0 \, \, \mbox{[simple pole]}, \, \, \pi i \, \, \mbox{[double pole]}, \, \, ...\, ,\nonumber\\ 
&&u-\theta_2 = ... - 2 \pi i \, \, \mbox{[double pole]}, \, \, 0 \, \, \mbox{[simple pole]}, \, \, \pi i \, \, \mbox{[simple pole]}, \, \,  ... \, ,\nonumber\\
&&u-\theta_3 = ... - 2 \pi i \, \, \mbox{[double pole]}, \, \, 0 \, \, \mbox{[simple pole]}, \, \, \pi i \, \, \mbox{[simple pole]}, \, \,  ... .  
\end{eqnarray}

\end{enumerate}

\end{enumerate}

We can capture the analytic structure graphically in the complex $u$-plane. The contours $Ci;123$, where $i$ stands for the numbering of the form factors, namely $1$ is $bbf$, $2$ is $bfb$ and $3$ is $fbb$, and $123$ stands for the arguments $G(\theta_1,\theta_2,\theta_3)$, avoid all the poles, and are more easily specified by drawing them directly onto the graphs rather than in words. For simplicity, we will draw the case when all $\theta_i$, $i=1,2,3$, are real, so that the three vertical tower of poles and zeros each have real part $\theta_i$, respectively, and imaginary part at multiples of $\pi$. In reality the $\theta_i$ are of course complex, and in fact later on we shall be interested in the case when some of these poles coalesce. The symbols P1, P2 and P3 will be explained in the next section in the precise occasion of poles coalescing.

\subsection{Graphs\label{graphs}}

\begin{figure}
\centerline{\includegraphics[width=18cm]{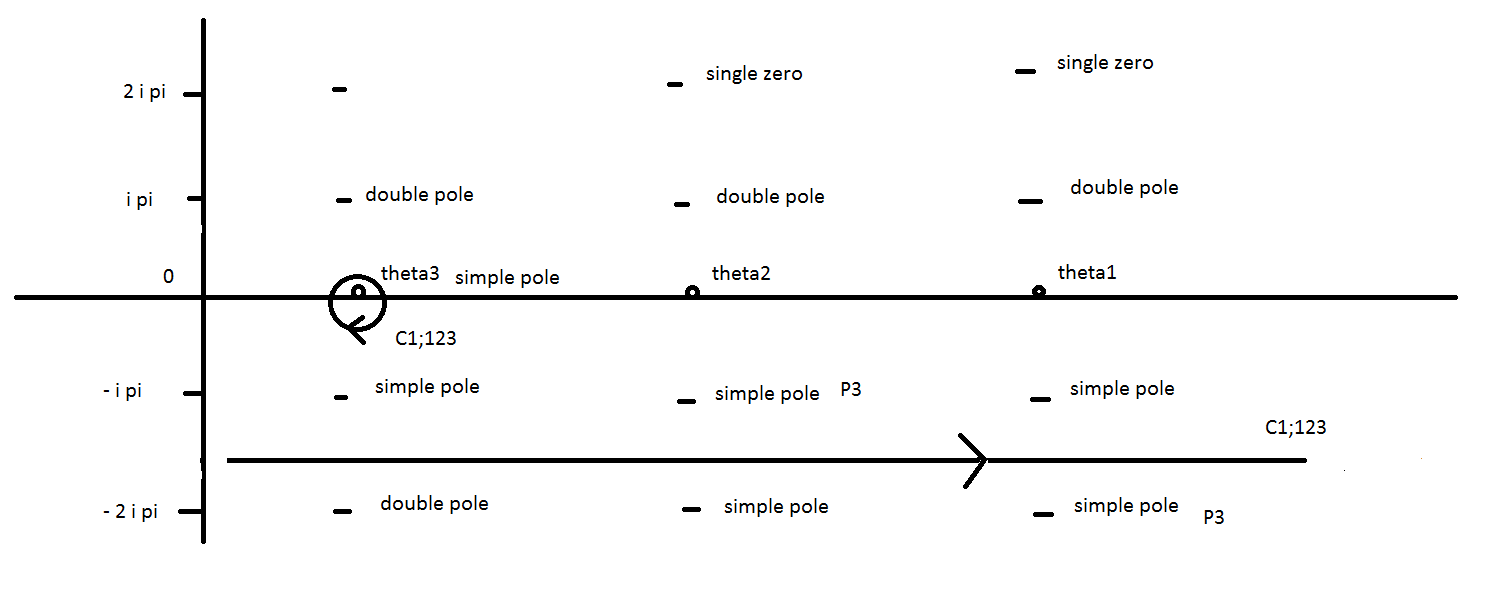}}
\caption{A portion of the portrait of the poles and zeros of $G_{bbf}$ and the contour $C1;123$.}
\label{fig1}
\end{figure}

\begin{figure}
\centerline{\includegraphics[width=18cm]{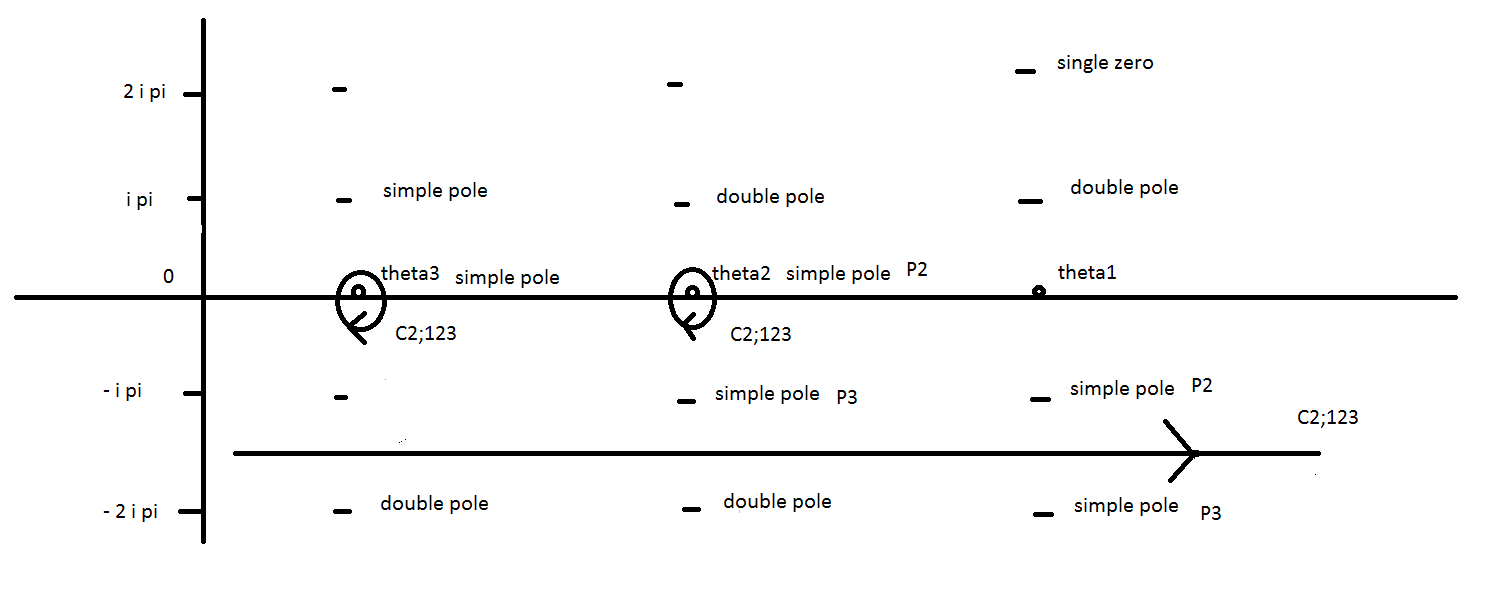}}
\caption{A portion of the portrait of the poles and zeros of $G_{bfb}$ and the contour $C2;123$.}
\label{fig2}
\end{figure}

\begin{figure}
\centerline{\includegraphics[width=18cm]{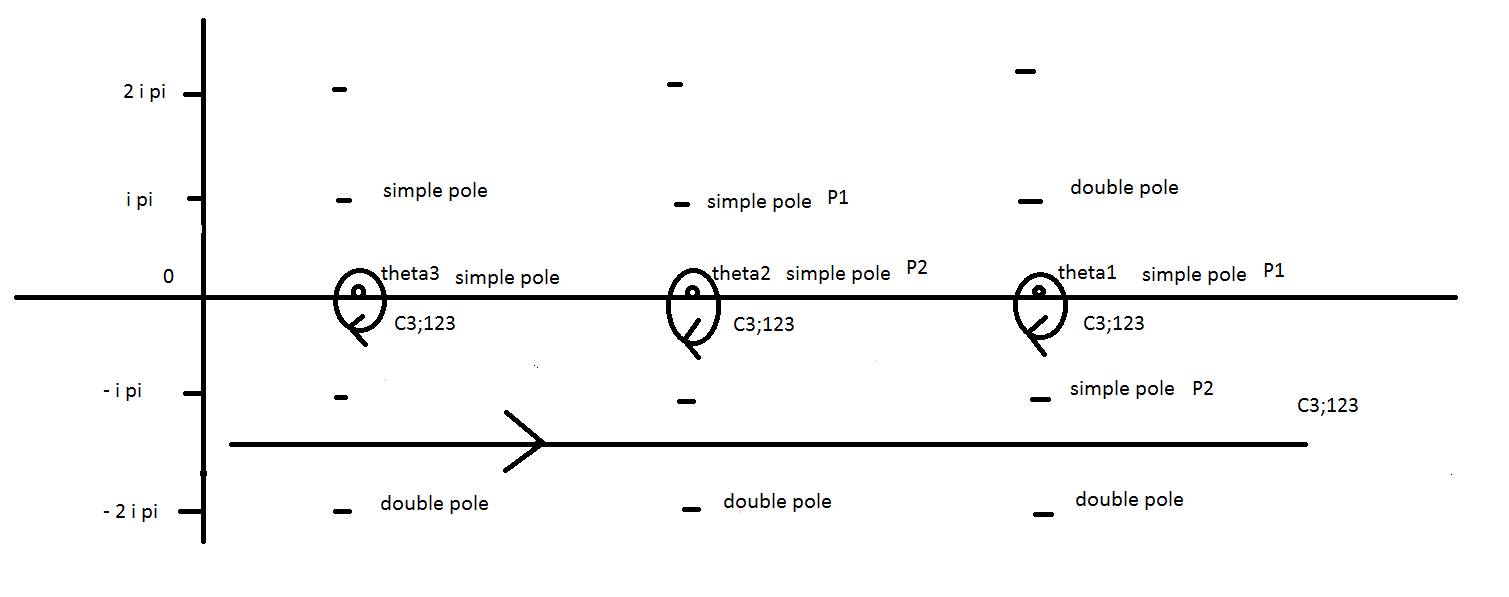}}
\caption{A portion of the portrait of the poles of $G_{fbb}$ and the contour $C3;123$.}
\label{fig3}
\end{figure}

We start with $G_{bbf}$ in figure \ref{fig1}. We then continue with $G_{bfb}$ in figure \ref{fig2}. Finally we have $G_{fbb}$ in figure \ref{fig3}.

Notice that the integrals around the poles could be done explicitly and have the term appearing in the formula. However, this would make the expressions bulkier and less prone to generalisation, but most importantly it would hinder the proofs to follow rather than facilitate them.

\subsection{\label{conv}Convergence}

We have - in part experimentally - verified the following asymptotics, relevant for the convergence of the integrals:

\begin{enumerate}

\item Zamolodchikov's dressing factor:
\begin{eqnarray}
\Phi(x + i y) \to e^{\pm i \frac{\pi}{4}}, \qquad x \, \, \mbox{real} \to \pm \infty, \qquad y \in [-2\pi,2\pi]
\end{eqnarray}
This can be explicitly verified using the integral representation (\ref{integrorep}): for real and large $\theta$ one has in fact, upon rescaling the integration variable,
\begin{eqnarray}
\Phi(\theta) = \exp \mp \int_0^\infty dy \, \frac{\sinh \frac{y}{i\pi}}{2 y \cosh^2 \frac{y}{2\theta}} \to \exp \pm i \int_0^\infty dy \, \frac{\sin \frac{y}{\pi}}{2 y} =  
\exp \pm i \frac{\pi}{4},
\end{eqnarray}
where the sign is the sign of $\theta$ (going to $\pm \infty$).

\item The function $Z(u)$: this function appears to be going to zero, for fixed imaginary part $\mbox{Im}(u) \in [-2\pi,2\pi]$ and large positive real part $\mbox{Re}(u)$, with great speed. As an example, the numerics already gives a magnitude of about $10^{-15}$ for $x\sim \pm 50$, with $l,m$ in the product (\ref{pro}) both truncated to $100$.   

By using the integral representation (\ref{eff}) we can actually be more accurate in our estimation: by a rescaling of the integration variable we get 
\begin{eqnarray}
F(\theta) = \exp \mp \int_0^\infty \frac{1-\cos \Big[y\big( \frac{1}{\pi} - \frac{i}{\theta}\big)\Big]}{2 y \sinh \frac{y}{\theta}+y\sinh \frac{2y}{\theta}} \sim \exp \pm \int_0^\infty \frac{- \theta(1-\cos \frac{y}{\pi})+i \sin \frac{y}{\pi}}{4 y^2} \sim  \exp -\frac{|\theta|}{8},\nonumber
\end{eqnarray}
where the sign is again the sign of $\theta$ (going to $\pm \infty$).

\item The remainder of the integrand goes like
\begin{eqnarray}
\frac{e^{(s -3\ell) u}}{\cosh\frac{u}{2}} 
\end{eqnarray}
which, for a specific example of $s=\frac{1}{2}$, $\ell = \frac{1}{4}$ we will encounter later on, goes at worst like $e^{\frac{|u|}{4}}$. This appears to fight quite strongly against the decay of $Z$, but numerically the latter still appears to be winning. In fact, one can try by excess of caution the combination of $Z(u)e^{\frac{|u|}{2}}$, which gives a magnitude of about $10^{-6}$ for $x\sim \pm 50$, with $l,m$ in the product (\ref{pro}) again both truncated to $100$. The same numerics at $x\sim \pm 100$ gives a magnitude of $10^{-11}$, which means that one single $Z$ of the three available can already on its own overcome the exponential. 

With our more accurate estimates we can finally say that convergence is guaranteed for all the integrals if
\begin{eqnarray}
|s-3\ell| -\frac{1}{2} + \frac{6}{8} -3 = |s-3\ell| - \frac{11}{4}<0. \label{bound}
\end{eqnarray}
We will see how feeble this condition still is though, as we may encounter cases where it will not be respected and a suitable regularisation may be necessary, as commented upon in particular in \cite{Babu2}, and as we will discuss in detail in a later section.  

\end{enumerate}

\section{\label{prop}Properties of the three $G$ functions}

We will now explicitly show that the expressions given in the previous section satisfy a series of properties which will allow to eventually prove the form factor requirements. Since the construction has followed \cite{BabuF}, the proofs also largely follow that paper, adapted to our particular situation. 

\subsection{\label{permutalo}Permutation}

The permutation properties of the functions (\ref{fermiFF}) relies on the RTT relations, as engineered by \cite{BabuF}, the main ingredient being that they satisfy
\begin{eqnarray}
\label{all}
&&G_{bfb}(\theta_1,\theta_2,\theta_3) = \Phi(\theta_1-\theta_2)\Big[-G_{fbb}(\theta_2,\theta_1,\theta_3) \, \tanh \frac{\theta_1 - \theta_2}{2} +G_{bfb}(\theta_2,\theta_1,\theta_3) \, \mbox{sech} \frac{\theta_1 - \theta_2}{2}\Big],\nonumber\\
&&G_{fbb}(\theta_1,\theta_2,\theta_3) = \Phi(\theta_1-\theta_2)\Big[G_{bfb}(\theta_2,\theta_1,\theta_3) \, \tanh \frac{\theta_1 - \theta_2}{2} +G_{fbb}(\theta_2,\theta_1,\theta_3) \, \mbox{sech} \frac{\theta_1 - \theta_2}{2}\Big],\nonumber\\
&&G_{bbf}(\theta_1,\theta_2,\theta_3) = \Phi(\theta_1-\theta_2)G_{bbf}(\theta_2,\theta_1,\theta_3),\qquad G_{fbb}(\theta_1,\theta_2,\theta_3) = \Phi(\theta_2-\theta_3)G_{fbb}(\theta_1,\theta_3,\theta_2)\nonumber\\
&&G_{bbf}(\theta_1,\theta_2,\theta_3) = \Phi(\theta_2-\theta_3)\Big[-G_{bfb}(\theta_1,\theta_3,\theta_2) \, \tanh \frac{\theta_2 - \theta_3}{2} +G_{bbf}(\theta_1,\theta_3,\theta_2) \, \mbox{sech} \frac{\theta_2 - \theta_3}{2}\Big],\nonumber\\
&&G_{bfb}(\theta_1,\theta_2,\theta_3) = \Phi(\theta_2-\theta_3)\Big[G_{bbf}(\theta_1,\theta_3,\theta_2) \, \tanh \frac{\theta_2 - \theta_3}{2} +G_{bfb}(\theta_1,\theta_3,\theta_2) \, \mbox{sech} \frac{\theta_2 - \theta_3}{2}\Big],\nonumber\\
\end{eqnarray}
provided we choose
\begin{eqnarray}
\label{nochoice}
N_{bbf} = N_{bfb} = N_{fbb} \equiv N_3.
\end{eqnarray}
This is verified in four steps. 

\begin{enumerate}
\item First, the part $\prod_{1\leq i<j\leq3}F(\theta_i-\theta_j)$, common to all three functions, is a standard construct precisely designed to produce the dressing factor under permutation. Therefore, the presence of $\Phi$ in all the relations in (\ref{all}) is taken care of by the product of $F$'s outside the integrals. The normalisation constants are also common to all three functions with the choice (\ref{nochoice}), and the function $\sigma(\theta_1,\theta_2,\theta_3)$ is completely symmetric by construction.
\item Second, one notices that the product $\prod_{i=1}^3 Z(\theta_i -u)\Phi(\theta_i -u)$ inside the integrand of all the form factors is completely invariant under permutations, so is the exponential factor $e^{\big[\ell \sum_{i=1}^3 \theta_i + (s - 3 \ell) u\big]}$. 
\item Third, the contours can be (and in fact originally should be) extended to encircle all the points $u=\theta_i$, $i=1,..,n$, while in the graphs of the previous section we have only highlighted the non-zero contributions. This means that all the contours - including the infinitely extended parts - can also be recast in a way which is manifestly common to all the permutations of the rapidities $\theta_i$. 
\item Fourth, what remains therefore to be shown is that the various components of the off-shell Bethe bra-vector (equivalently, the remaining parts of the integrands of the form factors) satisfy the relations (\ref{all}) without the dressing phase. This can be verified explicitly, since all such relations reduce to the basic identity 
\begin{eqnarray}
-\frac{\tanh \frac{\theta_2 - u}{2}}{\cosh \frac{\theta_1 - u}{2}} = \frac{\tanh \frac{\theta_1 - \theta_2}{2}}{\cosh \frac{\theta_1 - u}{2}} - \frac{\tanh \frac{\theta_1 - u}{2}}{\cosh \frac{\theta_1 - \theta_2}{2} \cosh \frac{\theta_2 - u}{2}}
\end{eqnarray}
and its variants, which form the ground for how the RTT relations work for our $R$-matrix.
\end{enumerate}

\subsection{\label{perrio}Periodicity}

\begin{figure}
\centerline{\includegraphics[width=18cm]{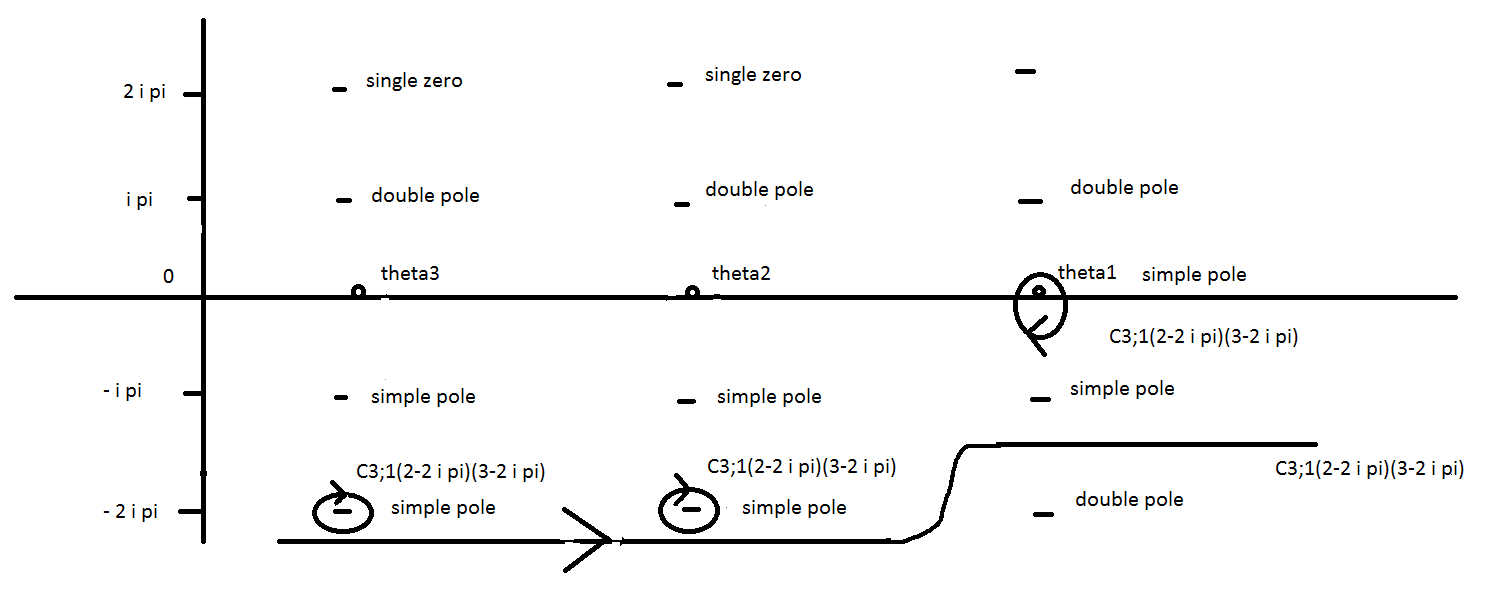}}
\caption{The contour $C3;1(2-2 i\pi)(3-2 i \pi)$.}
\label{fig4}
\end{figure}

Periodicity reads
\begin{eqnarray}
\label{perio}
\frac{G_{bbf}(\theta_1+2 i \pi,\theta_3,\theta_2)}{G_{bfb}(\theta_2,\theta_3,\theta_1)}= \frac{G_{bfb}(\theta_1+2 i \pi,\theta_3,\theta_2)}{G_{fbb}(\theta_2,\theta_3,\theta_1)} = 1 = -\frac{G_{fbb}(\theta_1+2 i \pi,\theta_3,\theta_2)}{G_{bbf}(\theta_2,\theta_3,\theta_1)},
\end{eqnarray}
where in the third instance we have taken into account a fermionic particle cycling through the fermionic operator. This is an argument based on the bare statistics, and in a future section we will refine it further - (\ref{perio}) will not be the final version of the periodicity condition. For the moment this may be taken as a mere toy-model calculation to display the properties of the integrals. The fact that the expressions (\ref{fermiFF}) satisfy these three conditions can be verified as follows. Let us start with the case of $G_{fbb}$. We first write the integral formula again in full, using the notation
\begin{eqnarray}
\label{nota}
\theta_i - \theta_j \equiv \theta_{ij}, \qquad i,j=0,1,2,3, \qquad u \equiv \theta_0. 
\end{eqnarray}
We have
\begin{eqnarray}
&&G_{fbb}(\theta_1, \theta_2,\theta_3) = N_{fbb} \, \sigma(\theta_1,\theta_2,\theta_3) \, F(\theta_{12})F(\theta_{13})F(\theta_{23})\int_{C3;123} d\theta_0 \frac{e^{\big[\ell \sum_{i=1}^3 \theta_i + (s - 3 \ell) \theta_0\big]}}{\cosh \frac{\theta_{10}}{2}}\nonumber\\
&&\qquad \times\frac{\Phi(\theta_{10})\Phi(\theta_{20})\Phi(\theta_{30})}{F(\theta_{10})F(\theta_{20})F(\theta_{30})F(\theta_{10}+i\pi)F(\theta_{20}+i\pi)F(\theta_{30}+i\pi)\sinh \theta_{10} \sinh \theta_{20} \sinh \theta_{30}}.\nonumber
\end{eqnarray}
Let us first shift $\theta_1$ by $2 \pi i$. Notice that the function $\sigma$ is invariant under permutations, and the variables it depends on are $x_i = e^{\theta_i}$, hence invariant under shift of $2 \pi i$ of any rapidity. We get
\begin{eqnarray}
&&G_{fbb}(\theta_1+2\pi i, \theta_2,\theta_3) = N_{fbb} \, \sigma(\theta_1,\theta_2,\theta_3) \,F(\theta_{12}+2 \pi i)F(\theta_{13}+2 \pi i)F(\theta_{23}) \int_{C3;(1+2\pi i)23} d\theta_0\nonumber\\
&&\frac{(-1)e^{\big[\ell \sum_{i=1}^3 \theta_i + (s - 3 \ell) \theta_0\big]} \, e^{2\pi i \ell}\, \Phi(\theta_{10}+2\pi i)\Phi(\theta_{20})\Phi(\theta_{30})}{\cosh \frac{\theta_{10}}{2}F(\theta_{10}+2\pi i)F(\theta_{20})F(\theta_{30})F(\theta_{10}+3i\pi)F(\theta_{20}+i\pi)F(\theta_{30}+i\pi)\sinh \theta_{10} \sinh \theta_{20} \sinh \theta_{30}}.\nonumber
\end{eqnarray}
where the contour $C3;(1+2\pi i)23$ is now pushed upwards of $2 \pi i$ with respect to $C3;123$ in all its sections pertaining to the tower of $\theta_1$. By using
\begin{eqnarray}
\label{lafor}
F(\theta+2\pi i) = F(-\theta), \qquad F(-\theta) = \frac{F(\theta)}{\Phi(\theta)},
\end{eqnarray}
we obtain
\begin{eqnarray}
&&G_{fbb}(\theta_1+2\pi i, \theta_2,\theta_3) = -e^{2 \pi i \ell} \, N_{fbb} \, \sigma(\theta_1,\theta_2,\theta_3) \,F(\theta_{21})F(\theta_{31})F(\theta_{23})\int_{C3;(1+2\pi i)23} d\theta_0 \times\nonumber\\
&&\frac{e^{\big[\ell \sum_{i=1}^3 \theta_i + (s - 3 \ell) \theta_0\big]}}{\cosh \frac{\theta_{10}}{2}}\frac{\Phi(\theta_{10})\Phi(\theta_{10}+i\pi)\Phi(\theta_{10}+2\pi i)\Phi(\theta_{20})\Phi(\theta_{30})}{F(\theta_{10})F(\theta_{20})F(\theta_{30})
F(\theta_{10}+i\pi)F(\theta_{20}+i\pi)F(\theta_{30}+i\pi)\sinh \theta_{10} \sinh \theta_{20} \sinh \theta_{30}}.\nonumber
\end{eqnarray}
Now we use 
\begin{eqnarray}
\label{formu}
\Phi(-\theta) = \frac{1}{\Phi(\theta)}, \qquad \Phi(\theta)\Phi(\theta + i \pi) = i \tanh\frac{\theta}{2}, \qquad \Phi(\theta+2 i \pi) = \Phi(\theta) \coth^2 \frac{\theta}{2}, 
\end{eqnarray}
the first two equations being the braiding unitarity and crossing symmetry of Zamolodchikov's dressing factor, and the third being obtained by applying crossing twice to the meromorphic expression $\Phi(\theta)$, to obtain
\begin{eqnarray}
\label{conto}
&&G_{fbb}(\theta_1+2\pi i, \theta_2,\theta_3) = -iN_{fbb} \, \sigma(\theta_1,\theta_2,\theta_3) \,F(\theta_{21})F(\theta_{31})F(\theta_{23})\int_{C3;(1+2\pi i)23} d\theta_0 \times\\
&&\frac{e^{\big[\ell \sum_{i=1}^3 \theta_i + (s - 3 \ell) \theta_0\big]}}{\cosh \frac{\theta_{10}}{2}}\frac{e^{2 \pi i \ell} \, \coth \frac{\theta_{10}}{2}\Phi(\theta_{10})\Phi(\theta_{20})\Phi(\theta_{30})}{F(\theta_{10})F(\theta_{20})F(\theta_{30})
F(\theta_{10}+i\pi)F(\theta_{20}+i\pi)F(\theta_{30}+i\pi)\sinh \theta_{10} \sinh \theta_{20} \sinh \theta_{30}}.\nonumber
\end{eqnarray}
Let us now compare this expression with 
\begin{eqnarray}
\label{bla}
&&G_{bbf}(\theta_2,\theta_3,\theta_1) = N_{bbf} \, \sigma(\theta_1,\theta_2,\theta_3) \,F(\theta_{21})F(\theta_{31})F(\theta_{23})\int_{C1;231} d\theta_0 \times\\
&&\frac{e^{\big[\ell \sum_{i=1}^3 \theta_i + (s - 3 \ell) \theta_0\big]}}{\cosh \frac{\theta_{10}}{2}}\frac{\tanh \frac{\theta_{20}}{2}\tanh \frac{\theta_{30}}{2}\Phi(\theta_{10})\Phi(\theta_{20})\Phi(\theta_{30})}{F(\theta_{10})F(\theta_{20})F(\theta_{30})
F(\theta_{10}+i\pi)F(\theta_{20}+i\pi)F(\theta_{30}+i\pi)\sinh \theta_{10} \sinh \theta_{20} \sinh \theta_{30}}.\nonumber
\end{eqnarray}
In order to bring them into a similar form, we can change variable $\theta_0 \to \tilde{\theta}_0 = \theta_0 - 2 i \pi$ in (\ref{conto}). The contour in the new variable is now $C3;1(2-2\pi i)(3-2\pi i)$. We obtain the pattern described in figure \ref{fig4}.

At the same time the change of variables transforms the integrand by turning all the $\theta_i - \theta_0$ into $\theta_i - \tilde{\theta_0} -2 i \pi$. By virtue of formulas (\ref{formu}) and (\ref{lafor}), combined with the second equation in (\ref{Watson}), we can perform a series of manipulations which bring the result into the form
\begin{eqnarray}
\label{contos}
&&G_{fbb}(\theta_1+2\pi i, \theta_2,\theta_3) = - N_{fbb} \, \sigma(\theta_1,\theta_2,\theta_3) \,F(\theta_{21})F(\theta_{31})F(\theta_{23}) \int_{C3;1(2-2\pi i)(3-2\pi i)} d\tilde{\theta}_0\nonumber\\
&&\frac{e^{\big[\ell \sum_{i=1}^3 \theta_i + (s - 3 \ell) \tilde{\theta}_0\big]}}{\cosh \frac{\theta_{1\tilde{0}}}{2}}\frac{e^{2 \pi i (s-2\ell)} \,\coth \frac{\theta_{1\tilde{0}}}{2}\Phi(\theta_{1\tilde{0}})\Phi(\theta_{2\tilde{0}})\Phi(\theta_{3\tilde{0}})\prod_{i=1}^3 \tanh \frac{\theta_i - \tilde{\theta}_0}{2}}{F(\theta_{1\tilde{0}})F(\theta_{2\tilde{0}})F(\theta_{3\tilde{0}})
F(\theta_{1\tilde{0}}+i\pi)F(\theta_{2\tilde{0}}+i\pi)F(\theta_{3\tilde{0}}+i\pi)\sinh \theta_{1\tilde{0}} \sinh \theta_{2\tilde{0}} \sinh \theta_{3\tilde{0}}},\nonumber\\
\end{eqnarray}
which is precisely what is needed to obtain the integrand of (\ref{bla}). In order to match with (\ref{perio}) we require
\begin{eqnarray}
\label{rel1}
e^{2\pi i (s - 2\ell)}N_{fbb} = N_{bbf}.
\end{eqnarray}
Given that $N_{fbb} = N_{bbf}$ from the analysis of the permutation requirement, we conclude that in this case
\begin{eqnarray}
e^{2\pi i (s - 2\ell)} = 1.
\end{eqnarray}

\begin{figure}
\centerline{\includegraphics[width=18cm]{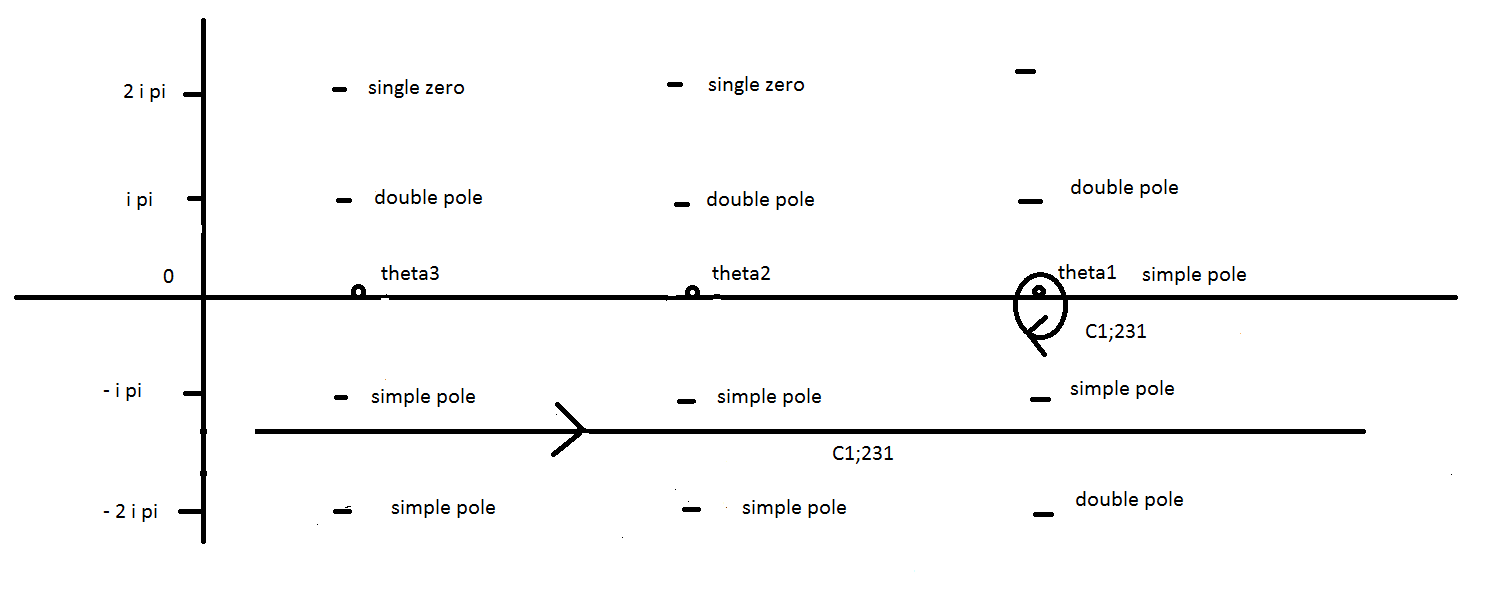}}
\caption{The contour $C1;231$.}
\label{fig5}
\end{figure}

The contour $C3;1(2-2\pi i)(3-2\pi i)$ superficially differs from $C1;231$, as shown in figure \ref{fig5}.
One can however see that the two are in fact exactly the same - after the extended part has been move across the two poles and two residues have been picked up. No contributions at infinity are received since the integrands decay very fast in the entire region of interest, as previously discussed (this consideration will be subject to a later caveat). 

The other two conditions in (\ref{perio}) work very similarly, although they are in fact easier to deal with. This is because one can notice that the shift $\theta_1 \to \theta_1 + 2 i \pi$, combined with the usual manipulations (\ref{formu}), (\ref{lafor}) and the second equation in (\ref{Watson}), is already sufficient to bring the integrand of $G_{bfb}$ into the integrand of $G_{fbb}(\theta_2,\theta_3,\theta_1)$, and the integrand of $G_{bbf}$ into the integrand of $G_{bfb}(\theta_2,\theta_3,\theta_1)$. In the process one collects a number of minuses, and the part outside the integral also rearranges nicely. In the end the actual expressions match if we require
\begin{eqnarray}
\label{rel2}
-i e^{2 i \pi \ell} N_{bfb} = N_{fbb}, \qquad i e^{2 i \pi \ell} N_{bbf} = - N_{bfb}.
\end{eqnarray}
Considering that $N_{bbf}=N_{bfb}=N_{fbb}$, we conclude that both conditions (\ref{rel2}) reduce to the single 
\begin{eqnarray}
e^{2 i \pi \ell} = i.
\end{eqnarray}    
As one can see, the relations (\ref{rel1}) and (\ref{rel2}) are compatible if the spin is half-integer, say $s=\frac{1}{2}$, and $\ell$ is chosen to be $\frac{1}{4}$, which is consistent with a fermionic operator for the current purposes. In this toy-exercise we also get a convergent integral which this particular choice of $\ell$.

The contours also match in a similar fashion as we have seen above, via a straightforward shift which picks up the correct residues. We have drawn them here below.

\begin{figure}\centerline{\includegraphics[width=18cm]{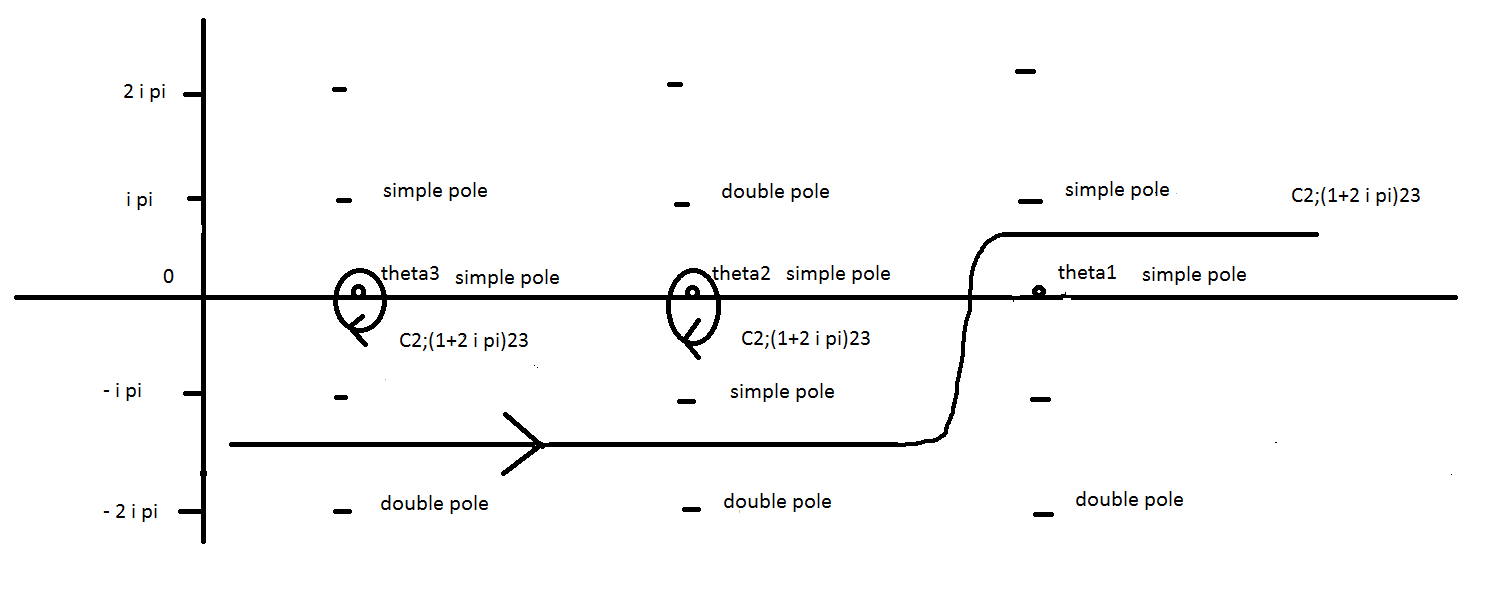}}
\caption{The contour $C2;(1+2 i \pi)23$.}
\label{fig6}
\end{figure}

\begin{figure}
\centerline{\includegraphics[width=18cm]{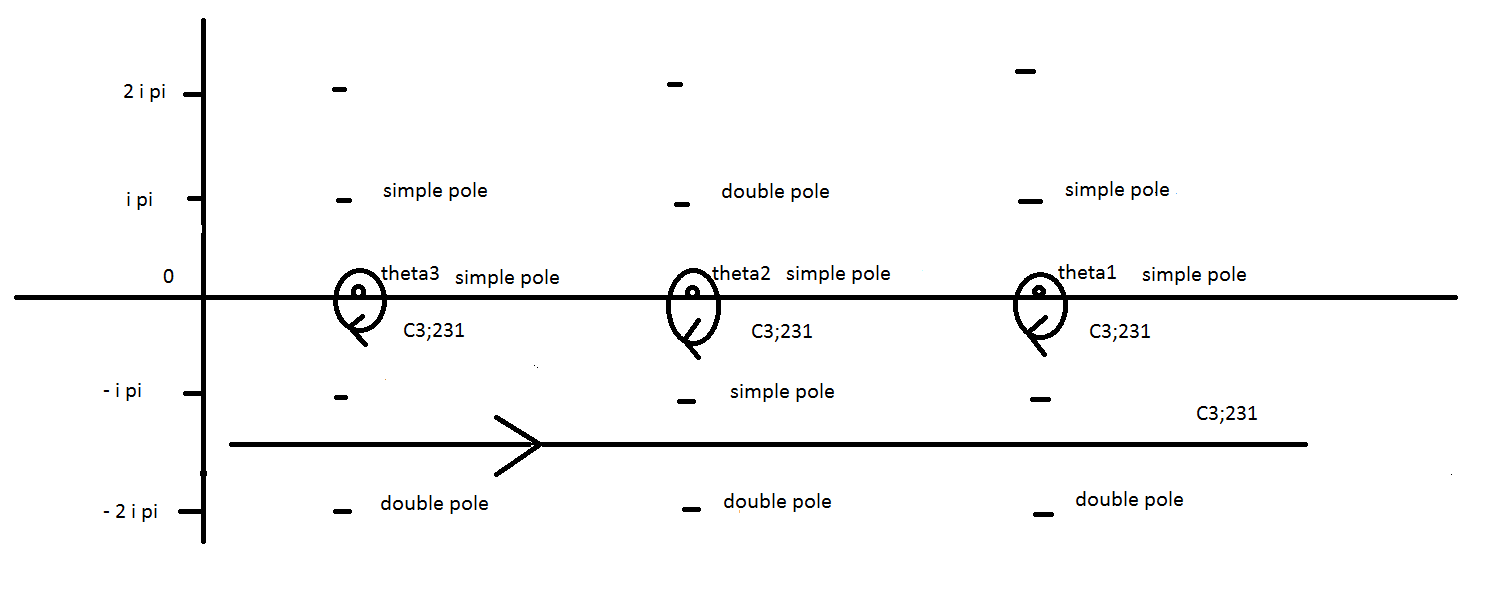}}
\caption{The contour $C3;231$.}
\label{fig7}
\end{figure}

The contour $C2;(1+2\pi i)23$ relevant to $G_{bfb}(\theta_1+2 i \pi,\theta_2,\theta_3)$ is depicted in figure \ref{fig6}.
This contour can be seen to match the contour $C3;231$ relevant to $G_{fbb}(\theta_2,\theta_3,\theta_1)$, which is depicted in figure \ref{fig7}. 

\begin{figure}
\centerline{\includegraphics[width=18cm]{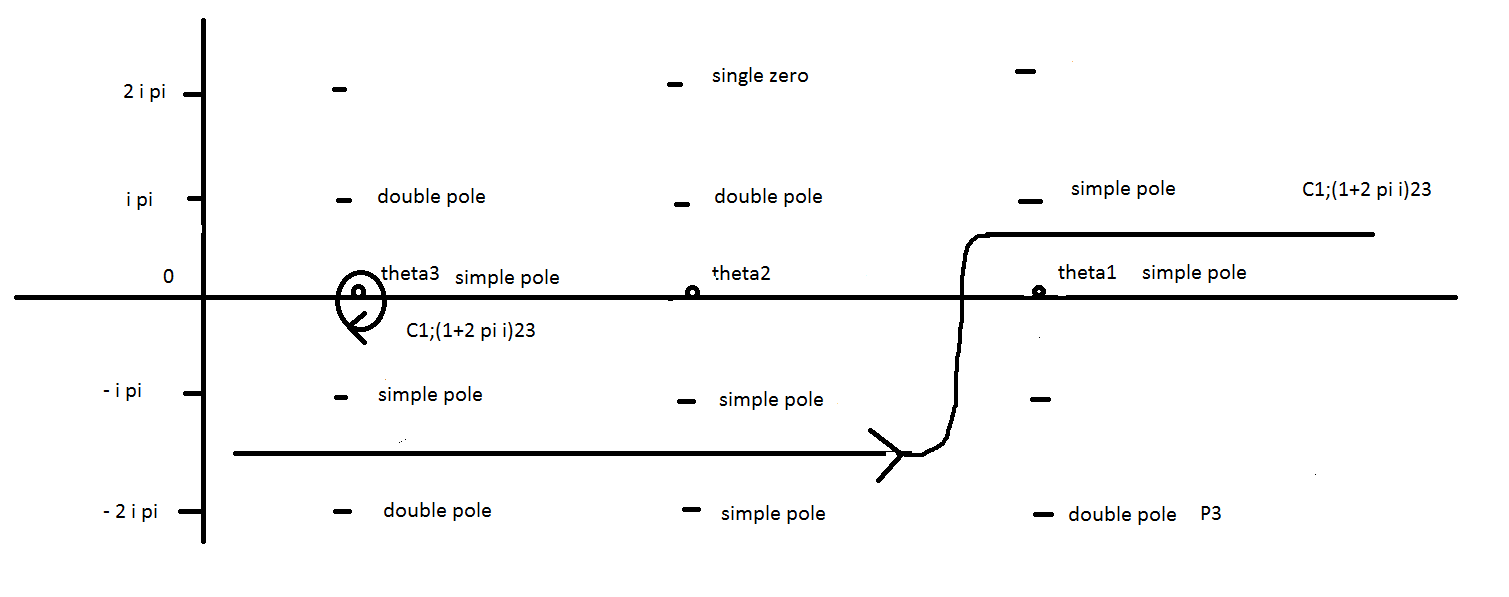}}
\caption{The contour $C1;(1+2 i\pi)23$.}
\label{fig8}.
\end{figure}

\begin{figure}
\centerline{\includegraphics[width=18cm]{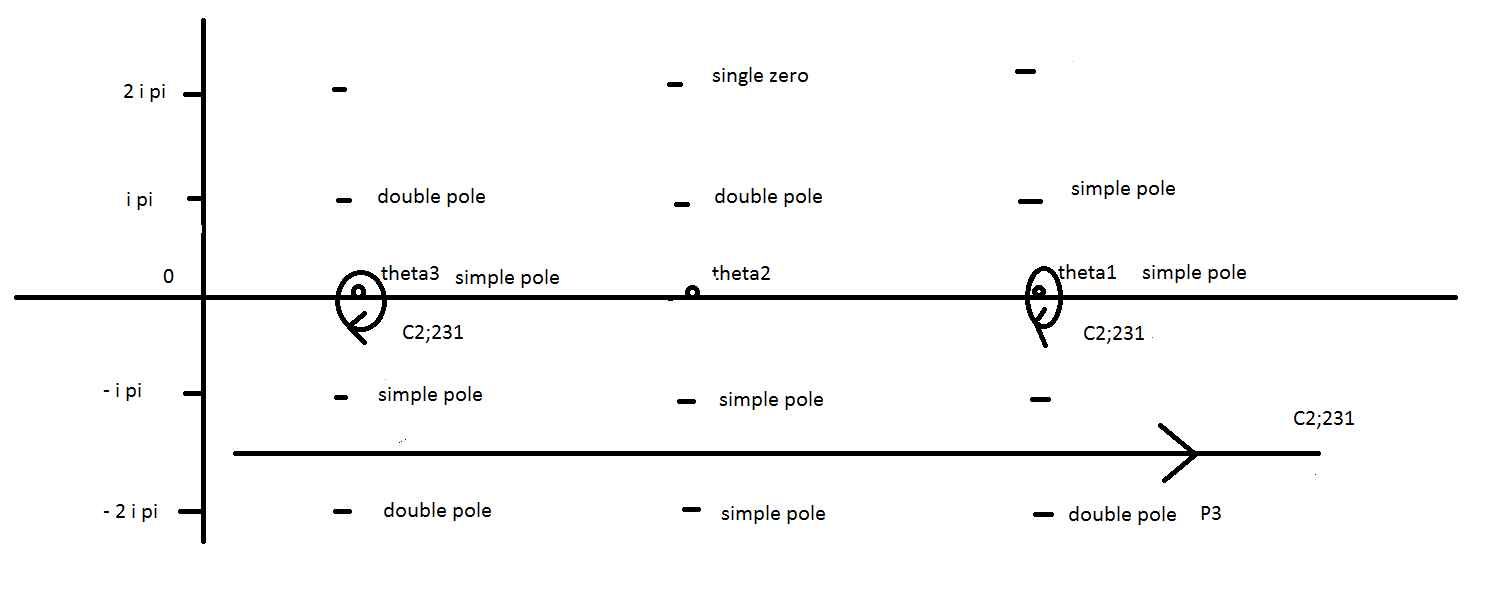}}
\caption{The contour $C2;231$.}
\label{fig9}
\end{figure}

The contour $C1;(1+2\pi i)23$ relevant to $G_{bbf}(\theta_1+2 i \pi,\theta_2,\theta_3)$ is portrayed in figure \ref{fig8}.
It can be seen to match the contour $C2;231$ relevant to $G_{bfb}(\theta_2,\theta_3,\theta_1)$, portrayed in figure \ref{fig9}.
This concludes the proof of periodicity for what concerns the functions and the contour, leaving the constants to a more detailed later discussion\footnote{ We wish to remark how rather non-trivial the re-adjustment of contours is in our case. The proof in the appendix B of \cite{BabuF} suggests that in their situation the re-adjustment of contours is much more straightforward and there is no need to cross poles. It is our shortcoming that we do entirely grasp their argument of why that is so in their case, since it naively appears that their poles are in fact more densely scattered across the region than ours. We believe that it must be since they approach the proof in a slightly different way, by first rearranging the requirement of the axioms. We plan to go back to the model of \cite{BabuF} and understand their rather impressive proof more fully for our own completeness, also in view of potentially trying one day to come up with an equally general proof for our multi-particle conjecture of one of our later sections.} 

\subsection{\label{boos}Boost}

The boost requirement involves the shift of all the rapidities $\theta_i$ by the same real constant $\Lambda$. If we apply this transformation to the expressions (\ref{fermiFF}) we see that the pre-factor outside the integral is invariant, as it only depends on the difference of the rapidities. For the function $ \sigma(\theta_1,\theta_2,\theta_3)$ (\ref{sigma}) invariance follows from the fact that the homogeneous polynomials $\sigma^{(3)}_2$ at the numerator and $\Big[\sigma^{(3)}_1\Big]^2$ at the denominator have the same degree, hence the function is overall scale-invariant. The arguments of all the functions appearing in the integrand, with the exception of the exponential factor 
\begin{eqnarray}
\label{expo}
e^{\big[\ell \sum_{i=1}^3 \theta_i + (s - 3 \ell) \tilde{\theta}_0\big]},
\end{eqnarray}
become functions of $\theta_{i0} +\Lambda$. We therefore perform a change of integration variable $\theta_0 \to \theta_0' = \theta_0 - \Lambda$ to reabsorb the shift. The contour is unaffected by this change of variables for what concerns its infinitely extended part (see graphs in section \ref{graphs}). In the old variable $\theta_0$ the small circles by construction surround the points $\theta_0 = \theta_i + \Lambda$ (after having performed the boost). This means that following the change of variable, the circles in the $\theta_0'$ plane surround again the points $\theta_i$. We have therefore reconstructed the same integral as we had before the boost, the only remaining difference is now that the exponential factor (\ref{expo}) becomes
\begin{eqnarray}
e^{\big[\ell \sum_{i=1}^3 \theta_i +3\ell\Lambda+ (s - 3 \ell) \theta'_0 +(s-3\ell)\Lambda\big]}.
\end{eqnarray} 
Hence the form factors simply change in total by multiplication by
\begin{eqnarray}
e^{s\Lambda},
\end{eqnarray}
as it should be to conform with Lorentz symmetry.

\section{\label{singo}Singularities}

Since we are in a massless situation (in truth conformal, as explained in the introduction), we must not have bound state singularities. This is respected by our integrals - it is easy to see that no pinching of poles can occur in our integrals across a span of less than $i\pi$, which is the size of the physical region. Pinching across $i \pi$, for instance at $\theta_1 \to \theta_2 + i \pi$ which is the subject of (\ref{reso}), does occur and this will be the kinematical singularities we shall now focus on. Because we are studying the scattering of left-left and right-right movers, we might indeed on the one hand expect kinematical singularities, as remarked by \cite{DelfinoMussSimo}. The massive form factors have of course to satisfy the specific relation which involves its residue at the kinematical singularities, which can be formulated as (\ref{reso}). Massless pure right-right and left-left form factors are expected on general grouds to follow the massive axiomatics except for the bound state requirement, however in this case the massive theory is non-relativistic and we might not expect the same requirements for the kinematical singularities. Nevertheless, (\ref{reso}) will be satisfied by the formulas which we have written down in the previous section, where the residue is taken at the simple pole  $\theta_1 \to \theta_2 + i \pi$. By permutation and periodicity this is then guaranteed in all channels. 

\subsection{\label{antipo}Anti-particles}

First of all, in order to speak of residue at a kinematical singularity we need to be able to introduce anti-particles. In the case of \cite{BabuF}, the soliton and anti-soliton are anti-particle of each other. Here, we have a notion of anti-particle whereby for each (boson,fermion) particle of type $(b,f)$ in the representation we have dealt with so far - which we will call $\rho$ - one has a (boson,fermion) antiparticle of type $(\bar{b},\bar{f})$ in the representation $\bar{\rho}$. There are several mixed-representation $S$- (and, upon permutation, $R$-) matrices, and also an $S$-matrix for the scattering of the $\bar{\rho}$ representation with $\bar{\rho}$. In the massless case these representations can be chosen to be isomorphic, such that the $S$-matrices will always be the same {\it mutatis mutandis}\footnote{To describe the full spectrum one needs in reality two copies of the multiplets we describe here, connected by the symmetry dubbed $\mathfrak{su}(2)_\circ$ in \cite{CompleteT4}. The two copies however will behave identically, and it is sufficient to focus on one copy in our study the same way it was done in \cite{DiegoBogdanAle}.}. In the particularly simple case of the BMN limit we are treating here, this implies that the functions associated with the various form factors with anti-particles present are always the same three functions we have studied so far, taken with different assignment of labels. As a rule of thumb - which we will make more precise shortly - it is sufficient to replace the label $b$ with $\bar{f}$, respectively $f$ with $\bar{b}$ (and switch the operator in the multiplet accordingly, in order to respect the fermion-number conservation), plus some adjustments of constant factors which we will describe below. We shall focus in particular on $G_{\bar{b}bb}$ for a bosonic operator (which has the same functional form as $G_{fbb}$ for a fermionic operator), and on $G_{\bar{f}fb}$ for a bosonic operator (which has the same functional form as $G_{bfb}$ for a fermionic operator). We will see that $G_{\bar{f}bf}$, which has the same functional form as $G_{bbf}$, will not have a residue contribution, consistently with the fact that $\bar{f}$ is not the anti-particle of $b$. 

Let us begin by listing the $S$-matrices for all the types of states in our conventions, and their crossing-symmetric and unitary dressing factors:
\begin{eqnarray}
\label{fasin}
\Phi_{\rho\rho}(\theta) = \Phi_{\bar{\rho}\bar{\rho}}(\theta)= \Phi(\theta), \qquad \Phi_{\bar{\rho}\rho}(\theta) = -\Phi_{\rho\bar{\rho}}(\theta)= i \Phi(\theta),
\end{eqnarray}
where $\Phi(\theta)$ is Zamolodchikov's dressing factor (\ref{zamo}), and
\begin{eqnarray}
&&S_{bb}^{bb}(\theta) = \Phi_{\rho \rho}(\theta), \qquad S_{ff}^{ff}(\theta) = \Phi_{\rho \rho}(\theta),\qquad S_{bf}^{fb}(\theta) = -\Phi_{\rho \rho}(\theta)\tanh \frac{\theta}{2}, \qquad S_{bf}^{bf}(\theta) = \Phi_{\rho \rho}(\theta) \mbox{sech}\frac{\theta}{2},\nonumber\\
&&S_{fb}^{bf}(\theta) = \Phi_{\rho \rho}(\theta)\tanh \frac{\theta}{2}, \qquad S_{fb}^{fb}(\theta) = \Phi_{\rho \rho}(\theta) \mbox{sech}\frac{\theta}{2},\qquad S_{\bar{b}\bar{b}}^{\bar{b}\bar{b}}(\theta) = \Phi_{\bar{\rho}\bar{\rho}}(\theta), \qquad S_{\bar{f}\bar{f}}^{\bar{f}\bar{f}}(\theta) = \Phi_{\bar{\rho}\bar{\rho}}(\theta),\nonumber\\
&&S_{\bar{b}\bar{f}}^{\bar{f}\bar{b}}(\theta) = -\Phi_{\bar{\rho}\bar{\rho}}(\theta)\tanh \frac{\theta}{2}, \qquad S_{\bar{b}\bar{f}}^{\bar{b}\bar{f}}(\theta) = \Phi_{\bar{\rho}\bar{\rho}}(\theta) \mbox{sech}\frac{\theta}{2},\nonumber\\
&&\qquad S_{\bar{f}\bar{b}}^{\bar{b}\bar{f}}(\theta) = \Phi_{\bar{\rho}\bar{\rho}}(\theta)\tanh \frac{\theta}{2}, \qquad S_{\bar{f}\bar{b}}^{\bar{f}\bar{b}}(\theta) = \Phi_{\bar{\rho}\bar{\rho}}(\theta) \mbox{sech}\frac{\theta}{2},\nonumber\\
&&S_{\bar{b}f}^{f\bar{b}}(\theta) = \Phi_{\bar{\rho}\rho}(\theta), \qquad S_{\bar{f}b}^{b\bar{f}}(\theta) = -\Phi_{\bar{\rho}\rho}(\theta),\qquad S_{\bar{b}b}^{b\bar{b}}(\theta) = -\Phi_{\bar{\rho}\rho}(\theta)\tanh \frac{\theta}{2}, \quad S_{\bar{b}b}^{f\bar{f}}(\theta) = \Phi_{\bar{\rho}\rho}(\theta) \mbox{sech}\frac{\theta}{2},\nonumber\\
&&S_{\bar{f}f}^{f\bar{f}}(\theta) = -\Phi_{\bar{\rho}\rho}(\theta)\tanh \frac{\theta}{2}, \qquad S_{\bar{f}f}^{b\bar{b}}(\theta) = -\Phi_{\bar{\rho}\rho}(\theta) \mbox{sech}\frac{\theta}{2},\quad S_{f\bar{b}}^{\bar{b}f}(\theta) = \Phi_{\rho\bar{\rho}}(\theta), \quad S_{b\bar{f}}^{\bar{f}b}(\theta) = -\Phi_{\rho\bar{\rho}}(\theta),\nonumber\\
&&S_{b\bar{b}}^{\bar{b}b}(\theta) = \Phi_{\rho\bar{\rho}}(\theta)\tanh \frac{\theta}{2}, \qquad S_{b\bar{b}}^{\bar{f}f}(\theta) = -\Phi_{\rho\bar{\rho}}(\theta) \mbox{sech}\frac{\theta}{2},\nonumber\\
&&\qquad S_{f\bar{f}}^{\bar{f}f}(\theta) = \Phi_{\rho\bar{\rho}}(\theta)\tanh \frac{\theta}{2}, \qquad S_{f\bar{f}}^{\bar{b}b}(\theta) = \Phi_{\rho\bar{\rho}}(\theta) \mbox{sech}\frac{\theta}{2}.
\end{eqnarray}
Notice that this setting is completely simmetric $\rho \leftrightarrow \bar{\rho}$: one can easily see that, as matrices on their respective spaces, the matrices of $R_{\rho\rho}$ and of $R_{\bar{\rho}\bar{\rho}}$ coincide with each other in form, and the matrices of $R_{\rho\bar{\rho}}$ and of $R_{\bar{\rho}\rho}$ coincide with each other in form.

One can check that the associated $R$-matrices (obtained by $S = \Pi \circ R$, $\Pi$ being the graded permutation), satisfy all the eigth possible direct and mixed Yang-Baxter equations, and satisfy crossing and braiding unitarity \cite{DiegoBogdanAle}. In Hopf algebra langauge one has
\begin{eqnarray}
\label{cruni}
&&R_{\rho\rho}(\theta)R_{\rho\rho}^{op}(-\theta)=\mathfrak{1}\otimes \mathfrak{1}, \qquad R_{\bar{\rho}\bar{\rho}}(\theta)R_{\bar{\rho}\bar{\rho}}^{op}(-\theta)=\mathfrak{1}\otimes \mathfrak{1}, \nonumber\\
&&R_{\bar{\rho}\rho}(\theta)R_{\rho\bar{\rho}}^{op}(-\theta)=\mathfrak{1}\otimes \mathfrak{1}, \qquad R_{\rho\bar{\rho}}(\theta)R_{\bar{\rho}\rho}^{op}(-\theta)=\mathfrak{1}\otimes \mathfrak{1}, \nonumber\\
&&R_{\rho\rho}(\theta)[{\bf C}^{-1}\otimes \mathfrak{1}] R_{\bar{\rho}\rho}^{st_1}(\theta+i \pi)[{\bf C}\otimes \mathfrak{1}] = \mathfrak{1}\otimes \mathfrak{1}, \qquad R_{\rho\bar{\rho}}(\theta)[{\bf C}^{-1}\otimes \mathfrak{1}] R_{\bar{\rho}\bar{\rho}}^{st_1}(\theta+i \pi)[{\bf C}\otimes \mathfrak{1}] = \mathfrak{1}\otimes \mathfrak{1},\nonumber\\
&&R_{\rho\rho}(\theta)[\mathfrak{1} \otimes {\bf C}] R_{\rho\bar{\rho}}^{st_2}(\theta+i \pi)[\mathfrak{1}\otimes {\bf C}^{-1}] = \mathfrak{1}\otimes \mathfrak{1}, \qquad R_{\bar{\rho}\rho}(\theta)[\mathfrak{1}\otimes{\bf C}] R_{\bar{\rho}\bar{\rho}}^{st_2}(\theta+i \pi)[\mathfrak{1}\otimes{\bf C}^{-1}] = \mathfrak{1}\otimes \mathfrak{1},\nonumber
\end{eqnarray}
where $st_i$ denotes supertransposition in space $i$ of the the tensor product, $i=1,2$, the apex ${X^{op}}$ denotes $\Pi \circ X$, and $\bf C$ is the charge-conjugation matrix, with non-zero entries ${\bf C}_b^{\bar{b}}=1$, ${\bf C}_f^{\bar{f}}=i$. We use the conventions 
\begin{eqnarray}
[E_{ij}]^{st} = (-)^{[i][j]+[i]} E_{ji}, 
\end{eqnarray}
if $E_{ij}$ are the matrix unities with all zeros but $1$ in row $i$, column $j$, and $[i]$ is the bare fermionic grading of an index: $[b]=0$, $[f]=1$. All the relations above in essence follow from the two fundamental properties of Zamolodchikov's dressing factor which we have already encountered before:
\begin{eqnarray} 
\label{zamol}
\Phi(\theta)\Phi(-\theta)=1, \qquad \Phi(\theta)\Phi(\theta + i \pi)=i \tanh \frac{\theta}{2},
\end{eqnarray}
the second one being (\ref{crosszamo}). We recall here that both properties (\ref{zamol}) are explicitly proven using the representation (\ref{zamo}) and standard identities of gamma functions. We also remarks that the charge-conjugation matrix does not square to the identity. When it comes to the anti-particle of an anti-particle one can prove the following relations:
\begin{eqnarray}
\label{feature}
&&R_{\bar{\rho}\rho}(\theta)[{\bf \bar{C}}^{-1}\otimes \mathfrak{1}] R_{\rho\rho}^{st_1}(\theta+i \pi)[{\bf \bar{C}}\otimes \mathfrak{1}] = \mathfrak{1}\otimes \mathfrak{1},\qquad R_{\bar{\rho}\bar{\rho}}(\theta)[{\bf \bar{C}}^{-1}\otimes \mathfrak{1}] R_{\rho\bar{\rho}}^{st_1}(\theta+i \pi)[{\bf \bar{C}}\otimes \mathfrak{1}] = \mathfrak{1}\otimes \mathfrak{1}\nonumber\\
&&R_{\rho\bar{\rho}}(\theta)[\mathfrak{1} \otimes {\bf \bar{C}}] R_{\rho\rho}^{st_2}(\theta+i \pi)[\mathfrak{1}\otimes {\bf \bar{C}}^{-1}] = \mathfrak{1}\otimes \mathfrak{1}, \qquad R_{\bar{\rho}\bar{\rho}}(\theta)[\mathfrak{1}\otimes{\bf \bar{C}}] R_{\bar{\rho}\rho}^{st_2}(\theta+i \pi)[\mathfrak{1}\otimes{\bf \bar{C}}^{-1}] = \mathfrak{1}\otimes \mathfrak{1},\nonumber\\
\end{eqnarray}
with ${\bf \bar{C}}_{\bar{b}}^b=-i$ and ${\bf \bar{C}}_{\bar{f}}^f=1$.

We need to slightly upgrade our formulas for the form factors to account for the plurality of states. Before we had no need to distinguish the various representations, since we were only dealing with particles, but now we should. Let us diplay one particular upgraded formula and discuss its components - the logic will then be clear for how the other ones should go. For example, let us consider the form factor $G_{\bar{b}bb}$. We shall write 
\begin{eqnarray}
\label{assem}
&&G_{\bar{b}bb}(\theta_1, \theta_2,\theta_3) = N_{\bar{b}bb} \, \sigma_{\bar{\rho}\rho\rho}(\theta_1,\theta_2,\theta_3) \, F_{\bar{\rho}\rho}(\theta_1-\theta_2)F_{\bar{\rho}\rho}(\theta_1-\theta_3)F_{\rho\rho}(\theta_2-\theta_3)\int_{C3;123} du \, e^{\ell \sum_{i=1}^3 \theta_i}\nonumber\\
&&e^{(s - 3 \ell) u}  \, \frac{1}{\cosh \frac{\theta_1-u}{2}}Z_{\bar{\rho}\rho}(\theta_1-u) \Phi_{\bar{\rho}\rho}(\theta_1 - u)Z_{\rho\rho}(\theta_2-u) \Phi_{\rho\rho}(\theta_2 - u)Z_{\rho\rho}(\theta_3-u) \Phi_{\rho\rho}(\theta_3 - u).
\end{eqnarray}
What we have done is to introduce separate functions $F$ for each pairs of representations, to keep track of the various constants in our conventions, satisfying
\begin{eqnarray}
F_{\rho\rho}(-\theta) = \frac{F_{\rho\rho}(\theta)}{\Phi_{\rho\rho}(\theta)}, \qquad F_{\bar{\rho}\bar{\rho}}(-\theta) = \frac{F_{\bar{\rho}\bar{\rho}}(\theta)}{\Phi_{\bar{\rho}\bar{\rho}}(\theta)}, \qquad F_{\bar{\rho}\rho}(-\theta) = \frac{F_{\rho\bar{\rho}}(\theta)}{\Phi_{\rho\bar{\rho}}(\theta)},
\end{eqnarray}
and used them to assemble (\ref{assem}). A solution to these relations consistent with (\ref{fasin}) is
\begin{eqnarray}
\label{fasinn}
F_{\rho\rho}(\theta) = F_{\bar{\rho}\bar{\rho}}(\theta) = F(\theta), \qquad F_{\bar{\rho}\rho}(\theta) = e^{i \frac{\pi}{4}} F(\theta), \qquad F_{\rho\bar{\rho}}(\theta) = e^{-i \frac{\pi}{4}} F(\theta).
\end{eqnarray}
with $F$ the familiar function (\ref{eff}). The first property in (\ref{Watson}) allows to prove (\ref{fasinn}) in conjunction with (\ref{fasin}). We also immediately have the property
\begin{eqnarray}
F_{xy}(i \pi + \theta) = F_{xy}(i \pi - \theta), 
\end{eqnarray}
with $x,y$ any combination of $\rho$ and $\bar{\rho}$, thanks to the second property in (\ref{Watson}) for the function $F$.

We have respected the identity of each particle in building (\ref{assem}), while at the same time thinking of the variable $u$ as an auxiliary space in the $\rho$ representation for definiteness, in the spirit of the algebraic Bethe ansatz. 
It is not obvious that the Bethe-vector part should be $\frac{1}{\cosh\frac{(\theta_1-u)}{2}}$, which used to pertain to $G_{fbb}$, but it is this assignment which satisfies the permutation requirement, and similarly one finds that all the other choices are also uniquely determined by the permutation requirement.  

Permutation itself acquires a slightly more complicated form. Let us exemplify one case, which should again be paradigmatic of the general procedure. Consider
\begin{eqnarray}
&&G_{\bar{b}bb}(\theta_1,\theta_2,\theta_3) = \Phi_{\bar{\rho}\rho}(\theta_1-\theta_2)\Big[G_{b\bar{b}b}(\theta_2,\theta_1,\theta_3) \, S_{\bar{b}b}^{b\bar{b}}(\theta_1-\theta_2) +G_{f\bar{f}b}(\theta_2,\theta_1,\theta_3) \, S_{\bar{b}b}^{f\bar{f}}(\theta_1-\theta_2)\Big].\nonumber
\end{eqnarray}
Since we are working with conventions where the difference between the dressing factors in different representations amounts to constants, with the Zamolodchikov part $\Phi(\theta)$ being common to all, after a little playing one finds how to distribute these constants in the formula to satisfy the requirement. Here one has to fix
\begin{eqnarray}
\label{t}
G_{f\bar{f}b} = -\beta \, \tilde{G}_{fbb}, \qquad G_{b\bar{b}b} = \beta \, \tilde{G}_{bfb}, \qquad G_{\bar{b}bb} = -i\beta \, \tilde{G}_{fbb},
\end{eqnarray}
with $\beta$ an arbitrary constant, and where
\begin{eqnarray}
\tilde{G}_{bbf} = \frac{G_{bbf}}{N_{bbf}}, \qquad \tilde{G}_{bfb} = \frac{G_{bfb}}{N_{bfb}}, \qquad \tilde{G}_{fbb} = \frac{G_{fbb}}{N_{fbb}}. 
\end{eqnarray}

Repeating this exercise for all the form factors, one can fix all the functions. If we suppress the arguments, them being everywhere $(\theta_1,\theta_2,\theta_3)$, and define
\begin{eqnarray}
&&\tilde{G}_{xyz} = G_{xyz}, \qquad \tilde{G}_{\bar{x}\bar{y}\bar{z}} = G_{\bar{x}\bar{y}\bar{z}}, \nonumber\\
&&\tilde{G}_{\bar{x}yz} = -i G_{\bar{x}yz}, \qquad \tilde{G}_{x\bar{y}z} = G_{x\bar{y}z},\qquad  \tilde{G}_{xy\bar{z}} = i G_{xy\bar{z}},\nonumber\\
&&\tilde{G}_{\bar{x}\bar{y}z} = -i G_{\bar{x}\bar{y}z}, \qquad \tilde{G}_{\bar{x}y\bar{z}} = G_{\bar{x}y\bar{z}},\qquad  \tilde{G}_{x\bar{y}\bar{z}} = i G_{x\bar{y}\bar{z}},
\end{eqnarray}
we have for a fermionic operator
\begin{eqnarray}
\label{perman1}
&&\tilde{G}_{\bar{b}\bar{b}\bar{f}} = n_2 \tilde{G}_{bbf}, \qquad \tilde{G}_{\bar{b}\bar{f}\bar{b}} = n_2 \tilde{G}_{bfb}, \qquad \tilde{G}_{\bar{f}\bar{b}\bar{b}} = n_2 \tilde{G}_{fbb},  \nonumber\\
&&\tilde{G}_{\bar{f}b\bar{b}} = -n_3 \tilde{G}_{bbf}, \qquad \tilde{G}_{\bar{f}f\bar{f}} = n_3 \tilde{G}_{bfb}, \qquad \tilde{G}_{\bar{b}b\bar{f}} = n_3 \tilde{G}_{fbb},  \nonumber\\
&&\tilde{G}_{\bar{f}\bar{f}f} = n_3 \tilde{G}_{bbf}, \qquad \tilde{G}_{\bar{f}\bar{b}b} = n_3 \tilde{G}_{bfb}, \qquad \tilde{G}_{\bar{b}\bar{f}b} = -n_3 \tilde{G}_{fbb},  \nonumber\\
&&\tilde{G}_{b\bar{f}\bar{b}} = n_3 \tilde{G}_{bbf}, \qquad \tilde{G}_{b\bar{b}\bar{f}} = -n_3 \tilde{G}_{bfb}, \qquad \tilde{G}_{f\bar{f}\bar{f}} = n_3 \tilde{G}_{fbb},  \nonumber\\
&&\tilde{G}_{ff\bar{f}} = -n_4 \tilde{G}_{bbf}, \qquad \tilde{G}_{fb\bar{b}} = -n_4 \tilde{G}_{bfb}, \qquad \tilde{G}_{bf\bar{b}} = n_4 \tilde{G}_{fbb},  \nonumber\\
&&\tilde{G}_{f\bar{b}b} = -n_4 \tilde{G}_{bbf}, \qquad \tilde{G}_{f\bar{f}f} = n_4 \tilde{G}_{bfb}, \qquad \tilde{G}_{b\bar{b}f} = n_4 \tilde{G}_{fbb},  \nonumber\\
&&\tilde{G}_{\bar{b}fb} = -n_4 \tilde{G}_{bbf}, \qquad \tilde{G}_{\bar{b}bf} = n_4 \tilde{G}_{bfb}, \qquad \tilde{G}_{\bar{f}ff} = -n_4 \tilde{G}_{fbb},
\end{eqnarray}
with $n_2$, $n_3$ and $n_4$ arbitary constants. For a bosonic operator we likewise have
\begin{eqnarray}
\label{perman2}
&&\tilde{G}_{ffb} = m_1 \tilde{G}_{bbf}, \qquad \tilde{G}_{fbf} = - m_1 \tilde{G}_{bfb}, \qquad \tilde{G}_{bff} = m_1 \tilde{G}_{fbb},  \nonumber\\
&&\tilde{G}_{\bar{f}\bar{f}\bar{b}} = m_2 \tilde{G}_{bbf}, \qquad \tilde{G}_{\bar{f}\bar{b}\bar{f}} = -m_2\tilde{G}_{bfb}, \qquad \tilde{G}_{\bar{b}\bar{f}\bar{f}} = m_2 \tilde{G}_{fbb},  \nonumber\\
&&\tilde{G}_{\bar{b}f\bar{f}} = -m_3 \tilde{G}_{bbf}, \qquad \tilde{G}_{\bar{b}b\bar{b}} = -m_3 \tilde{G}_{bfb}, \qquad \tilde{G}_{\bar{f}f\bar{b}} = m_3 \tilde{G}_{fbb},  \nonumber\\
&&\tilde{G}_{\bar{b}\bar{b}b} = -m_3 \tilde{G}_{bbf}, \qquad \tilde{G}_{\bar{b}\bar{f}f} = m_3 \tilde{G}_{bfb}, \qquad \tilde{G}_{\bar{f}\bar{b}f} = m_3 \tilde{G}_{fbb},  \nonumber\\
&&\tilde{G}_{f\bar{b}\bar{f}} = -m_3 \tilde{G}_{bbf}, \qquad \tilde{G}_{f\bar{f}\bar{b}} = -m_3 \tilde{G}_{bfb}, \qquad \tilde{G}_{b\bar{b}\bar{b}} = -m_3 \tilde{G}_{fbb},  \nonumber\\
&&\tilde{G}_{bb\bar{b}} = -m_4 \tilde{G}_{bbf}, \qquad \tilde{G}_{fb\bar{f}} = m_4 \tilde{G}_{fbb}, \qquad \tilde{G}_{bf\bar{f}} = -m_4 \tilde{G}_{bfb},  \nonumber\\
&&\tilde{G}_{f\bar{f}b} = -m_4 \tilde{G}_{fbb}, \qquad \tilde{G}_{b\bar{b}b} = m_4  \tilde{G}_{bfb}, \qquad \tilde{G}_{b\bar{f}f} = m_4 \tilde{G}_{bbf},  \nonumber\\
&&\tilde{G}_{\bar{b}bb} = -m_4 \tilde{G}_{fbb}, \qquad \tilde{G}_{\bar{f}fb} =- m_4 \tilde{G}_{bfb}, \qquad \tilde{G}_{\bar{f}bf} = -m_4 \tilde{G}_{bbf},
\end{eqnarray}
with $m_i$ arbitrary constants (having renamed $m_4$ the constant $\beta$ in (\ref{t}) for convenience.

Let us also remark that all the form factors multiplied by $n_2$ (the same holds for all those multiplied by $n_3$, and, respectively, $n_4$, and also of course for different clusters of $m_i$), only mix among themselves under permutation by the action of the $S$-matrix, but not across groups: a form factor, say, multiplied by $n_3$ cannot mix under the $S$-matrix action with one multiplied by $n_4$. It also means that technically we can allow three more independent values $\ell_i$, $i=2,3,4$, in the respective integrands. Form factors for fermionic operators with the same constant $n_i$ in front will have the same $\ell_i$. We may also rename the $N_3$ of the all-particle (and no anti-particle) case as $n_1$, with $\ell$ becoming $\ell_1$ in that case, for homogeneity of notation. For bosonic operators it will be the same, this time with $\ell'_i$, $i=1,2,3,4$ in correspondence with the $m_i$s.  

Periodicity now also becomes marginally more complicated. As an example we can take
\begin{eqnarray}
\label{rhs}
G_{\bar{b}bb}(\theta_1+2 i \pi,\theta_2,\theta_3) = +G_{bb\bar{b}}(\theta_2,\theta_3,\theta_1),
\end{eqnarray}
where the putative $+$ sign is initially there because we are cycling a bosonic state (across a bosonic operator in this case).
The proof we gave in the case where no anti-particle were present needs to be slightly upgraded. The difference with respect to that situation is that we now have
\begin{enumerate}
\item pre-factors which now read
\begin{eqnarray}
F_{\bar{\rho}\rho}(\theta_1+ 2 i \pi-\theta_2)F_{\bar{\rho}\rho}(\theta_1-\theta_3)F_{\rho\rho}(\theta_2-\theta_3) \qquad \mbox{and} \qquad F_{\rho\rho}(\theta_2-\theta_3)F_{\rho\bar{\rho}}(\theta_2-\theta_1)F_{\rho\bar{\rho}}(\theta_3-\theta_1)\nonumber
\end{eqnarray}
respectively for the lhs and the rhs of (\ref{rhs}).
\item The integrand also cycles accordingly:
\begin{eqnarray}
Z_{\bar{\rho}\rho}(\theta_1+2\pi i-u) \Phi_{\bar{\rho}\rho}(\theta_1 + 2 \pi i- u)Z_{\rho\rho}(\theta_2-u) \Phi_{\rho\rho}(\theta_2 - u)Z_{\bar{\rho}\rho}(\theta_3-u) \Phi_{\bar{\rho}\rho}(\theta_3 - u)
\end{eqnarray}
and 
\begin{eqnarray}
Z_{\rho\rho}(\theta_2-u) \Phi_{\rho\rho}(\theta_2 - u)Z_{\rho\rho}(\theta_3-u) \Phi_{\rho\rho}(\theta_3 - u)Z_{\bar{\rho}\rho}(\theta_1-u) \Phi_{\bar{\rho}\rho}(\theta_1 - u),
\end{eqnarray}
for the lhs and rhs of (\ref{rhs}), respectively.
\end{enumerate}
The rest of the proof goes the same way as before, since the functions from the Bethe state and the exponential are always the same, provided one keeps track, using (\ref{perman1}) and (\ref{perman2}), of which labels one is cycling. The assignments (\ref{perman1}) and (\ref{perman2}) in fact do respect the cyclic pattern of the three basis functions $G_{bbf}$, $G_{bfb}$ and $G_{fbb}$ on which they rely.

Given the fact that the Zamolodchikov factor is always common, we simply accumulate some extra constants with respect to the case with no anti-particle, when writing down the periodicity constraints. This turns out to be consistent with the fact that the relative statistical factors of the particles/anti-particles and the associated creation operators in the algebraic Bethe ansatz is determined by the values of the $S$-matrices at equal rapidities\footnote{It is easy to see for instance that the two $B$ operators, creating magnons with an $f$ out of a pseudovacuum of $b$s, commute with one another \cite{DiegoBogdanAle,JuanMiguelAle}, despite they original assignment of bare statistics of the one-particle states. This is because the $R$-matrix has a minus (hence, the $S$-matrix has a plus) in the $ff \to ff$ entry, and the RTT relation connected to two $B$'s has a relative plus instead of a minus. This is well-known in integrable systems: for example, the $2D$ non-linear Schroedinger (NLS) equation has bosonic particles whose $S$-matrix equals $-1$ at equal momenta, hence the NLS bosons effectively behave like fermions. The two-particle incoming momentum-eigenstate
\begin{eqnarray}
&&|k_1,k_2\rangle \equiv \int_{-\infty}^\infty\int_{-\infty}^\infty dx_1 dx_2 \Big[\Theta(x_1 - x_2) + S(k_1,k_2) \Theta(x_2 - x_1) \Big] e^{i k_1 x_1 + i k_2 x_2} \psi^\dagger(x_1) \psi^\dagger(x_2)|0\rangle
\end{eqnarray}
taken for $k_1=k_2$ vanishes, due to the fact that the $S$-matrix $S(k_1,k_2) = \frac{k_1 - k_2 - i}{k_1-k_2+i}$ satisfies $S(k,k)=-1$ and the creation operators are bosonic. This states satisfies Pauli's exclusion principle as a net effect \cite{Thacker}. 
The wave functions for the protected spectrum of $AdS_3$, involving massless states at zero momentum, have recently been explicitly computed in \cite{Majumder:2021zkr}.}. In our conventions the non-zero entries read (given that $\Phi(0)=1$ from the (\ref{zamo}) representation)
\begin{eqnarray}
\label{equ}
&&S_{bb}^{bb}(0)= 1, \qquad S_{ff}^{ff}(0)=1, \qquad S_{bf}^{bf}(0) = 1 = S_{fb}^{fb}(0), \nonumber\\
&&S_{\bar{b}\bar{b}}^{\bar{b}\bar{b}}(0)= 1, \qquad S_{\bar{f}\bar{f}}^{\bar{f}\bar{f}}(0)=1, \qquad S_{\bar{b}\bar{f}}^{\bar{b}\bar{f}}(0) = 1 = S_{\bar{f}\bar{b}}^{\bar{f}\bar{b}}(0), \nonumber\\
&&S_{\bar{b}b}^{f\bar{f}}(0)= i, \qquad S_{\bar{f}f}^{b\bar{b}}(0)=-i, \qquad S_{\bar{b}f}^{f\bar{b}}(0) = i = - S_{\bar{f}b}^{b\bar{f}}(0),\nonumber\\
&&S_{b\bar{b}}^{\bar{f}f}(0)= i, \qquad S_{f\bar{f}}^{\bar{b}b}(0)=-i, \qquad S_{f\bar{b}}^{\bar{b}f}(0) = -i = -S_{b\bar{f}}^{\bar{f}b}(0).
\end{eqnarray}
These values are all consistent with braiding unitarity, as shown in (\ref{cruni}). For instance, one can check a few cases
\begin{eqnarray}
\label{exom}
&&S_{bb}^{bb}(0)S_{bb}^{bb}(0)=1 \times 1=1, \qquad S_{\bar{f}b}^{b\bar{f}}(0)S_{b\bar{f}}^{\bar{f}b}(0)=-i \times i=1, \nonumber\\
&&S_{\bar{b}b}^{b\bar{b}}(0)S_{b\bar{b}}^{\bar{b}b}(0)+S_{\bar{b}b}^{f\bar{f}}(0)S_{f\bar{f}}^{\bar{b}b}(0)=0 \times 0 + i \times (-i)=1,\nonumber\\
&&S_{fb}^{bf}(0)S_{bf}^{fb}(0)+S_{fb}^{fb}(0)S_{fb}^{fb}(0)=0 \times 0 + 1 \times 1=1.
\end{eqnarray}
This is in fact the analogue of the discussion around formulas (2.19) and (2.20) of \cite{zamoTBA}, where the effective statistics of the particles is discussed from the very same point of view, for the purposes of how these particles enter the TBA.

The periodicity rule which we have observed appears to remember this braided statistics\footnote{To continue with our NLS analogy, the bare (perturbative) statistics of the elementary annihilation operator of the NLS boson $\psi(x)$ is $(-)^0$, but the operator annihilating (non-perturbative) eigenstates of the quantum transfer matrix is
\begin{eqnarray}
&&B(\lambda) = - i \, e^{i \frac{\lambda}{2} (s_+ + s_-)} \sum_{n=1}^\infty \kappa^n \int_{s_+ > \eta_{n+1} > \xi_n > \eta_n > \xi_{n-1}...>\eta_1 > s_-} d\xi_1 ... d\xi_n \, d\eta_1 ... d\eta_{n+1} \nonumber \\
&&\qquad \qquad \qquad \qquad \qquad \qquad \qquad \qquad \qquad e^{i \lambda (\xi_1 + ... + \xi_n - \eta_1 - ... - \eta_{n+1})} \psi^\dagger(\xi_1) ... \psi^\dagger(\xi_n) \, \psi(\eta_1) ... \psi(\eta_{n+1}),\nonumber
\end{eqnarray}
with a spectral parameter $\lambda$ and on a finite interval $(s_-,s_+)$, where $\kappa$ is the coupling constant of the NLS model \cite{Thacker,Evgeny}. It is these non-perturbative dressed operators which have the statistics dictated by the $S$-matrix at equal rapidities.}. Focusing on the form factors which we will be considering for the purposes of the residue, namely the ones with two particles and one anti-particle, we have found a solution to all the constraints where a boson $b$ picks up a $-1$ when cycling the bosonic operator, $f$ picks up an $i$, $\bar{b}$ a $1$ and $\bar{f}$ a $i$. Based on the above argument, this seems to suggest that the bosonic operator may be the operator creating a $\bar{b}$. This appears vaguely consistent since there is an excess of particles over anti-particles in the amount of $1$, and it is the matrix element with the operator associated with this particle which has the likelihood of being non-zero. In the absence of a more precise description (in fact of any description at this stage) of the operators of the theory, we take this intuition as a tentative working hypothesis to proceed on, at least in the three-particle case, as we will cautiously do in a few more instances in the next section. 

Keeping this working hypothesis in mind, the relative statistics is then determined by the diagonal entries at equal rapidities (\ref{equ}) $S_{\bar{b}\bar{b}}^{\bar{b}\bar{b}}(0)=1$, $S_{\bar{b}f}^{f\bar{b}}(0) = i$, which is consistent with the values we find from periodicity. We remark that for the $S$-matrix it is the process $S_{xy}^{yx}$ to be the diagonal one (transmission). The statistical factors for $f$ and $\bar{b}$ are harder to read, as there is no non-zero {\it diagonal} entry at equal rapidities involving only $\bar{f}$ with $\bar{b}$ or $b$ with $\bar{b}$. The values we find from periodicity for these seem at least reminiscent of the entries (\ref{equ}). 

The possible values of the constants $\ell_i$ and $\ell'_i$ are all determined by the procedure. We remark here that these values could potentially lead to divergent integrals if violating the bound (\ref{bound}). Divergences of the integral over $u$ are actually contemplated and do occur in the general analysis originating from \cite{BabuF}. They seem to be hard to avoid even in the massive case, particularly when considering higher-rank operators such as the energy-momentum tensor. For such operators one has in fact to introduce in the integrand certain additional $p$-functions which are basically exponentials of the rapidities, and which produce unbalances in the convergence count. One can refer in particular to \cite{Babu2}, where it is remarked that for large values of (an analogue of the number) $\ell$ the form factors have to be defined as analytic continuations of the integral representation from (in their case) a sufficiently small value of their parameter $\nu$ (which is a function of the coupling of the model) to the desired  values. The difference is that we do not have a coupling to potentially tune here, as a consequence of our dealing with a critical point CFT which already sits at a special value of the coupling. One way to see this is that the natural $AdS_3$ coupling constant $h$ has been scaled to infinity in the BMN limit. Any divergences we may observe may therefore be another consequence of the (massless) conformality of our situation. The most sensible thing which we could do here, in order not to break any symmetry or any other carefully constructed property of the integral with a brutal regularisation, would be in the end to go to a region of complex spin $s$ where convergence is restored, and then attempt an analytic continuation in the complex variable $s$ back to our desired value of $s=0$. This is the closest approach to what \cite{Babu2} suggest for their problem, and we will adopt this route as possible in all the cases which should require it.

A similar solution occurs when we have two anti-particles and one particle: the bosonic operator in this case could be the one creating a $b$, and the statistical factors we pick up are now a $1$ when $b$ is cycling through (consistent with $S_{bb}^{bb}(0)=1$), a $i$ when $f$ is cycling through, a $-1$ when $\bar{b}$ cycles through and an $i$ when $\bar{f}$ cycles through (consistent with $S_{b\bar{f}}^{\bar{f}b}(0)=i$). 

We have not analysed the full extent of all possible solutions to the periodicity constraints, and will be content that there seems to be solutions which are consistent with our physical arguments.

\subsection{\label{resd}Residue calculation}

\begin{figure} 
\centerline{\includegraphics[width=8cm]{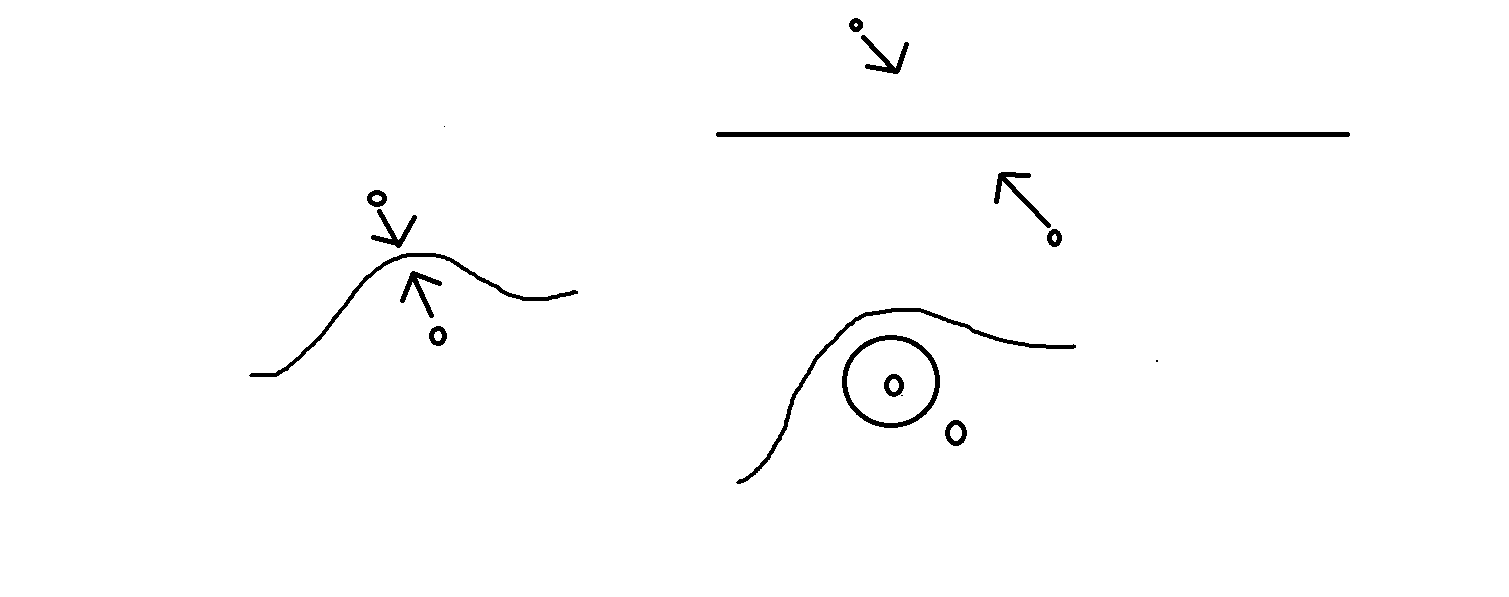}}
\caption{\it Two poles coalesce and pinch the contour in between.}
\label{fig10}
\end{figure}

We now need to compute the residue of the expression which we have written down in the previous section at the singularity $\theta_1 \to \theta_2 + i \pi$. We continue with our example of two particles and one anti-particle with a bosonic operator - the case with two anti-particles and a particle, involving a bosonic operator of a different type, will follow along similar lines. Since the contour can always be deformed to avoid all the poles of the integrand, such singularities of the integral can only occur when two poles of the integrand coalesce as $\theta_1 \to \theta_2 + i \pi$ and pinch the contour, forcing it to run over a pole \cite{Eden:1966dnq}. A sketch of this pinching process is drawn in figure \ref{fig10}.
There are three possibilities for poles to pinch the contour:
\begin{enumerate}

\item P1: a pole at $u=\theta_1$ coalesces with a pole at $u = \theta_2 + i \pi$ as $\theta_1 \to \theta_2 + i \pi$;

\item P2: a pole at $u=\theta_1-i \pi$ coalesces with a pole at $u = \theta_2$ as $\theta_1 \to \theta_2 + i \pi$;

\item P3: a pole at $u=\theta_1-2i \pi$ coalesces with a pole at $u = \theta_2 - i \pi$ as $\theta_1 \to \theta_2 + i \pi$;

\end{enumerate} 

The label P1, P2, P3, to distinguish the types of pinching is marked on the graphs in section \ref{graphs}. Two poles indicated by P1, P2 or P3 will pinch in the limit $\theta_1 \to \theta_2 + i \pi$ in the corresponding way listed above. The resulting pole of the integral is simple in all cases.

The most difficult part is to actually extract the residue from the pinching of the contour. For $G_{\bar{b}bb}$ the residue receives two contributions: P1 and P2. 

For this calculation, we write once more the corresponding integral formula in full, using the same notation (\ref{nota}). We have
\begin{eqnarray}
&&G_{\bar{b}bb}(\theta_1, \theta_2,\theta_3) = N_{\bar{b}bb}\, \sigma(\theta_1,\theta_2,\theta_3) \, F(\theta_{12})F(\theta_{13})F(\theta_{23})\int_{\cal{C}} d\theta_0 \frac{e^{\big[\ell'_4 \sum_{i=1}^3 \theta_i + (s - 3 \ell'_4) \tilde{\theta}_0\big]}}{\cosh \frac{\theta_{10}}{2}}\nonumber\\
&&\qquad \qquad \qquad \times \frac{\Phi(\theta_{10})\Phi(\theta_{20})\Phi(\theta_{30})}{F(\theta_{10})F(\theta_{20})F(\theta_{30})F(\theta_{10}+i\pi)F(\theta_{20}+i\pi)F(\theta_{30}+i\pi)\sinh \theta_{10} \sinh \theta_{20} \sinh \theta_{30}},\nonumber
\end{eqnarray}
where $N_{\bar{b}bb} = -m_4$ already takes into account all the constant factors from the various components of the formula. From our general analysis we know that the $\frac{1}{\sinh \theta_{10}}$ is responsible for the pole inside the small circle around $\theta_1$. We then send $\theta_2 \to \theta_1-i \pi$, and get 
\begin{eqnarray}
\label{contri1}
&&G_{\bar{b}bb}(\theta_1, \theta_2,\theta_3)_{\mbox{P1}} \to i\pi N_{\bar{b}bb} \Phi(\theta_{23}) \frac{e^{\theta_2-\theta_3} e^{(s-\ell'_4)\theta_2 +\ell'_4 \theta_3 +i\pi(s-2\ell'_4)} \, }{F(0)^2 \big(\theta_1 - (\theta_2 + i \pi)\big)},
\end{eqnarray}  
where the pole comes from a $\cosh \frac{\theta_{10}}{2} \to i \sinh \frac{\theta_1 - \theta_2 - i \pi}{2}$ at the denominator.
To obtain this result we have made repeated use of the ever useful relations:
\begin{eqnarray}
\Phi(0)=1, \qquad F(i \pi) = 1, \qquad F(-\theta) = \frac{F(\theta)}{\Phi(\theta)}, \qquad F(i\pi+\theta) = F(i\pi - \theta), \qquad \Phi(\theta) = \frac{1}{\Phi(-\theta)}.
\end{eqnarray}
The first two relations can easily be seen from the gamma function representations (\ref{zamo}) and (\ref{pro}), respectively. Finally, we have verified numerically that
\begin{eqnarray}
F(0) = \exp[- 7 \zeta'(-2)] \sim 1.23756...
\end{eqnarray}
where $\zeta(s)$ is the Riemann zeta function. Finally, we have cancelled a $\sinh \theta_{23}$ at the denominator coming from the integral, against a similar factor at the denominator coming from the function $\sigma$ (this being in fact the very reason for us introducing the function $\sigma$ in the first place). This procedure is especially different from the one in \cite{BabuF}, since they could use a product of $\sinh$ functions as a multiplier. That product is however antisymmetric under individual transpositions of two rapidities, accounting for the fermionic nature of the (anti-)solitons they were considering, while here the statistics is already incorporated in the $S$-matrix and we need a completely symmetric function upfront. Since we use the $S$-matrix to incorporate the relative statistics, the case of a fermionic operator should also follow along similar lines once we will have worked out the bosonic case.  

The other residue is computed analogously. One extra ingredient comes from the fact that the simple pole in this case is the result of a cancellation of, for instance, a double pole and a zero. One therefore needs to compute 
\begin{eqnarray}
\label{conc}
\lim_{\theta_1 \to \theta_2 + i\pi} (\theta_1 - \theta_2-i\pi)\frac{\Phi(\theta_{12})}{\cosh^2 \frac{\theta_{12}}{2}} = -2i.
\end{eqnarray} 
This can be proven as follows:
\begin{eqnarray}
\frac{\Phi(\theta_{12})}{\cosh^2 \frac{\theta_{12}}{2}} = \frac{\Phi(\theta_{12}-i \pi + i \pi)}{\cosh^2 \frac{\theta_{12}}{2}} = \frac{\Phi(\theta_{21}+i \pi) \, i \tanh \frac{\theta_{12}-i \pi}{2}}{\cosh^2 \frac{\theta_{12}}{2}} \to \Phi(0) \frac{2i}{\sinh \theta_{12}} 
\end{eqnarray}
as $\theta_1 \to \theta_2 + i\pi$. Since $\Phi(0)=1$, we conclude (\ref{conc}).
 
Taking this into account we obtain
\begin{eqnarray}
\label{contri2}
&&G_{\bar{b}bb}(\theta_1, \theta_2,\theta_3)_{\mbox{P2}} \to i\pi N_{\bar{b}bb} \frac{e^{\theta_2 - \theta_3} e^{(s-\ell'_4)\theta_2+\ell'_4 \theta_3 + i\pi \ell'_4} \, }{F(0)^2 \big(\theta_1 - (\theta_2 + i \pi)\big)}.
\end{eqnarray}
By combining the two contributions (\ref{contri1}) and (\ref{contri2}) we obtain
\begin{eqnarray}
\label{fine}
-\frac{i}{2} \, \mbox{Res}_{\theta_1 \to \theta_2+i\pi}G_{\bar{b}bb} = \frac{\pi}{2} N_{\bar{b}bb} \frac{e^{\theta_2 - \theta_3}e^{(s-\ell'_4)\theta_2 +\ell'_4 \theta_3 + i\pi \ell'_4}}{F(0)^2} \Big(1 +e^{i \pi (s-3\ell'_4)} \Phi(\theta_{23}) \Big).
\end{eqnarray}

%The mixed $LR$ $R$-matrix has entries which are simply a reshuffling of the those of (\ref{eq:RLLlimtheta}). In terms of $R$-matrices one has (having suitably chosen the normalisation of the matrix part)
%\begin{equation}\label{eq:RLLlimthetaa}
  %\begin{aligned}
   % &R_c |\bar{\phi}\rangle \otimes |\phi\rangle\ = -\tanh\frac{\theta}{2} |\bar{\phi}\rangle \otimes |\phi\rangle - {\rm sech}\frac{\theta}{2}|\bar{\psi}\rangle \otimes |\psi\rangle, \\
%&R_c|\bar{\phi}\rangle \otimes |\psi\rangle\ = \,  |\bar{\phi}\rangle \otimes |\psi\rangle, \\
% &R_c |\bar{\psi}\rangle \otimes |\phi\rangle\ =   \, -|\bar{\psi}\rangle \otimes |\phi\rangle,\\
   % &R_c |\bar{\psi}\rangle \otimes |\psi\rangle\ = - \, {\rm sech}\frac{\theta}{2} |\bar{\phi}\rangle \otimes |\phi\rangle + \tanh\frac{\theta}{2} |\bar{\psi}\rangle \otimes |\psi\rangle,
%\end{aligned}
%\end{equation}
%and similarly for the $LR$ scattering. We finally multiply the matrix part by the Zamolodchikov dressing factor $\tau^{-1} \Phi(\theta)$, where $\tau$ is a relative normalisation between the mixed and the direct $S$-matrix.    
 
We will shortly see that the right hand side of (\ref{fine}) satisfies the condition (\ref{reso}), since the S-matrix for particle $2$ ($b$) and particle $3$ ($b$) is simply the dressing factor $\Phi(\theta_{23})$ (diagonal scattering). We will also obtain a relation between the normalisation $N_{\bar{b} bb}$ and the normalisation of the one point function $N_b$, from 
\begin{eqnarray}
\label{finel}
-\frac{i}{2} \, \mbox{Res}_{\theta_1 \to \theta_2+i\pi}G_{\bar{b}bb} = \frac{\pi}{2} N_{\bar{b}bb} \frac{e^{\theta_2 - \theta_3}e^{(s-\ell'_4)\theta_2 +\ell'_4 \theta_3 + i\pi \ell'_4}}{F(0)^2} \Big(1 +e^{i \pi (s-3\ell'_4)} [S_{23}]_{bb}^{bb} \Big),
\end{eqnarray}
with $[S_{23}]_{bb}^{bb} = \Phi(\theta_{23})$, and where the one-point function will contain $e^{s \theta_3}$ for a consistent boost transformation. We will discuss further the various parts of this formula later on in conjunction with the second residue. 

We are now in a position to check the other residue which satisfies the requirement (\ref{reso}), namely $G_{\bar{f}fb}$, which we can compute from the expression for $G_{bfb}$. We again have two pinching contributions, this time P2 and P3. We have repeated the entire procedure and we only report the result. First, we have the P2 contribution, which gives
\begin{eqnarray}
G_{\bar{f}fb}(\theta_1,\theta_2,\theta_3)_{P2} \to - \frac{\pi N_{\bar{f}fb} \, e^{\theta_2 - \theta_3}e^{(s-\ell'_4)\theta_2 +\ell'_4 \theta_3 +i \pi \ell'_4}}{F(0)^2 (\theta_1 - \theta_2 - i \pi)},
\end{eqnarray}
where $N_{\bar{f}fb}$ again equals $-m_4$.
The P3 contribution produces a similar result, multiplied by one factor of $\Phi$ at a shifted point:
\begin{eqnarray}
G_{\bar{f}fb}(\theta_1,\theta_2,\theta_3)_{P3} \to  -\frac{\pi N_{\bar{f}fb} \, e^{\theta_2 - \theta_3}e^{(s-\ell'_4)\theta_2 + \ell'_4 \theta_3 +i\pi(4\ell'_4-s)}\Phi(\theta_3-\theta_2+i \pi)}{F(0)^2 (\theta_1 - \theta_2 - i \pi)}.
\end{eqnarray} 
Using the property of the Zamolodchikov dressing factor 
\begin{eqnarray}
\Phi(-\theta+i \pi) = - i \, \Phi(\theta) \tanh \frac{\theta}{2},
\end{eqnarray}
and again the fact that $s$ is half-integer for the fermionic operators, we can recast the total result from the two contributions as
\begin{eqnarray}
\label{finet}
-\frac{i}{2} \, \mbox{Res}_{\theta_1 \to \theta_2+i\pi}G_{\bar{f}fb} = -\frac{i}{2} \pi N_{\bar{f}fb} e^{\theta_2 - \theta_3}\frac{e^{(s-\ell'_4)\theta_2 + \ell'_4 \theta_3 +i\pi \ell'_4} \, }{F(0)^2} \Big(1-ie^{i\pi(3\ell'_4 -s)} [S_{23}]_{fb}^{bf} \Big),
\end{eqnarray}
with $[S_{23}]_{fb}^{bf} = \Phi(\theta_{23}) \tanh \frac{\theta_{23}}{2}$. 
%The extra factor of $i$ in (\ref{finet}) might be interpreted in view of the fact the $S$-matrix for two particles satisfies the condition
%\begin{eqnarray}
%S(i \pi) = \begin{pmatrix}0&0&0&0\\0&i&-1&0\\0&1&i&0\\0&0&0&0\end{pmatrix},
%\end{eqnarray}
%as a result of the compensation of the zero occurring in the dressing factor and the hyperbolic cosine developing a zero at the denominator. This gives an extra factor of $i$ in the $fb$ contribution with respect to $bb$, as opposed to the situation in \cite{BabuF}. 

We also see that the function $G_{bbf}$, which has the same functional form as $G_{\bar{f}bf}$, does not have any pinching contour in the limit $\theta_1 \to \theta_2 + i \pi$, consistently with the assignment of labels. 

Since the one-point function must be proportional to $e^{s \theta_3}$ for boost symmetry, given that $s$ is integer - which we will be $0$ for the operator annihilating $b$ - and $\ell'_4=1$ with the regularisation caveat which we have discussed - we obtain for the residues (where $\theta_1 \to \theta_2+i\pi$)
\begin{eqnarray}
\label{alla}
&&-\frac{i}{2} \, \mbox{Res}_{\theta_1 \to \theta_2+i\pi}G_{\bar{b}bb} = \frac{\pi}{2} N_{\bar{b}bb} \frac{e^{\theta_2 - \theta_3}e^{(s-\ell'_4)\theta_2 +\ell'_4 \theta_3 + i\pi \ell'_4}}{F(0)^2} \Big(1 +e^{i \pi (s-3\ell'_4)} [S_{23}]_{bb}^{bb} \Big),\nonumber\\
&&\qquad \qquad \qquad \qquad =  \frac{\pi}{2} \, {\bf C}_{\bar{b}}^b \, m_4 \frac{1}{F(0)^2} \Big(1 - [S_{23}]_{bb}^{bb} \Big), \nonumber\\
&&-\frac{i}{2} \, \mbox{Res}_{\theta_1 \to \theta_2+i\pi}G_{\bar{f}fb} = -\frac{i}{2} \pi N_{\bar{f}fb} e^{\theta_2 - \theta_3}\frac{e^{(s-\ell'_4)\theta_2 + \ell'_4 \theta_3 +i\pi \ell'_4} \, }{F(0)^2} \Big(1-ie^{i\pi(3\ell'_4 -s)} [S_{23}]_{fb}^{bf} \Big)\nonumber\\
&&\qquad \qquad \qquad \qquad =\frac{\pi}{2} \, {\bf C}_{\bar{f}}^f \,m_4 \frac{1}{F(0)^2} \Big(1 + i[S_{23}]_{fb}^{bf} \Big),
\end{eqnarray}
where ${\bf C}_{\bar{b}}^b=1$ and ${\bf C}_{\bar{f}}^f=-i$. This quite happily coincides with the requirement (\ref{reso}), including the statistical factors. Based on the discussion at the end of the previous subsection, we expect a $1$ - hence a minus in front of $[S_{23}]_{bb}^{bb}$ - when the $\bar{b}$ cycles the operator annihilating $b$, and a $-i$ - hence a $+i$ in front of $[S_{23}]_{fb}^{bf}$ - when $\bar{f}$ is cycling. 

We have also checked that the other two relevant instances of kinematical pole with two particles present work as well. From the assignment (\ref{perman2}) it is straightforward to re-utilise the above results and convert them accordingly. We get in this way
\begin{eqnarray}
\label{alla2}
&&-\frac{i}{2} \, \mbox{Res}_{\theta_1 \to \theta_2+i\pi}G_{f\bar{f}b} =   \frac{\pi}{2} \, {\bf \bar{C}}_f^{\bar{f}} \, m_4 \frac{1}{F(0)^2} \Big(1 - i [S_{23}]_{\bar{f}b}^{b\bar{f}} \Big), \nonumber\\
&&-\frac{i}{2} \, \mbox{Res}_{\theta_1 \to \theta_2+i\pi}G_{b\bar{b}b} = \frac{\pi}{2} \, {\bf \bar{C}}_b^{\bar{b}} \,m_4 \frac{1}{F(0)^2} \Big(1 -[S_{23}]_{\bar{b}b}^{b\bar{b}} \Big),
\end{eqnarray}
where now the matrix ${\bf \bar{C}}_f^{\bar{f}}=1$ and ${\bf \bar{C}}_b^{\bar{b}}=i$ features due to the opposite ordering of particle and anti-particle, in agreement with the discussion around (\ref{feature}). The statistical factors quite remarkably again agree the ones discussed at the end of the previous section, respectively for an $f$ and a $\bar{b}$ cycling through the operators annihilating the $b$.

Based on these results we are brought to argue a relationship between the constant $m_4$ and the (constant) form factor $N_1$ of the operator annihilating $b$ with the state $b$:
\begin{eqnarray}
\label{fr}
\frac{\pi}{2} \frac{m_4}{F(0)^2} = N_1.
\end{eqnarray}
These types of recursions are typical of the form factor programme, and are often used to determine (together with the other constraints and, in the massive case, with the bound-state residue axiom, the general form of the form factors \cite{Mussardo}.

For completeness, we provide here the other residue relevant to the bosonic operator annihilating $\bar{b}$:
\begin{eqnarray}
\label{alla3}
&&-\frac{i}{2} \, \mbox{Res}_{\theta_1 \to \theta_2+i\pi}G_{\bar{b}b\bar{b}}  = -\frac{\pi}{2} \, i  {\bf C}_{\bar{b}}^b \, m_3 \frac{1}{F(0)^2} \Big(1 - [S_{23}]_{b\bar{b}}^{\bar{b}b} \Big), \nonumber\\
&&-\frac{i}{2} \, \mbox{Res}_{\theta_1 \to \theta_2+i\pi}G_{\bar{f}f\bar{b}} = -\frac{\pi}{2} \, i {\bf C}_{\bar{f}}^f \,m_3 \frac{1}{F(0)^2} \Big(1 - i[S_{23}]_{f\bar{b}}^{\bar{b}f} \Big),\nonumber\\
&&-\frac{i}{2} \, \mbox{Res}_{\theta_1 \to \theta_2+i\pi}G_{f\bar{f}\bar{b}} =  - \frac{\pi}{2} \, i {\bf \bar{C}}_f^{\bar{f}} \, m_3 \frac{1}{F(0)^2} \Big(1 + i [S_{23}]_{\bar{f}\bar{b}}^{\bar{b}\bar{f}} \Big), \nonumber\\
&&-\frac{i}{2} \, \mbox{Res}_{\theta_1 \to \theta_2+i\pi}G_{b\bar{b}\bar{b}} = -\frac{\pi}{2} \, i {\bf \bar{C}}_b^{\bar{b}} \,m_3 \frac{1}{F(0)^2} \Big(1 -[S_{23}]_{\bar{b}\bar{b}}^{\bar{b}\bar{b}} \Big).
\end{eqnarray}
Here we have also fixed $\ell'_3=1$ under the usual caveat. All of these formulas (\ref{alla3}) are again precisely respecting the assignment of statistical factors of each left-most state respectively cycling through the operator annihilating $\bar{b}$, as discussed at the end of the previous section. This suggests that 
\begin{eqnarray}
\label{om}
-\frac{\pi}{2} i \frac{m_3}{F(0)^2} = \bar{N}_1,
\end{eqnarray}
with $\bar{N}_1$ the one-point function of the operator annihilating $\bar{b}$ with the state $\bar{b}$. We in fact expect, on the grounds of $\rho \leftrightarrow \bar{\rho}$ symmetry, that $N_1=\bar{N}_1$, hence we expect from (\ref{fr}) and (\ref{om}) that this should imply $m_4 = -i m_3$.

Finally, fermionic operators work just the same with the assignment (\ref{perman1}), always using the same calculations for the residues of the basic functions $\tilde{G}_{fbb}$ and $\tilde{G}_{bfb}$. Once more the functions with non-zero residue are exactly those which have a particle and an anti-particle in the correct position to trigger the requirement (\ref{reso}). Focusing on $|s|=\frac{1}{2}$ for instance, the values of $\ell$ one should choose are now $\ell_3=\ell_4 = \frac{3}{2}$ for the spin $s=\frac{1}{2}$ component of the spinor, and $\ell_3=\ell_4 = \frac{1}{2}$ for the spin $s=-\frac{1}{2}$ component (still applying the caveat of the previous section about the region of convergence of the integrals). We list the results here below for the example of $s=-\frac{1}{2}$ operators\footnote{For spin $+\frac{1}{2}$ we get the correct residue with $e^{+\frac{\theta_3}{2}}$, with all the appropriate $S$-matrix entries and all the prefactors very nicely arranged, except for a systematic $-1$ in front of the $S$-matrix inside the square bracket, which we do not entirely understand at the moment. We shall need to return to these expressions again in the future, when we will have a more clear idea of how the fermionic operators actually precisely look like. In fact all our formulas shall have to be eventually revisited in that light.}:
those pertaining to the field annihilating an $f$ satisfy
\begin{eqnarray}
\label{alla6}
&&-\frac{i}{2} \, \mbox{Res}_{\theta_1 \to \theta_2+i\pi}G_{f\bar{f}f}  =  \frac{\pi}{2} \, i {\bf \bar{C}}_f^{\bar{f}} \, n_4 \frac{1}{F(0)^2} \Big(1 - [S_{23}]_{\bar{f}f}^{f\bar{f}} \Big)e^{- \frac{\theta_3}{2}}, \nonumber\\
&&-\frac{i}{2} \, \mbox{Res}_{\theta_1 \to \theta_2+i\pi}G_{\bar{b}bf} = \frac{\pi}{2} \, i {\bf C}_{\bar{b}}^b \,n_4 \frac{1}{F(0)^2} \Big(1 - i[S_{23}]_{bf}^{fb} \Big)e^{- \frac{\theta_3}{2}},\nonumber\\
&&-\frac{i}{2} \, \mbox{Res}_{\theta_1 \to \theta_2+i\pi}G_{b\bar{b}f} =  \frac{\pi}{2} i  \, {\bf \bar{C}}_b^{\bar{b}} \, n_4 \frac{1}{F(0)^2} \Big(1 + i [S_{23}]_{\bar{b}f}^{f\bar{b}} \Big)e^{- \frac{\theta_3}{2}}, \nonumber\\
&&-\frac{i}{2} \, \mbox{Res}_{\theta_1 \to \theta_2+i\pi}G_{\bar{f}ff} = \frac{\pi}{2} i  \, {\bf C}_{\bar{f}}^f \,n_4 \frac{1}{F(0)^2} \Big(1 -[S_{23}]_{ff}^{ff} \Big)e^{- \frac{\theta_3}{2}},
\end{eqnarray}
while those pertaining to the field annihilating an $\bar{f}$ satisfy
\begin{eqnarray}
\label{alla7}
&&-\frac{i}{2} \, \mbox{Res}_{\theta_1 \to \theta_2+i\pi}G_{\bar{f}f\bar{f}}  =  -\frac{\pi}{2}  \, {\bf C}_{\bar{f}}^f \, n_3 \frac{1}{F(0)^2} \Big(1 - [S_{23}]_{f\bar{f}}^{\bar{f}f} \Big) e^{- \frac{\theta_3}{2}}, \nonumber\\
&&-\frac{i}{2} \, \mbox{Res}_{\theta_1 \to \theta_2+i\pi}G_{b\bar{b}\bar{f}} = -\frac{\pi}{2}   \, {\bf \bar{C}}_b^{\bar{b}} \,n_3 \frac{1}{F(0)^2} \Big(1 - i[S_{23}]_{\bar{b}\bar{f}}^{\bar{f}\bar{b}} \Big)e^{- \frac{\theta_3}{2}},\nonumber\\
&&-\frac{i}{2} \, \mbox{Res}_{\theta_1 \to \theta_2+i\pi}G_{\bar{b}b\bar{f}} =  -\frac{\pi}{2}  \, {\bf C}_{\bar{b}}^b \, n_3 \frac{1}{F(0)^2} \Big(1 +i[S_{23}]_{b\bar{f}}^{\bar{f}b} \Big)e^{- \frac{\theta_3}{2}}, \nonumber\\
&&-\frac{i}{2} \, \mbox{Res}_{\theta_1 \to \theta_2+i\pi}G_{f\bar{f}\bar{f}} = -\frac{\pi}{2}   \, {\bf \bar{C}}_f^{\bar{f}} \,n_3 \frac{1}{F(0)^2} \Big(1 -[S_{23}]_{\bar{f}\bar{f}}^{\bar{f}\bar{f}} \Big)e^{- \frac{\theta_3}{2}}.
\end{eqnarray}
The relative statistical factors $(-)^\sigma$ in (\ref{reso}) are again consistent with the values of the $S$-matrices at equal rapidities from (\ref{equ}), considering the left-most particle cycling through the corresponding operators: when cycling through the operator annihilating $f$, a state $b$ picks up a $-i$, $f$ picks up a $1$, $\bar{b}$ picks up an $i$ and $\bar{f}$ a $1$; likewise, when cycling through the operator annihilating $\bar{f}$, a state $b$ picks up an $i$, $f$ picks up a $1$, $\bar{b}$ picks up a $-i$ and $\bar{f}$ a $1$. The recursions we obtain are now
\begin{eqnarray}
\frac{\pi}{2} i \frac{n_4}{F(0)^2} = M_1, \qquad -\frac{\pi}{2}   \frac{n_3}{F(0)^2} = \bar{M}_1, 
\end{eqnarray} 
with the one-point functions of the operator annihilating $f$ with the state $f$ (equal to $M_1 e^{-\frac{\theta_3}{2}}$) and of the operator annihilating $\bar{f}$ with the state $\bar{f}$ (equal to $\bar{M}_1 e^{-\frac{\theta_3}{2}}$). We expect once again that, on the grounds of the $\rho \leftrightarrow \bar{\rho}$ symmetry, we shall have $n_4=i n_3$. 

We find it rather astonishing that the map of all the charge conjugation entries falls into the expected places, as they appear to be rather stringently dictated by the formulas. The same holds for the statistics factors, whose pattern appears to give a very strong hint that we are going in the right direction. We still believe that this will need to be experimentally verified in some way before we can put complete confidence in all the factors. It is nevertheless comforting that these factors, which came out of these involved multi-step derivations and seem at first to assemble slightly precariously, appear to be quite favourably conspiring in the final results.  

We conclude by recalling that we have described the three-particle form factors with the most involved stricture - those with one $C$ in the off-shell algebraic Bethe ansatz. Those states with no $C$ correspond to the pseudovacua (only $b$ or $\bar{f}$, for a total of of $2^3 =8$ combinations). These states can only scatter diagonally by inspecting the respective $S$-matrix entries - this means that they are eigenvectors of the $R$-matrix. Those with two $C$s (only $f$ or $\bar{b}$, another $8$ possible combinations) are not more complicated, in fact by symmetry they are of the same simplicity as the pseudo-vacua - one may just have chosen the highest instead of the lowest weigth, an although a choice needed to be made, the top or the bottom of the final-dimensional representation ought to be of an equal level of complication. These particular form factors are just proportional to $\prod_{1\leq i<j\leq 3}F(\theta_i-\theta_j)e^{s\sum_{i=1}^3 \theta_i}$ by some constants $N_{\alpha_1\alpha_2\alpha_3}$, with no integral (which is something that is carried along by the $C$ operator of the algebraic Bethe ansatz). These $16$ simpler form factors clearly never present a configuration which may be subject to the kinematical-singularity requirement (\ref{reso}). The permutation requirement works very simply according to the diagonal $S$-matrix processes. For instance, the first $8$ states have $S_{bb}^{bb}=S_{\bar{f}\bar{f}}^{\bar{f}\bar{f}}=1$, and $S_{b\bar{f}}^{\bar{f}b} =i = - S_{\bar{f}b}^{b\bar{f}}$, therefore the only constraints on the constants are
\begin{eqnarray}
\label{coin}
N_{bb\bar{f}} = i N_{b\bar{f}b}, \qquad N_{\bar{f}\bar{f}b} = i N_{\bar{f}b\bar{f}}, \qquad N_{b\bar{f}\bar{f}} = i N_{\bar{f}b\bar{f}}, \qquad N_{\bar{f}bb}=-iN_{b\bar{f}b},
\end{eqnarray}
from scattering the second and third space in $G_{bb\bar{f}}$ and $G_{\bar{f}\bar{f}b}$, and the first and second in $G_{b\bar{f}\bar{f}}$ and $G_{\bar{f}bb}$. There are in principle four more conditions, from scattering the first two spaces of $G_{b\bar{f}b}$ and $G_{\bar{f}b\bar{f}}$, and the second and third of $\bar{f}b\bar{f}$ and $G_{b\bar{f}b}$, but they turn out to coincide with (\ref{coin}). A completely analogous situation occurs for the other $8$ form factors, since $S_{ff}^{ff}=1=S_{\bar{b}\bar{b}}^{\bar{b}\bar{b}}$, and $S_{f\bar{b}}^{\bar{b}f} = -i = - S_{\bar{b}f}^{f\bar{b}}$. This produces the four independent relations
\begin{eqnarray}
N_{ff\bar{b}}=-i N_{f\bar{b}f}, \qquad N_{f\bar{b}f} = -i N_{\bar{b}ff}, \qquad  N_{f\bar{b}\bar{b}} = -i N_{\bar{b}f\bar{b}}, \qquad N_{\bar{b}\bar{b}f} = i N_{\bar{b}f\bar{b}}. 
\end{eqnarray}
Periodicity is again following the statistical factors from the $S$-matrix at equal rapidities, taking into account fermion-number conservation. These $16$ form factors join the $48$ more complicated which we have studies so far (including those without a kinematical-pole requirement), to give the necessary $4^3 = 64$ possible form factors.

%When comparing with \cite{Babu}, we need to take into account that the $S$-matrix for two particles satisfies the condition
%\begin{eqnarray}
%S_{\bar{b}b}^{b\bar{b}}(i \pi) =-i, \qquad S_{\bar{f}f}^{f\bar{f}}(i \pi) =-i, \qquad S_{\bar{b}b}^{f\bar{f}}(i \pi) =1, \qquad S_{\bar{f}f}^{b\bar{b}}(i \pi) =-1, 
%\end{eqnarray}
%as a result of the compensation of the zero occurring in the dressing factor and the hyperbolic cosine developing a zero at the denominator. This enters the splitting of lines in the residue formula and may give an additional factor of $i$ in the $fb$ contribution with respect to $bb$, as opposed to the situation in \cite{BabuF}. In our case it seems to combine with the $i$ expected from $S_{\bar{f}b}^{b\bar{f}}(0)=-i$ for the statistical factor in the second formula (see the discussion at the end of the previous subsection), hence we get a $-(-)^\sigma = - i \times (-i)$. 

%The relations one might argue from (\ref{alla}) between the three-point function with two particles and one anti-particle, and the one-point function $N_1$ of the particle $b$ (which is just a constant), for the bosonic operator annihilating $b$ is therefore
%\begin{eqnarray}
%2 \pi \, m_4 \frac{1}{F(0)^2} = N_1. 
%\end{eqnarray}

\section{\label{specola}Speculations on a general formula}

It is tempting to speculate that a general formula exists along the example of \cite{BabuF}. It is natural to postulate the form factor formula
\begin{eqnarray}
\label{gene}
&&G_{\alpha_1 ... \alpha_n}(\theta_1, ...,\theta_n) = N_{\vec{\alpha}}  \, \Sigma_{\vec{\theta}} \, \prod_{1\leq i<j\leq n}F_{IJ}(\theta_i-\theta_j)\times\\
&&\times \int_{\cal{C}} du_1 ... du_m \, e^{\big[\ell \sum_{i=1}^n \theta_i +\frac{s-n\ell}{m}\sum_{j=1}^m u_j\big]} \, \prod_{i=1}^n Z_{I\rho}(\theta_i-u) \Phi_{I\rho}(\theta_i - u) \, \tau(\vec{u}) \, p_{\vec{\theta},\vec{u}} \, \langle \Omega| \prod_{i=1}^m C(\vec{\theta}, u_i),\nonumber
\end{eqnarray}    
where the $m=0$ case is understood as having no integration at all (hence no issue with the $\frac{1}{m}$) in the exponent. We understand again that the $\alpha_1...\alpha_n$ component needs to be extracted from the bra-vector on the r.h.s. of (\ref{gene}) in much the same way as we showed in the three-particle case (in this sense (\ref{gene}) constains a slight abuse of notation). In this formula, we have as many integrations as we have $C$-operators from the algebraic Bethe ansatz, with $|\Omega\rangle$ being the $n$-particle pseudo-vacuum 
\begin{eqnarray}
|\Omega\rangle = |b\rangle \otimes |b\rangle \otimes ... \otimes |b\rangle.
\end{eqnarray}
Fermionic number conservation imposes odd $m$ for fermionic operators and even $m$ for bosonic operators, since each $C$ operator carries fermion number $1$. As discussed in the three-particle case, and as we will reiterate shortly with an example in the two-particle case, one could either use $C$ from mixed monodromy matrices and create states over all the pseudovacua (including anti-particles), or just always the same functions from one Bethe ansatz, say, over $|\Omega\rangle$, and utilise the fact that the $R$-matrix entries are always very similar, to adjust the formula correctly. Either way the procedure does become quite rapidly very cumbersome. The issue of convergence of the integrals with its perilous caveat follow along the same lines of the discussion of the three-particle case. 

By $F_{IJ}$, $Z_{I\rho}$ and $\Phi_{I\rho}$ we intend to recall that the auxiliary space is always thought of as being in the $\rho$ representation, but the state with rapidity $\theta_i$ carries a representation $I \in \{\rho,\bar{\rho}\}$. For the functions involved we know that this amounts to different constant factors. The function $\Sigma (\vec{\theta}) = \Sigma (\theta_1,...,\theta_n)$ will surely be needed to satisfy the axioms, however, given how non-trivial it has been to fix its three-particle case, we are not sure how easy it will be to obtain a closed formula for it. This is reminiscent of the standard situation \cite{Mussardo}, where a recursion for such functions of the symmetric polynomials is typically setup and in notable cases explicitly solved. Periodicity is also quite non-trivial and it should inform us about the various values of $\ell$.
 
The contour $\cal{C}$ is always formed of $n$ small clockwise circles around each point $u_j = \theta_i$, for all $j=1,...,m$ and $i=1,..,n$, plus an infinite horizontal part avoiding all poles and lying within $\mbox{Im}(u_j) \in (-2\pi,\pi)$, for all $j=1,...,m$. The extra ingredient which is not a straightforward generalisation of the three-particle formulas is the function
\begin{eqnarray}
\tau({\vec{u}}) = \prod_{1\leq j< j\leq m} \frac{1}{Z(u_i-u_j)Z(u_j-u_i)},
\end{eqnarray}
which is absent in the case one integration only, and is borrowed from the formula of \cite{BabuF} with all the $u_j$ being thought of as attached to an auxiliary $\rho$ representation. Likewise, the function $p_{\vec{\theta},\vec{u}}$, which will depend on the operator, signals the fact that the description of generic operators will necessarily call for extra polynomials in the  
exponentials of the rapidities $\theta_i$ and the $u_j$ variables, as described in \cite{Babu2} and reviewed very thoroughly in \cite{GrinzaPonsot}. Such functions will automatically weaken the convergence properties of the integral and incur in the caveat which we have discussed in the three-particle case. The vector notation is everywhere used to denote the collections of all the $\theta_i$ or, respectively, $u_j$. Once again the arrangements of $\alpha_1,...,\alpha_n$ are read as a row vector, which matches the row (bra-)vector on the right hand side.

Although there is a complete control over the algebraic Bethe ansatz \cite{DiegoBogdanAle}, the explicit expression for the off-shell Bethe bra-vector appearing in (\ref{gene}) quickly becomes extremely bulky with growing $n$. This is of course with the exception of the case $\alpha_i = b$ for all $i=1,..,n$, in which case there is no integration at all and the result is just the prefactor. The same can presumably be said of the case 
$\alpha_i = f$ for all $i=1,..,n$, which in (\ref{gene}) appears as $m=n$. As we already commented upon, we expect there to be a sort of duality whereby one might simply construct states with more fermions than bosons by starting from an alternative pseudovacuum with all $f$ (highest weight as opposed to lowest), and using the $B$ generator instead of $C$. This is why the complication is probably really maximal when $m$ is of the order of $\frac{n}{2}$. There is a chance that the free-fermion formalism developed in \cite{deLeeuw:2020bgo} could in principle be adjusted to provide explicit general formulas for the off-shell Bethe vectors. 

The integrand is always still meromorphic with only poles and zeros on the imaginary axes of the $u_j$, but figuring out where the poles actually lie which are not cancelled by zeros is going to be exceedingly laborious. Performing the integration is probably only conceivable numerically for a small number of particles and in general going to a safe region of complex $s$, from which to start ideally moving back towards the wanted values. Linking all the constants to one another as we have managed to do quite surprisingly in the three-particle case, with a combination of periodicity and residue constraints, appears rather daunting. Nevertheless, there is a merit in conjecturing the structure of the formula, because it was observed in \cite{GrinzaPonsot} that even in the massless case the contribution to correlation functions might just be dominated by the first few form factors with a small number of particles. There seems to be no theoretical justification behind this, as it is a well-known feature of massless integrable theories the one of having any suppression of the higher-particle contributions \cite{MI,Abbott:2020jaa}, the same way as in performing the TBA one cannot truncate and one has to proceed exactly \cite{DiegoBogdanAle}. However, the empyrical observations of \cite{GrinzaPonsot} give us hope that something similar (and similarly rather surprising) might happen here as well. Such expectation might be dampened in the end by the fact that we do not have a natural coupling constant in the game, and our theory is at a critical point with $c=6$ \cite{DiegoBogdanAle}. Moreover, we should not forget that on general grounds one expects infrared singularities when constructing the correlators
\begin{equation}
\langle {\cal{O}}(i R,0) {\cal{O}}(0,0) \rangle =  \sum_{n=1}^\infty \frac{1}{n!} \int_{-\infty}^\infty \frac{d \theta_1}{2 \pi} ... \int_{-\infty}^\infty \frac{d \theta_n}{2 \pi} \big\vert F^{\cal{O}}_{a_1,...,a_n}(\theta_1,...,\theta_n) \big\vert^2 e^{- \frac{R}{2} \sum_{i=1}^n e^{\theta_i}},\nonumber
\end{equation}
due to the massless dispersion relation \cite{DelfinoMussSimo}. These divergencies will have to be regularised to extract the space-time dependence of the correlation function. One might use a circle as a regulator and compute finite-volume form-factors\footnote{We thank Sergey Frolov for suggestions on this point.} \cite{finite}.

\subsection{\label{dufe}Fermionic two-particle form factor}

We can now apply the general formula to obtain the fermionic two-particle form factor. For this we need to apply one $C$ operator - this time extracted from
\begin{eqnarray}
{\cal{M}} \equiv \prod_{i=1}^2 R_{i0}(\theta_i - \theta_0) = A \otimes E_{11} + B \otimes E_{12} + C \otimes E_{21} + D \otimes E_{22},
\end{eqnarray}
to the bra-vector
\begin{eqnarray}
\langle \Omega| = \langle b| \otimes \langle b|.
\end{eqnarray}
We obtain 
\begin{eqnarray}
\label{perme}
\langle \Omega|C(u) = \frac{-\tanh \frac{\theta_1-u}{2}}{\cosh \frac{\theta_2-u}{2}} \langle b| \otimes \langle f| + \frac{1}{\cosh \frac{\theta_1-u}{2}} \langle f| \otimes \langle b| \equiv x_{12} \, \langle b| \otimes \langle f| + y_{12} \,  \langle f| \otimes \langle b| 
\end{eqnarray}
whence the formulas become
\begin{eqnarray}
\label{fermiFFF}
&&G_{bf}(\theta_1, \theta_2) = -N_2 F(\theta_1-\theta_2) \int_{C1;12} du \, e^{\big[\ell \sum_{i=1}^2 \theta_i + (s - 2 \ell) u\big]} \, \frac{\tanh \frac{\theta_1-u}{2}}{\cosh \frac{\theta_2-u}{2}}\prod_{i=1}^2 Z(\theta_i-u) \Phi(\theta_i - u),\nonumber\\
&&G_{fb}(\theta_1, \theta_2) = N_2 F(\theta_1-\theta_2) \int_{C2;12} du \, e^{\big[\ell \sum_{i=1}^2 \theta_i + (s - 2 \ell) u\big]} \, \frac{1}{\cosh \frac{\theta_1-u}{2}}\prod_{i=1}^2 Z(\theta_i-u) \Phi(\theta_i - u),
\end{eqnarray}
where we have used the fact that there are no constant factors in either purely $\rho$ or purely $\bar{\rho}$ form factors.
In this simple case the function $\Sigma$ can be set to $1$. Once again, the signs have been kept (and not reabsorbed in a redefinition of the constants $N$) to make it easier to identify exactly the integrand with the entries of the Bethe bra-vector. Our notation for the contour is the same as we have used in the three-particle case.
It is easy to verify the permutation requirement from
\begin{eqnarray}
&&x_{12} = x_{21} \, \mbox{sech} \frac{\theta_1-\theta_2}{2} - y_{21}\tanh \frac{\theta_1-\theta_2}{2},\nonumber\\
&&y_{12} = y_{21} \, \mbox{sech} \frac{\theta_1-\theta_2}{2} + x_{21} \, \tanh \frac{\theta_1-\theta_2}{2},
\end{eqnarray}
for all values of $u$ using (\ref{perme}), as befitting the $R$-matrix (\ref{eq:RLLlimtheta}). Boost invariance is built-in the form of the exponential function inside the integral, which picks up $2s\Lambda$ from $\theta_i \to \theta_i + \Lambda$ minus $s\Lambda$ from shifting $u$ (which restores the contour to its original place), hence a global $e^{s\Lambda}$ is produced. Periodicity is also proven the same way as the three-particle case, in fact by borrowing most of the argument. In particular, one can easily see that the singularities of the integrands are the same as the $\theta_1$ and $\theta_2$ towers of poles and zeros of $G_{bff}$ (for what concerns $G_{bf}$) and $G_{bfb}$ (for what concerns $F_{fb}$). The shifting of the contour and picking up of the poles also works the same way, and a little exercise shows that one ends up with matching contours in the same way as shown in the three-particle case. The residue requirement does not apply to the two-particle situation.

The other $14$ form factors, to make up the necessary $4^2=16$, proceed now in completely analogous fashion to the way we have seen in the three-particle case. We will have $4$ made only of $b$ and $\bar{f}$ (pseudovacua), $4$ made only of $f$ and $\bar{b}$ (at the other end of the representation), without the integral over $u$ and scattering diagonally. They are just proportional to $F(\theta_1-\theta_2)e^{s (\theta_1+\theta_2)}$. The remaining $6$ have the integral over $u$, and coincide pairwise with the basic ones in (\ref{fermiFFF}). The two expressions $G_{\bar{b}\bar{f}}$ and $G_{\bar{f}\bar{b}}$ are exactly the same as $G_{bf}$ and, respectively, $
G_{fb}$, owing to $\rho \leftrightarrow \bar{\rho}$ symmetry. The remaining $4$ read
\begin{eqnarray}
&&G_{\bar{f}f} (\theta_1,\theta_2) = -r_1 F_{\bar{\rho}\rho}(\theta_{12}) \int_{C1;12} du \, e^{\big[\ell_1 \sum_{i=1}^2 \theta_i + (s - 2 \ell_1) u\big]} \, \frac{\tanh \frac{\theta_1-u}{2}}{\cosh \frac{\theta_2-u}{2}}\times \nonumber\\
&& \qquad \qquad\qquad \qquad \qquad \qquad \qquad \qquad \times Z_{\bar{\rho}\rho}(\theta_1-u) \Phi_{\bar{\rho}\rho}(\theta_1 - u)Z_{\rho\rho}(\theta_2-u) \Phi_{\rho\rho}(\theta_2 - u),\nonumber\\
&&G_{b\bar{b}} (\theta_1,\theta_2)= r_1 F_{\rho\bar{\rho}}(\theta_{12}) \int_{C1;12} du \, e^{\big[\ell_1 \sum_{i=1}^2 \theta_i + (s - 2 \ell_1) u\big]} \, \frac{\tanh \frac{\theta_1-u}{2}}{\cosh \frac{\theta_2-u}{2}}\times \nonumber\\
&&\qquad \qquad\qquad \qquad \qquad \qquad \qquad \qquad \times  Z_{\rho\rho}(\theta_1-u) \Phi_{\rho\rho}(\theta_1 - u)Z_{\bar{\rho}\rho}(\theta_2-u) \Phi_{\bar{\rho}\rho}(\theta_2 - u),\nonumber\\
&&G_{\bar{b}b} (\theta_1,\theta_2) = r_1 F_{\bar{\rho}\rho}(\theta_{12}) \int_{C2;12} du \, e^{\big[\ell_1 \sum_{i=1}^2 \theta_i + (s - 2 \ell_1) u\big]} \, \frac{1}{\cosh \frac{\theta_1-u}{2}}\times \nonumber\\
&&\qquad \qquad\qquad \qquad \qquad \qquad \qquad \qquad \times  Z_{\bar{\rho}\rho}(\theta_1-u) \Phi_{\bar{\rho}\rho}(\theta_1 - u)Z_{\rho\rho}(\theta_2-u) \Phi_{\rho\rho}(\theta_2 - u),\nonumber\\
&&G_{f\bar{f}} (\theta_1,\theta_2)= r_1 F_{\rho\bar{\rho}}(\theta_{12}) \int_{C2;12} du \, e^{\big[\ell_1 \sum_{i=1}^2 \theta_i + (s - 2 \ell_1) u\big]} \, \frac{1}{\cosh \frac{\theta_1-u}{2}}\times \nonumber\\
&&\qquad \qquad\qquad \qquad \qquad \qquad \qquad \qquad \times  Z_{\rho\rho}(\theta_1-u) \Phi_{\rho\rho}(\theta_1 - u)Z_{\bar{\rho}\rho}(\theta_2-u) \Phi_{\bar{\rho}\rho}(\theta_2 - u),\nonumber\\
\end{eqnarray}
with $\theta_{12} \equiv \theta_1 - \theta_2$ and where $r_1$ is a constant. These four expressions are all linked together by the $S$-matrix permutation requirement.
These formulas have also been cross-checked with using directly a mixed-representation monodromy matrix and pseudo-vacuum to construct the Bethe bra-vector, namely when basing the algebraic Bethe ansatz upon
\begin{eqnarray}
|\Omega\rangle = |\bar{f}\rangle \otimes |b\rangle,
\end{eqnarray}
\begin{eqnarray}
{\cal{M}} \equiv R_{10}(\theta_1 - \theta_0)R_{c,20}(\theta_2-\theta_0) = A \otimes E_{11} + B \otimes E_{21} + C \otimes E_{12} + D \otimes E_{22},
\end{eqnarray} 
the auxiliary space $0$ being taken in the particle representation.

The periodicity constraints on the constants in front of the integrals do admit a solution which in principle determines all the $\ell$s. With respect to the three-particle case, it seems slightly less clear what a principle could be to identify the operators, therefore it is of no real use to discuss the solution here until one can have a better understanding of which operators are likely to be described by these form factors.

From observation of the two- and three-particle case it appears natural to postulate that the $n$-particle form factor with only one fermionic particle among all bosonic particles should be of the form
\begin{eqnarray}
\label{gene1}
&&G_{bb...bfb...bb}(\theta_1, ...,\theta_3) = N_{\alpha_1 ... \alpha_n} \, \Sigma_{\vec{\theta}} \,\prod_{1\leq i<j\leq n}F(\theta_i-\theta_j) \int_{\cal{C}} du\, \\
&&\qquad \qquad e^{-s\big[(n-1)u - \sum_{i=1}^n \theta_i\big]}  \, (-)^{n-m} \, \cosh \frac{\theta_m-u}{2}\prod_{i=1}^{m-1} \tanh\frac{\theta_i-u}{2}\prod_{i=1}^3 Z(\theta_i-u) \Phi(\theta_i - u) \, p_{\vec{\theta},u},\nonumber
\end{eqnarray}
where the $G_{bb...bfb...bb}$ has $n$ labels of which the $m$-th only is an $f$. Finally, let us comment that we have systematically taken the integral $u$ to be thought of as associated with a particle rather than an anti-particle. We have not been able so far to organise an explicit formula for the most generic combinations of states. In the spirit of the algebraic Bethe ansatz, where one expects that a difference choice of auxiliary-space representation corresponds to a mere rearrangements of the conserved charges, we do believe that the end result should have been the same had we everywhere chosen an antiparticle instead, and presumably even more colorful assortments. 

\section{\label{conclu}Conclusions and Future Perspectives}

In this paper we have attempted to provide an exact solution to the form factor programme applied to the situation of massless relativistic (BMN) $AdS_3$ scattering of \cite{DiegoBogdanAle}. Although exact solutions in this particular approach to $AdS/CFT$ form factor integrability are not frequent, the relative simplicity of the present situation allows for traditional methods to be employed, which we have done. We have found formulas which are inspired by \cite{BabuF} and satisfy the basic requirements. The most delicate axiom, involving the residue at the kinematical singularities, might not have to hold since we are in a massless and conformal situation strictly speaking without a massive theory to compare with, as the massive $AdS_3$ theory is non-relativistic. Nevertheless, the kinematical residue axiom seems to hold for our formulas quite non-trivially.

\subsection{Future Investigation}

There are several issues which are important for future investigation.

\subsubsection{Testing specific operators and truncation schemes}

It will be paramount for the future to establish a way to test these formulas. At the moment we only know the central charge of the theory, and any attempt to regain it from the correlation function of the stress-energy tensor is facing an uphill battle against the complication of the form factors. Not only that, but in the massless case we cannot in principle truncate over the number of particles as it is traditionally done in the massive case. However, the situation might not be so desperate, since truncation has been working in other context even in the massless case \cite{GrinzaPonsot}. This is something which we would like to pursue in the future. It is only when we will be confident that some experimental verification sustains our proposal beyond the checks of internal consistency we have performed here, that it will make sense to proceed to a systematic  phenomenology of all the form factors for an arbitrary number and assortment of particles and anti-particles. For this purpose, it will be intrumental to develop some description of the operators of the theory, which may either corroborate or force us to revise our considerations on the relative statistics, as they seemed to emerge from the analysis. A promising way forward could be to establish a possible connection with the formulas obtained for the norms and the scalar products of the Bethe states in \cite{JuanMiguelAle}, and attempt in this way a verification of our results.  

\subsubsection{Extension away from BMN using pseudo-relativistic invariance}

Another interesting question is the one of pseudo-relativistic invariance. Using the $\gamma$ variable of \cite{gamma1,gamma2}, it is possible to recast the entire non-relativistic massless left-left and right-right moving sector of the $AdS_3$ scattering theory in exactly identical terms as the BMN limit which we have analysed here, except for a different (non-relativistic) dispersion relation. Much of the formalism which we have constructed here should hold straight away, but integrating the form factors to obtain the correlation functions will be tough. To give an idea, let us take a sketch of the general formula for the two-point function of a scalar operator 
\begin{eqnarray}
&&\langle {\cal{O}}(i R,0) {\cal{O}}(0,0) \rangle \sim \nonumber\\
&&\qquad \qquad \qquad \sum_{n=1}^\infty \frac{1}{n!} \int_{-\infty}^\infty \frac{d \theta_1}{2 \pi} ... \int_{-\infty}^\infty \frac{d \theta_n}{2 \pi} F^{\cal{O}}_{\alpha_1...\alpha_n}(\theta_1,...,\theta_n) [F^{{\cal{O}},\alpha_1...\alpha_n}]^*(\theta_1,...,\theta_n) e^{- \frac{R}{2} \sum_{i=1}^n e^\theta_i},\nonumber
\end{eqnarray}
where reeated indices are summed over, and restrict it to the two-particle contribution, whose form factors depend only on the difference of the rapidities. For the purposes of this discussion, let us imagine that the sum over indices just reduces to a numerical factor in front, such that in the end we are left with computing 
\begin{equation}
\langle {\cal{O}}(i R,0) {\cal{O}}(0,0) \rangle \sim \int_{-\infty}^\infty \frac{d \theta_1}{2 \pi} \int_{-\infty}^\infty \frac{d \theta_2}{2 \pi} |F(\theta_1 - \theta_2)|^2 e^{- \frac{R}{2} (e^{\theta_1} + e^{\theta_2})},
\end{equation}
where $F$ is the function (\ref{pro}).
With a change of variables this can be recast into
\begin{equation}
\langle {\cal{O}}(i R,0) {\cal{O}}(0,0) \rangle \sim \int_{-\infty}^\infty \frac{d \theta}{2 \pi} \int_{-\infty}^\infty \frac{d \sigma}{2 \pi} |F(\theta)|^2 e^{- R e^{\frac{\sigma}{2}} \cosh \frac{\theta}{2}}.
\end{equation}
This integral is IR divergent, which comes from the fact that this is a massless theory. Regularising in a not particularly subtle way we obtain
\begin{equation}
\langle {\cal{O}}(i R,0) {\cal{O}}(0,0) \rangle \sim \int_{-\infty}^\infty \frac{d \theta}{2 \pi} \int_{-\mu}^\infty \frac{d \sigma}{2 \pi} |F(\theta)|^2 e^{- R e^{\frac{\sigma}{2}} \cosh \frac{\theta}{2}} = \int_{-\infty}^\infty \frac{d \theta}{2 \pi^2} |F(\theta)|^2 \, \Gamma(0,m R \cosh \frac{\theta}{2}),
\end{equation}
where
\begin{equation}
m \equiv e^{- \frac{\mu}{2}},
\end{equation}
and $\Gamma(0,x)$ is the incomplete gamma function. Notice that
\begin{equation}
\label{maybe}
\frac{\partial}{\partial R} \langle {\cal{O}}(i R,0) {\cal{O}}(0,0) \rangle \sim - \frac{1}{R}\int \frac{d \theta}{2 \pi^2} |F(\theta)|^2 \, e^{- m R \cosh \frac{\theta}{2}}.
\end{equation}
If we now we export this to the massless non-relativistic case, we should write
\begin{equation}
\label{noway}
\langle {\cal{O}}(i R,0) {\cal{O}}(0,0) \rangle \sim \int_{-\infty}^\infty \frac{d \gamma_1}{2 \pi} \int_{-\infty}^\infty \frac{d \gamma_2}{2 \pi} |F(\gamma_1 - \gamma_2)|^2 e^{- R h \Big(\frac{1}{\cosh \gamma_1} + \frac{1}{\cosh \gamma_2}\Big)},
\end{equation}
where $h$ is the coupling which has reappeared away from the BMN limit, and the massless non-relativistic dispersion relation with the sech function features in the exponent. If the hope of exactly evaluating (\ref{maybe}) is feeble, computing (\ref{noway}) appears like a significantly harder task. 

\subsubsection{Generalisation to massive particles}

It is clear in this connection that the natural question of how this could be transferred to the Zhukovski variables which one typically uses in massive $AdS/CFT$ integrability is a very difficult one if we were to strictly follow this route. So much of what we have done here depends on being able to manipulate meromorphic functions with no issues of branch cuts, and on the relative simplicity of the $S$-matrix entries. This is also why the hexagon approach is certainly more appropriate for the massive case, the massless situation ideally playing the role of a potential point of contact between the two \cite{Eden:2021xhe}. It is a very interesting question to try and connect our analysis with the hexagon approach initiated for $AdS_3$ in \cite{Eden:2021xhe}. A version of that procedure should ideally lead to a {\it conformal} hexagon, tailored to the BMN limit of the massless purely left-left and right-right moving $AdS_3$ $S$-matrix which we have been considering here. Although they unfold in a different directions, we believe that it should be possible to connect the two routes. We should like in the future to fix a general recursion for our formulas and possibly solve it, in order to assess how the two procedures might be related. 

\subsubsection{Convergence issues and analytic continuation}

An additional related issue is the one of the convergence of the integrals directly involved in the form factors expressions. By following the ideas of \cite{Babu2} we may adopt the analytic continuation in the spin parameter $s$ wherever necessary, in a spirit akin to the dimensional regularisation of Feynman integrals. Although similar operations do appear to be rather routinely, it will be very important in the future to carry this out explicitly in some simple examples, for instance along the lines of what is done in \cite{Palmai}. It will be necessary to see whether this prescription is likely to produce sensible results in our case as well, since it is possible that it may still produce divergencies when trying to analytically continue back. If such divergencies are present, one will have to subtract them in some way eventually. It would also be interested to study what other types of regularisation one might adopt.  

\subsubsection{Application to mixed-flux backgrounds}

A similar argument applies when trying to adapt our results to the relativistic theory developed for mixed fluxes \cite{gamma2}. We believe that it must be possible to repeat our analysis substituting the mixed-flux $R$-matrix, although it will probably require a good deal of work to include the general $k$-dependence. This  is because the functions become more complicated, and we do not have (discovered yet) all the nice properties of the (mixed-flux analog of the) Zamolodchikov dressing factor \cite{progress}. We are planning to study this problem in future work.

\subsubsection{Role of the Yangian charges and Faddeev-Zamolodchikov algebra}

A stimulating question would be to try and understand whether the higher non-abelian charges have any role to play. The Yangian \cite{Florian}, which in the case of this $S$-matrix acquires typical conformal traits, has very much remained in the back seat here. It would be very nice to investigate whether one can organise the formulas using the Yangian representation theory, which would give us some more confidence in their structural connections. It would equally be very nice to investigate the free-field representation of the Faddeev-Zamolodchikov algebra as analysed in \cite{Sergey} for the chiral Gross-Neveu model, and see whether this formalism can be applied to corroborate our results or otherwise. The presence of supersymmetry should make for a very interesting and convoluted problem which we are planning to take on in the future.

\subsubsection{Connection with the free-fermion formulation}

It would also be extremely interesting to see whether the free-fermion formalism developed for this and related models in \cite{deLeeuw:2020bgo} can really be pushed to reveal even more powerful consequences for the theory. We do wonder whether the existence of a free-fermion description, which in fact underlies all the $AdS_3$ matrices in particular, has any bearance on the correlation functions, as we believe that it should. We plan to attack this fascinating problem in the future.

\section*{\label{sec:Ackn}Acknowledgments}

We would like to thank Zoltan Bajnok, Marius de Leeuw, Sergey Frolov, Suvajit Majumder, Tristan McLoughlin, Fabrizio Nieri, Chiara Paletta, Anton Pribytok, Ana Retore, Bogdan Stefa\'nski and Istvan Sz\'ecs\'enyi for useful discussions. We thank Juan Miguel Nieto for discussions and for pointing out several typos in the first version of the manuscript. This work is supported by the EPSRC-SFI grant EP/S020888/1 {\it Solving Spins and Strings}. We thank the STFC for support under the Consolidated Grant project nr. ST/L000490/1. For the $83$rd birthday of my PhD supervisor Prof. Antonio Bassetto, unparallelled master of analytic functions. We thank the Galileo Galilei Institute for Theoretical Physics (GGI) for the hospitality and INFN for partial support within the programme {\it New Developments in $AdS_3/CFT_2$ Holography}, where this work started.

\section*{\label{sec:Data}Data management}

No data beyond those presented in this paper are needed to validate its results. 

\begin{appendix}

\section{\label{sec:Min3}Curious solution without kinematical singularities}

In this appendix we show how relaxing the requirement of kinematical singularities allows one to find an enormously simpler solution for the minimal form factors - with a gigantic residual freedom \cite{Mussardo}. We focus on a scalar operator for simplicity, but the same could be shown for a fermionic one. We also base this naive calculation on the bare statistics, since it is just for an exercise. 

After some manipulations, the permutation and periodicity equations for the 3-particle form factors for a scalar operator can be written in the following way. Due to the bosonic nature of the operator, only the form factors
\begin{eqnarray}
F^{\cal{O}}_{bff}(\theta_1,\theta_2,\theta_3), \qquad F^{\cal{O}}_{fbf}(\theta_1,\theta_2,\theta_3), \qquad F^{\cal{O}}_{ffb}(\theta_1,\theta_2,\theta_3), \qquad 
F^{\cal{O}}_{bbb}(\theta_1,\theta_2,\theta_3), \end{eqnarray} 
are non-zero. In all four cases we separate a ``dressing'' part $F_3$ and a ``matrix'' part $G'$
\begin{eqnarray}
F^{\cal{O}}_{\alpha_1 \alpha_2 \alpha_3}(\theta_1,\theta_2,\theta_3) \equiv F^{\cal{O}}_3(\theta_1,\theta_2,\theta_3) \, G'_{\alpha_1 \alpha_2 \alpha_3}(\theta_1,\theta_2,\theta_3),
\end{eqnarray}
respectively sensitive to the dressing factor and matrix part of the $S$-matrix. The dressing part can easily be determined based on the standard procedure:
\begin{eqnarray}
\label{diffo}
F^{\cal{O}}_3(\theta_1,\theta_2,\theta_3)  = \prod_{i<j} F(\theta_i - \theta_j),
\end{eqnarray}   
with $F$ given by (\ref{eff}).

The axioms for the function $G'_{bbb}(\theta_1,\theta_2,\theta_3)$ only depend on the $S$-matrix entry $S_{bb}^{bb}=1$, hence this form factor does not mix with the other three. Permutation implies that $G'_{bbb}(\theta_1,\theta_2,\theta_3)$ is symmetric under the exchange of rapidities $1$ and $2$, and also $2$ and $3$. Since these two moves generate the permutation group $S_3$, we conclude that  this function is completely symmetric in all three arguments. 

The axioms for the remaining three non-zero form factors, which mix among themselves, can be re-written here in the following equivalent form, using the special properties of the $R$-matrix \cite{JoakimAle}. Permutation reads equivalently
\begin{eqnarray}
&&G'_{bff}(\theta_2,\theta_1,\theta_3) = -G'_{fbf}(\theta_1,\theta_2,\theta_3) \, \tanh \frac{\theta_1 - \theta_2}{2} +G'_{bff}(\theta_1,\theta_2,\theta_3) \, \mbox{sech} \frac{\theta_1 - \theta_2}{2},\nonumber\\
&&G'_{fbf}(\theta_2,\theta_1,\theta_3) = G'_{bff}(\theta_1,\theta_2,\theta_3) \, \tanh \frac{\theta_1 - \theta_2}{2} +G'_{fbf}(\theta_1,\theta_2,\theta_3) \, \mbox{sech} \frac{\theta_1 - \theta_2}{2},\nonumber\\
&&G'_{ffb}(\theta_2,\theta_1,\theta_3) = G'_{ffb}(\theta_1,\theta_2,\theta_3),\qquad G'_{bff}(\theta_1,\theta_3,\theta_2) = G'_{bff}(\theta_1,\theta_2,\theta_3)\nonumber\\
&&G'_{fbf}(\theta_1,\theta_3,\theta_2) = -G'_{ffb}(\theta_1,\theta_2,\theta_3) \, \tanh \frac{\theta_2 - \theta_3}{2} +G'_{fbf}(\theta_1,\theta_2,\theta_3) \, \mbox{sech} \frac{\theta_2 - \theta_3}{2},\nonumber\\
&&G'_{ffb}(\theta_1,\theta_3,\theta_2) = G'_{fbf}(\theta_1,\theta_2,\theta_3) \, \tanh \frac{\theta_2 - \theta_3}{2} +G'_{ffb}(\theta_1,\theta_2,\theta_3) \, \mbox{sech} \frac{\theta_2 - \theta_3}{2},
\end{eqnarray} 
while periodicity implies
\begin{eqnarray}
\frac{G'_{bff}(\theta_1+2 i \pi,\theta_3,\theta_2)}{G'_{ffb}(\theta_2,\theta_3,\theta_1)}=1, \qquad  \frac{G'_{ffb}(\theta_1+2 i \pi,\theta_3,\theta_2)}{G'_{fbf}(\theta_2,\theta_3,\theta_1)} = \frac{G'_{fbf}(\theta_1+2 i \pi,\theta_3,\theta_2)}{G'_{bff}(\theta_2,\theta_3,\theta_1)} = -1.
\end{eqnarray}
By extensive experimentation we have managed to find a particularly curious solution to all these relations: 
%The procedure we have followed consists of first eliminating the functions $G'_{fbf}$ and $G'_{ffb}$ using the first two permutation relations, and then realise that all the remaining relations collapse to two single ones for the function $G'_{bff}$. One condition is clearly that $G'_{bff}(\theta_1,\theta_2,\theta_3)$ is symmetric under the exchange $\theta_2 \leftrightarrow \theta_3$, as this is one of the permutation relations. By using the fact that all the objects involved are purely functions of the differences of rapidities, one can define 
%\begin{eqnarray}
%G'_{bff}(\theta_1,\theta_2,\theta_3) \equiv g(\theta_1-\theta_2,\theta_1 -\theta_3),
%\end{eqnarray}
%whereby we conclude that $g(x,y)$ must be a symmetric function. We then switch to rational functions by introducing the variables
%\begin{eqnarray}
%s\equiv e^{\frac{\theta_1-\theta_2}{2}}, \qquad t\equiv e^{\frac{\theta_1-\theta_3}{2}},
%\end{eqnarray} 
%and redefine
%\begin{eqnarray}
%g(x,y)\equiv h(s,t).
%\end{eqnarray}
%The other condition then reads
%\begin{eqnarray}
%h(s,t) = \frac{(1+s^2)t^2 h(-s,t) - (1+t^2)s^2 h(s,-t)}{(s-t)(s+t)}. 
%\end{eqnarray} 
%\begin{eqnarray}
%h(s,t) = \frac{1+s^2 +t^2}{st}, 
%\end{eqnarray}
\begin{eqnarray}
&&G'_{bff}(\theta_1,\theta_2,\theta_3) = e^{-\frac{1}{2}(\theta_2+\theta_3)}\frac{N_{bff}(e^{\theta_1},e^{\theta_2},e^{\theta_3})}{D_{bff}(e^{\theta_1},e^{\theta_2},e^{\theta_3})}\nonumber\\
&&G'_{fbf}(\theta_1,\theta_2,\theta_3) = e^{-\frac{1}{2}(\theta_1+\theta_3)}\frac{N_{fbf}(e^{\theta_1},e^{\theta_2},e^{\theta_3})}{D_{fbf}(e^{\theta_1},e^{\theta_2},e^{\theta_3})},\nonumber\\
&&G'_{ffb}(\theta_1,\theta_2,\theta_3) = e^{-\frac{1}{2}(\theta_1+\theta_2)}\frac{N_{ffb}(e^{\theta_1},e^{\theta_2},e^{\theta_3})}{D_{ffb}(e^{\theta_1},e^{\theta_2},e^{\theta_3})},
\end{eqnarray}
where the $N(x,y,z)$'s and $D(x,y,z)$'s are arbitrary {\it completely symmetric} polynomials of their variables and are a freedom of the solution (cf. \cite{Mussardo}).

The natural guess extended from the 3-particle case works as well. We obtain
\begin{eqnarray}
&&G'_{bbff}(\theta_1,\theta_2,\theta_3,\theta_4) = e^{-\frac{1}{2}(\theta_3+\theta_4)}\frac{N_{bbff}(e^{\theta_1},e^{\theta_2},e^{\theta_3},e^{\theta_4})}{D_{bbff}(e^{\theta_1},e^{\theta_2},e^{\theta_3},e^{\theta_4})}\nonumber\\
&&G'_{bfbf}(\theta_1,\theta_2,\theta_3,\theta_4) = e^{-\frac{1}{2}(\theta_2+\theta_4)}\frac{N_{bfbf}(e^{\theta_1},e^{\theta_2},e^{\theta_3},e^{\theta_4})}{D_{bfbf}(e^{\theta_1},e^{\theta_2},e^{\theta_3},e^{\theta_4})},\nonumber\\
&&G'_{bffb}(\theta_1,\theta_2,\theta_3,\theta_4) = e^{-\frac{1}{2}(\theta_2+\theta_3)}\frac{N_{bffb}(e^{\theta_1},e^{\theta_2},e^{\theta_3},e^{\theta_4})}{D_{bffb}(e^{\theta_1},e^{\theta_2},e^{\theta_3},e^{\theta_4})},\nonumber\\
&&G'_{fbbf}(\theta_1,\theta_2,\theta_3,\theta_4) = e^{-\frac{1}{2}(\theta_1+\theta_4)}\frac{N_{fbbf}(e^{\theta_1},e^{\theta_2},e^{\theta_3},e^{\theta_4})}{D_{fbbf}(e^{\theta_1},e^{\theta_2},e^{\theta_3},e^{\theta_4})}\nonumber\\
&&G'_{fbfb}(\theta_1,\theta_2,\theta_3,\theta_4) = e^{-\frac{1}{2}(\theta_1+\theta_3)}\frac{N_{fbfb}(e^{\theta_1},e^{\theta_2},e^{\theta_3},e^{\theta_4})}{D_{fbfb}(e^{\theta_1},e^{\theta_2},e^{\theta_3},e^{\theta_4})},\nonumber\\
&&G'_{ffbb}(\theta_1,\theta_2,\theta_3,\theta_4) = e^{-\frac{1}{2}(\theta_1+\theta_2)}\frac{N_{ffbb}(e^{\theta_1},e^{\theta_2},e^{\theta_3},e^{\theta_4})}{D_{ffbb}(e^{\theta_1},e^{\theta_2},e^{\theta_3},e^{\theta_4})},
\end{eqnarray}
again for $N$ and $D$ all completely symmetric polynomials of the arguments. This brings us to postulate the $N$-particle case simply as
\begin{eqnarray}
\label{axio}
G'_{\alpha_1 ... \alpha_n}(\theta_1,...,\theta_n)=e^{-\frac{1}{2}\sum_{\alpha_j = f}\theta_j} \, \frac{N_{\alpha_1 ... \alpha_n}(e^{\theta_1},...,e^{\theta_n})}{D_{\alpha_1 ... \alpha_n}(e^{\theta_1},...,e^{\theta_n})},
\end{eqnarray}
where the functions $N$ and $D$ are {\it completely symmetric} polynomials of their variables.

We can now give an analytic proof of this conclusion. 

\begin{enumerate}

\item {\it Proof of Periodicity} 

The factors of the type $\frac{N}{D}$ are clearly cyclic invariant, and insensitive to the shift by $2 i \pi$ in $\theta_1$. The factor $e^{-\frac{1}{2}\sum_{\alpha_j = f}\theta_j}$ is also cyclic-invariant, since it contains a sum over all rapidities of the fermionic particles. The shift of $2 i \pi$ in $\theta_1$ causes it to pick up a minus sign only if $\alpha_1=f$, which is correct: since $G'$ is globally a boson - due to the bosonic nature of the operator we are considering - then $\alpha_2 + ... + \alpha_n$ is a fermion, therefore having $\alpha_1$ being a fermion automatically always produces a minus sign. This proves the periodicity axiom for (\ref{axio}).

\item {\it Proof of Permutation}

The factors $\frac{N}{D}$ are invariant under permutation of each pair of adjacent indices, therefore they always drop out of the relation. The remaining condition only involves two adjacent indices, therefore can be dealt with locally. Let consider two adjacent indices $(i,i+1)$, $i=1,..,n-1$. We have three cases:

\begin{enumerate}

\item {\it The two indices are either both bosons or both fermions}. In this case the proof is immediate, since either the rapidities are both absent from this factor, or both present, hence the factor is invariant under permutation, as the axiom requires in the case of $S_{bb}=S_{ff}=1$.

\item {\it The index $\alpha_i$ is a boson and $\alpha_{i+1}$ is a fermion}. In this case, the $S$-matrix condition applied to this factor simply reduces to 
\begin{eqnarray}
\label{same}
e^{-\frac{\theta_i}{2}} = - \tanh \frac{\theta_i - \theta_j}{2} \, e^{- \frac{\theta_i}{2}} + \mbox{sech} \frac{\theta_i - \theta_j}{2} \, e^{- \frac{\theta_j}{2}} 
\end{eqnarray} 
which is an identity.

\item {\it The index $\alpha_i$ is a fermion and $\alpha_{i+1}$ is a boson}. In this case, the $S$-matrix condition applied to this factor simply reduces to 
\begin{eqnarray}
e^{-\frac{\theta_j}{2}} = \tanh \frac{\theta_i - \theta_j}{2} \, e^{- \frac{\theta_j}{2}} + \mbox{sech} \frac{\theta_i - \theta_j}{2} \, e^{- \frac{\theta_i}{2}} 
\end{eqnarray} 
which is the same identity as (\ref{same}) with $i$ and $j$ interchanged.

\end{enumerate} 

This proves the permutation axiom for for (\ref{axio}).

\end{enumerate}

The change in (\ref{axio}) under a Lorentz boost is given by
\begin{eqnarray}
G' \to e^{-m\Lambda}, \qquad N_f = 2m, \qquad m \, \, \mbox{non-negative integer},
\end{eqnarray}
where $N_f$ is the {\it even} number of fermionic particles (even, since the form factor corresponds to a bosonic operator). 
%The change in the factor (\ref{fatto}) is
%\begin{eqnarray}
%e^{-k \Lambda}.
%\end{eqnarray}
This means that we should need to choose the degrees of the completely symmetric polynomials $N$ and $D$ as 
\begin{eqnarray}
\label{fro}
\mbox{deg}(N) - \mbox{deg}(D) = s+m,
\end{eqnarray}
which is an integer for any integer spin $s$, to obtain the desired result. 
%Therefore the improved formula
%\begin{eqnarray}
%\label{axiom}
%G'_{\alpha_1 ... \alpha_N}(\theta_1,...,\theta_N)=\Bigg[\sum_{i=1}^Ne^{\Big(\frac{N_f}{2}+s-2\Big) \theta_i}\Bigg]e^{-\frac{1}{2}\sum_{\alpha_j = f}\theta_j} \sum_{i<j=1}^N e^{\theta_i + \theta_j}, \qquad N_f \, \, \mbox{even}, \qquad \nonumber \alpha_i = (b,f),
%\end{eqnarray}
%with $N_f$ the even number of fermionic ($f$) particles, satisfies all the axioms for a bosonic operator of spin $s$. 
Clearly the dressing part (\ref{diffo}) of the form factor satisfies the Lorentz-boost axiom with no spin contribution, since it only depends on the pairwise differences of the rapidities.

This curious solution appears to be too simple to really have a chance of playing a role in the physics of the model, but we thought that it was worth to report it, since ultimately the issue of kinematical singularities at the conformal point remains subtle. One point which this solution highlights in our opinion is once more the rather remarkable functional elegance of the $R$-matrix (\ref{eq:RLLlimtheta}), which keeps revealing an almost surprising algebraic simplicity - such as something like (\ref{same}) - and yet a great richness \cite{deLeeuw:2020bgo}.   

\end{appendix}

\end{document}